\documentclass[a4paper,fleqn]{cas-sc}

\usepackage[authoryear]{natbib}
\nocite{*}
\usepackage{url}

\usepackage{enumitem}
\usepackage{float}
\usepackage{amsmath}
\usepackage{booktabs}
\usepackage{enumitem}
\usepackage{amsmath}
\usepackage{makecell}
\usepackage{multirow}
\usepackage{adjustbox}
\usepackage{booktabs}
\usepackage{graphicx}
\usepackage{float}
\usepackage{xcolor}
\usepackage{mathtools}
\usepackage{etoolbox}
\usepackage{setspace}
\usepackage{subcaption}
\usepackage{comment}
\usepackage{listings}
\usepackage{bbm}
\usepackage{siunitx}
\ifdefined\unit\else
  \NewDocumentCommand\unit{m}{\si{#1}}
\fi
\usepackage{tikz}

\usepackage{amssymb}
\usepackage{amsthm}
\usepackage{booktabs}
\usepackage{subcaption}
\usepackage{tikz}
\newcommand{\lon}{\operatorname{lon}}
\newcommand{\latt}{\operatorname{lat}}
\newcommand{\hav}{\operatorname{hav}}
\newcommand{\sog}{\operatorname{sog}}
\newcommand{\cog}{\operatorname{cog}}
\newcommand{\unix}{\operatorname{unix}}
\newcommand{\mmsi}{\operatorname{mmsi}}

\usepackage[ruled,vlined]{algorithm2e}
\SetKw{Or}{\hspace{\algoskipindent}\itshape or}
\SetKw{And}{\hspace{\algoskipindent}\itshape and}
\SetKwBlock{Condition}{}{}

\usepackage{url}
\newcommand{\overbar}[1]{\mkern 1.5mu\overline{\mkern-1.5mu#1\mkern-1.5mu}\mkern 1.5mu}
\usepackage{wasysym}

\usepackage{newfloat}
\usepackage{chngcntr}
\usepackage{etoolbox} 
\DeclareFloatingEnvironment[
    fileext=loc,
    listname={List of Code Listings},
    name=Code Listing,
    placement=tbhp,
    within=none, 
]{codelisting}
\counterwithin{codelisting}{section}

\renewcommand{\thesection}{\arabic{section}}

\preto{\appendix}{%
    \renewcommand{\thesection}{\Alph{section}} 
    \counterwithin{codelisting}{section} 
}

\def\tsc#1{\csdef{#1}{\textsc{\lowercase{#1}}\xspace}}
\tsc{WGM}
\tsc{QE}

\definecolor{frgreen}{HTML}{D5E8D4}
\definecolor{frpurple}{HTML}{E1D5E7}
\definecolor{frblue}{HTML}{DAE8FC}
\definecolor{fryellow}{HTML}{FFF2CC}

\definecolor{frgreenbo}{HTML}{82B366}
\definecolor{frpurplebo}{HTML}{9673A6}
\definecolor{frbluebo}{HTML}{6C8EBF}
\definecolor{fryellowbo}{HTML}{D6B656}

\DeclareRobustCommand\frbox[2]{\tikz[baseline={([yshift=-.6ex]current bounding box.center)}]{
    \filldraw[fill=#1,draw=#2, very thick] (0,0) rectangle (0.7,0.2);}}

\begin{document}
\let\WriteBookmarks\relax
\def\floatpagepagefraction{1}
\def\textpagefraction{.001}

\shorttitle{$\alpha$-method for AIS trajectory extraction}    

\shortauthors{Paulig, Okhrin (2024)}  

\title [mode = title]{An open-source framework for data-driven trajectory extraction from AIS data - the $\alpha$-method}



%

\author[1]{Niklas Paulig}[type=editor,linkedin=niklaspaulig0,
  orcid=0000-0002-0220-6702]

\cormark[1]


\ead{niklas.paulig@tu-dresden.de}



\affiliation[1]{organization={Dresden University of Technology},
            addressline={Chair of Econometrics and Statistics, esp. in the
            Transport Sector}, 
            city={Dresden},
            postcode={01187}, 
            state={Saxony},
            country={Germany}}

\affiliation[2]{organization={Center for Scalable Data Analytics and Artificial Intelligence},
            addressline={(ScaDS.AI)}, 
            city={Dresden/Leipzig}}

\author[1,2]{Ostap Okhrin}[]


\ead{ostap.okhrin@tu-dresden.de}




\cortext[1]{Corresponding author}



\begin{abstract}
    Ship trajectories from Automatic Identification System (AIS) messages are important in maritime safety, domain awareness, and algorithmic testing. Although the specifications for transmitting and receiving AIS messages are fixed, it is well known that technical inaccuracies and lacking seafarer compliance lead to severe data quality impairment. This paper proposes an adaptable, data-driven, maneuverability-dependent, $\alpha$-quantile-based framework for decoding, constructing, splitting, and assessing trajectories from raw AIS records to improve transparency in AIS data mining. Results indicate the proposed filtering algorithm robustly extracts clean, long, and uninterrupted trajectories for further processing. An open-source Python implementation of the framework is provided.
\end{abstract}



\begin{keywords}
AIS \sep data-driven \sep trajectory extraction \sep big data \sep open-source
\end{keywords}

\maketitle

\section{Introduction and background}\label{intro}

The emergence of the AIS represents a crucial advancement in maritime technology with implications for navigational safety and maritime domain awareness. The International Maritime Organization (IMO) played a crucial role in shaping the global adoption of AIS, incorporating mandatory requirements into the Safety of Life at Sea (SOLAS) regulation V/19. The required use of AIS systems for all ships of more than 300 gross tonnages on international voyages and cargo ships of more than 500 gross tonnages on national voyages became effective on 31 December 2004 \citep{imo-ais}.

Since then, a large group of researchers analyzed trajectories extracted from AIS records for various research domains such as anomaly detection \citep{pallotta2013vessel,zhang2016advanced,zhen2017maritime,rong2020data,wolsing2022anomaly}, collision avoidance and risk assessment \citep{mou2010study,silveira2013use,chen2018pattern,rong2022ship,zhang2023interpretable}, pattern classification \citep{herrero2019ais,sanchez2020arch,duan2022asemi,luo2023anew} or path planning \citep{he2019ship,xu2019use,gu2023improved,waltz2023}. 

Trajectories, however, cannot be constructed directly from raw AIS messages as they contain notable inconsistencies \citep{bailey2005training,harati2007automatic} such as erroneous positions, time, speed or course, Maritime Mobile Service Identity (MMSI) misuse or sharing or intentional transceiver deactivation, rendering them useless for further analysis. Trajectory extraction usually consists of two consecutive steps. First, filtering the raw data on an individual message level to eliminate erroneous messages. Second, the cleaned messages are combined into meaningful, coherent trajectories. A shared challenge for these steps lies in determining the diverse thresholds crucial for filtering individual messages or entire trajectories, especially since the range of many of the possible parameters to consider, like velocity, acceleration or turning rate depends on the maneuvering capabilities of individual ships, of which many different share the national and international waters.

\begin{figure}
    \centering
    \includegraphics[width=.8\textwidth]{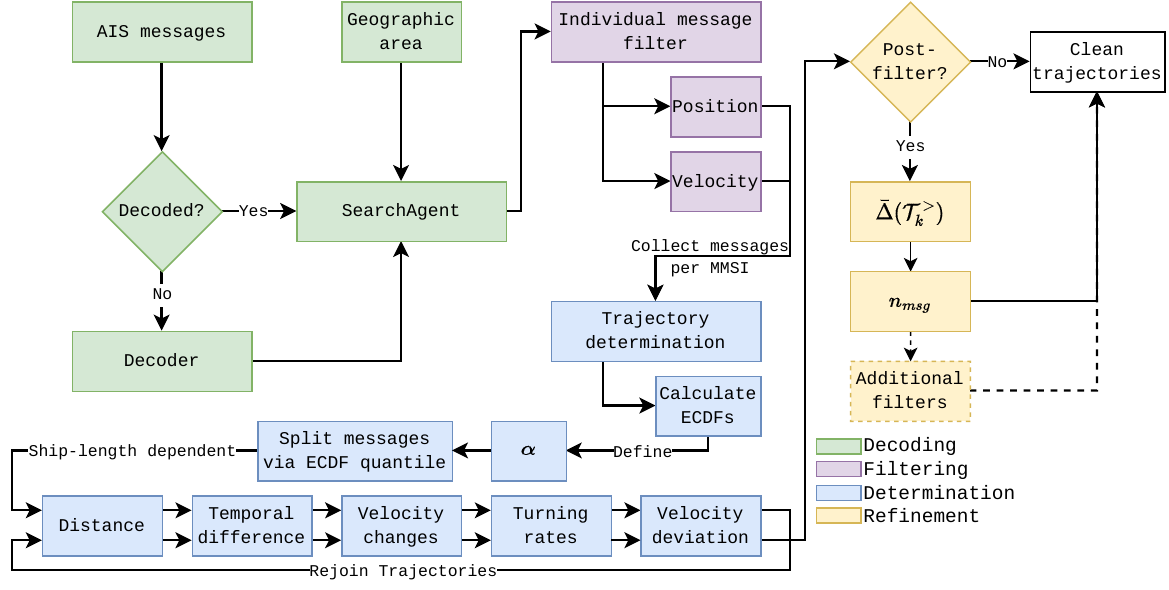}
    \caption{Flow chart diagram of the trajectory extraction framework presented in this article. Data collection and decoding is discussed in Section \ref{data-set}, Filtering in Section \ref{message-exclusion}, Determination in Section \ref{trex} and post-filtering in Section \ref{spatial-properties}.}
    \label{framework}
\end{figure}

In this paper, we introduce a framework (Figure \ref{framework}) for maneuverability-dependent trajectory extraction, characterized by its near-complete reliance on data-driven methodologies with minimal reliance on predefined threshold values. The framework aims to cover the entire process of trajectory extraction from raw AIS messages, beginning with the collection and decoding phase (marked with green in Figure \ref{framework}), MMSI grouping and individual message exclusion (marked with Purple), trajectory and sub-trajectory determination (Blue), and post-determination refinement (Yellow). In the following, relevant section headings will be marked with their corresponding framework color. In particular, our work contributes to the field as follows:

\begin{itemize}
    \item \emph{Generalizability:} Due to the pure data-driven approach, the framework is universally usable, regardless of the geographical region under consideration.
    \item \emph{Expanded methodology:} This article goes beyond the methodology of \citet{zhao2018ship}, which splits trajectories into sub-tracks based on velocity and time differences between individual messages by additionally considering turning rates, distances, and the maneuverability of the ship by using its length as a proxy.
    \item \emph{Trajectory assessment:} We introduce the concept of \emph{average absolute change of course}, which allows for flexible assessment of extracted trajectories in terms of maneuvering behavior.
    \item \emph{Implementation:} Our end-to-end framework is publicly available as an open-source package \citep{pytsa2024} for the Python programming language.
\end{itemize}

Central to our approach is utilizing the $\alpha$-quantiles of empirical distributions constructed from different features in the data, whose values will be used to distinguish coherent from incoherent trajectories. The parameter $\alpha$ can be adjusted to control the statistical power of the extracted trajectories.

To test the framework, a comprehensive analysis was conducted on a dataset spanning 912 days of AIS records from January 2020 to June 2022. The geographical scope of this investigation encompasses the North Sea and regions of the Baltic Sea, with the primary objective of determining robust methodologies for extracting long, uninterrupted, and clean trajectories from the dataset. By \emph{clean}, we understand the absence of erratic and highly volatile movement sequences of navigational parameters (SOG, COG, Position). Our experiment shows that the fully data-driven paradigm is suitable to effectively remove erroneous messages and construct meaningful trajectories out of the remaining messages.

The rest of the paper is organized as follows. Section \ref{related-work} recapitulates the different methodologies in trajectory extraction so far, Section \ref{ais} briefly revisits AIS records and their structure, and Section \ref{data-set} details the obtained data set and Section \ref{message-exclusion} explains the individual message exclusion processes. Applications and comparisons of the framework are conducted in Sections \ref{applications} and \ref{comparison}. Section \ref{trex} introduces the split-point procedure based on empirical distribution functions. Section \ref{spatial-properties} proposes tools for assessing spatial features of the extracted trajectories, while Section \ref{discussion} discusses the obtained results. Section \ref{conclusion} concludes. Additionally, Appendix \ref{implementation} provides exemplary usage of the accompanying Python package.

\section{Related works}
\label{related-work}

Concerning the methodology for trajectory extraction, a notable observation emerges. While a confluence of approaches exists, an agreed-upon standard within the scientific community is still absent.

Various procedures have been proposed to identify and eliminate abnormal individual AIS records. One filtering category applies to abnormal position reports, specifically those outside designated data areas. \citet{zhang2018novel}, \citet{sang2015novel}, \citet{yuan2019novel}, and \citet{chen2020ship} have all explicitly addressed this issue. Another set of rules revolves around abnormal velocities. Let $SOG$ be a message's speed over ground, then \citet{yuan2019novel} suggests to exclude messages outside of $2 kn < SOG < 20 kn$, while \citet{sang2015novel} drops the current message if the absolute difference between the last and the current velocity is outside the interval $[0.3,0.8]$. \citet{zhang2018novel} incorporates ship-type-specific maximum speeds, and \citet{chen2020ship} sets a threshold at 30 knots for velocity-based exclusion. Abnormal Course over Ground (COG) is addressed by \citet{yuan2019novel}, \citet{sang2015novel}, and \citet{chen2020ship}, each proposing unique criteria, such as valid COG ranges and comparisons with theoretical tactical diameters. Abnormal Rate of Turn is considered by \citet{zhang2018novel} and \citet{yuan2019novel}, with \citet{zhang2018novel} specifying a maximum rotational limit based on vessel characteristics. \citet{zhang2018novel} also introduces conditions for abnormal acceleration, restricting values within $[-1, 1]$ knots per second, while abnormal jerk, the derivative of acceleration, is constrained by \citet{zhang2018novel} to values beyond $\pm15 m/s^3$. Lastly, the issue of repeated data records is addressed by \citet{yuan2019novel}, albeit with unspecified criteria based on time judgments.

Following the initial filtration of individual messages, trajectories are typically constructed by sequentially linking messages emanating from the same source, identified by the MMSI \citep{mao2018automatic,zhao2018ship,yuan2019novel,guo2021improved,capobianco2021deep}. Subsequently, an additional layer of trajectory-based filtering is applied to the generated trajectories. Several criteria have been proposed for filtering and splitting trajectories to enhance the accuracy and granularity of maritime data. \citet{yuan2019novel}, \citet{zhao2018ship}, and \citet{mao2018automatic} have contributed specific conditions for these purposes, while \citet{yuan2019novel} suggests filtering trajectories based on the number of observations, proposing a threshold of 20 observations per track. \citet{zhao2018ship} introduces a more stringent criterion, recommending a minimum of 100 observations for a track to be considered, whereas \citet{mao2018automatic} sets a threshold of 500 observations for tracks where the $SOG$ is not equal to zero, thereby emphasizing the importance of trajectories during active movement. Additionally, \citet{mao2018automatic} proposes a condition to avoid high-complexity trajectories, employing the average cosine of the angles between three consecutive message points as an indicator, with a threshold set at $0.8$.

Moreover, the concept of splitting trajectories into sub-tracks is addressed by \citet{zhao2018ship}. This involves three key conditions: a speed change greater than 15 knots between consecutive messages, a time interval exceeding 10 minutes between messages, and a permissible deviation between reported and generated time within $[-5, 5]$ seconds, covering 96\% of the value range. These conditions collectively provide a comprehensive approach to delineating sub-tracks within longer trajectories, thereby extracting meaningful and discrete segments of vessel movement from AIS records. 

\section{AIS specifications}
\label{ais}

Over two decades ago, the \citet{itu2001} defined 27 different message types broadcast to serve many purposes, such as dynamic and static information about a voyage, safety-relevant information, or UTC inquiries. 

AIS broadcasts are sent from two types of transceivers (Class A and Class B), dependent on ship size, purpose, and local regulations; where Class A transceivers meet the performance standards and carriage requirements adopted by IMO, while Class B only partially fulfill them. Class A stations autonomously transmit their position (message types 1/2/3) every 2-10 seconds, depending on the vessel's speed and course changes, and at intervals of every three minutes or less when the vessel is anchored or moored. Additionally, these stations broadcast the vessel's static and voyage-related information (message type 5) every 6 minutes. 

Like Class A stations, Class B reports their position every three minutes or less when the vessel is anchored or moored. However, their position reports (messages 6 and 8) occur less frequently and with reduced power. Likewise, they transmit the vessel's static data (messages 18/24) every 6 minutes, excluding voyage-related information. Due to the limited range and reliability, Class B reports are responsible for the small minority of records collected by base stations.

\section{Raw data and decoding (\frbox{frgreen}{frgreenbo})}
\label{data-set}

\begin{figure}
    \centering
        \begin{tikzpicture}
            \node[anchor=south west,inner sep=0] (image) at (0,0) {\includegraphics[width=.4\textwidth]{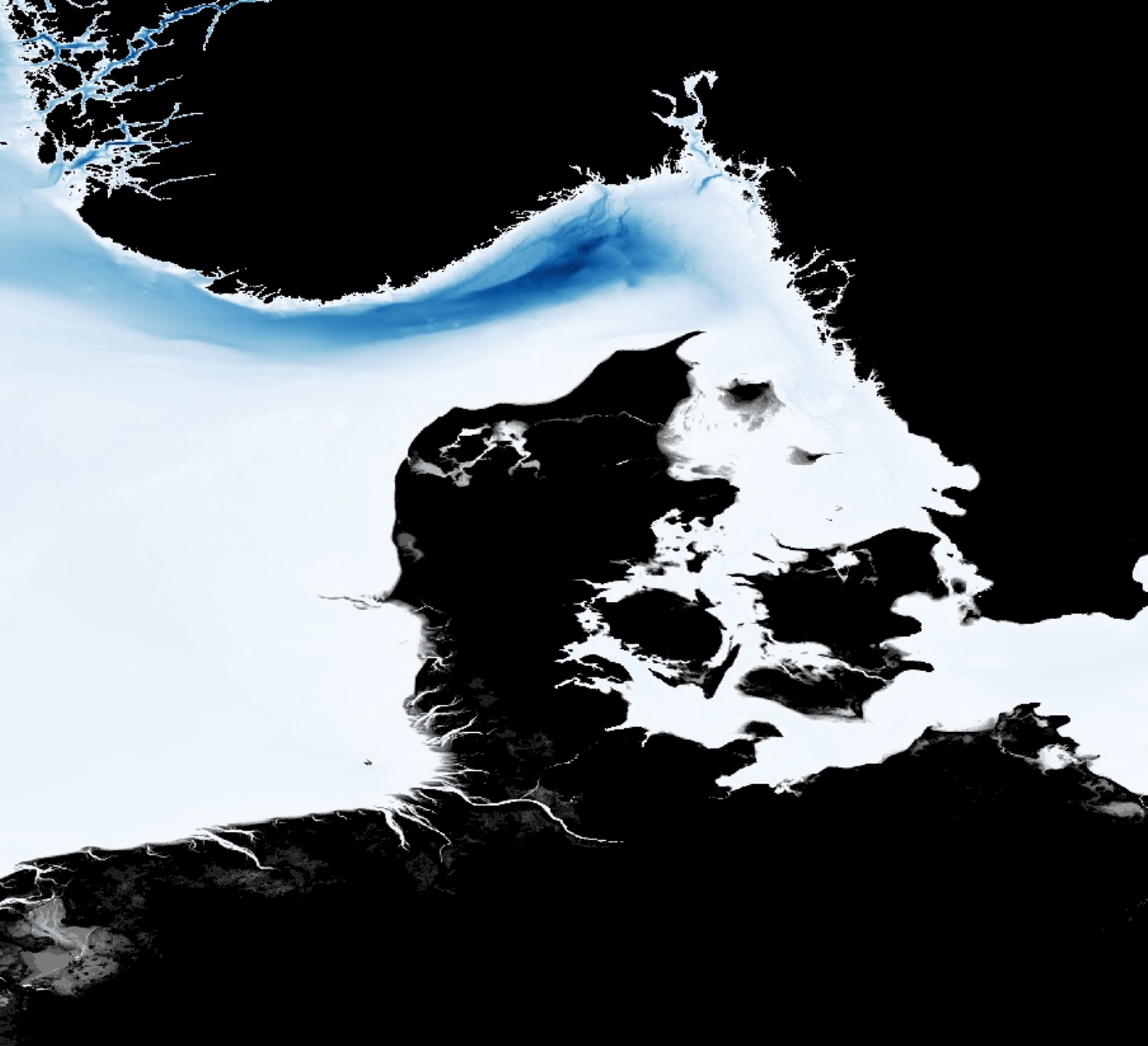}};
            \begin{scope}[x={(image.south east)},y={(image.north west)}]
                \node[anchor=south west] at (-0.4,-0.03) {{\footnotesize $51.85^{\circ}$N, $4.85^{\circ}$E}}; 
                \node[anchor=south east] at (1.4,-0.03) {{\footnotesize $51.85^{\circ}$N, $14.3^{\circ}$E}}; 
                \node[anchor=north west] at (-0.4,1.03) {{\footnotesize $60.49^{\circ}$N, $4.85^{\circ}$E}}; 
                \node[anchor=north east] at (1.4,1.03) {{\footnotesize $60.49^{\circ}$N, $14.3^{\circ}$E}}; 
            \end{scope}
        \end{tikzpicture}
    \caption{Geographical extent of the data set.}
    \label{area}
\end{figure}
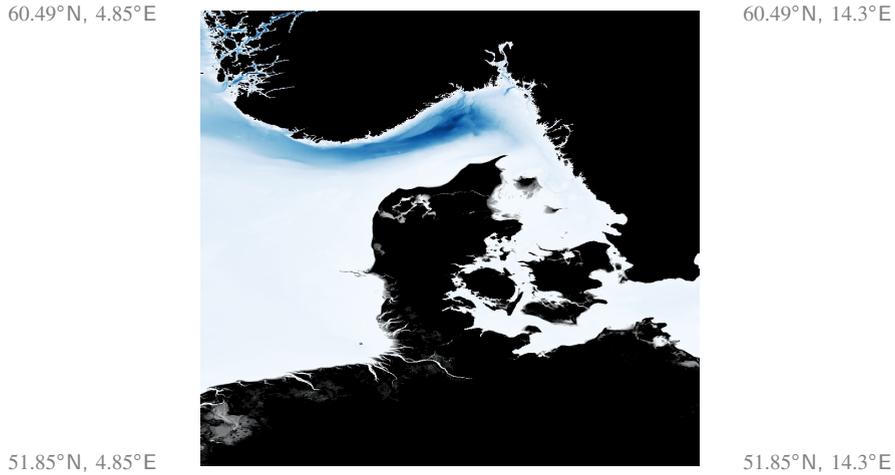

\begin{figure} 
\begin{subfigure}{.32\textwidth}
  \centering
  \includegraphics[width=\linewidth]{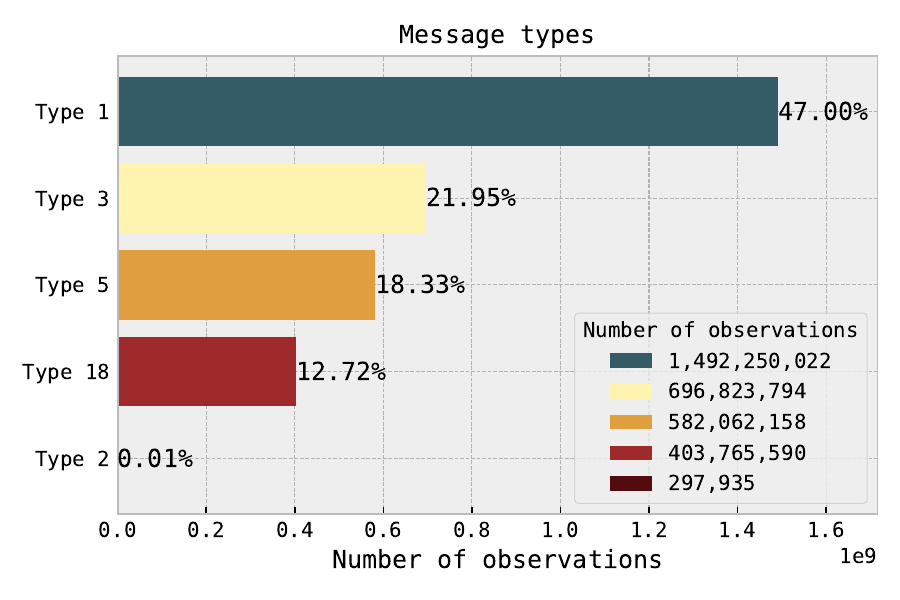}
  \caption{Count of messages by type.}
  \label{msgtype}
\end{subfigure}
\begin{subfigure}{.32\textwidth}
  \centering
  \includegraphics[width=\linewidth]{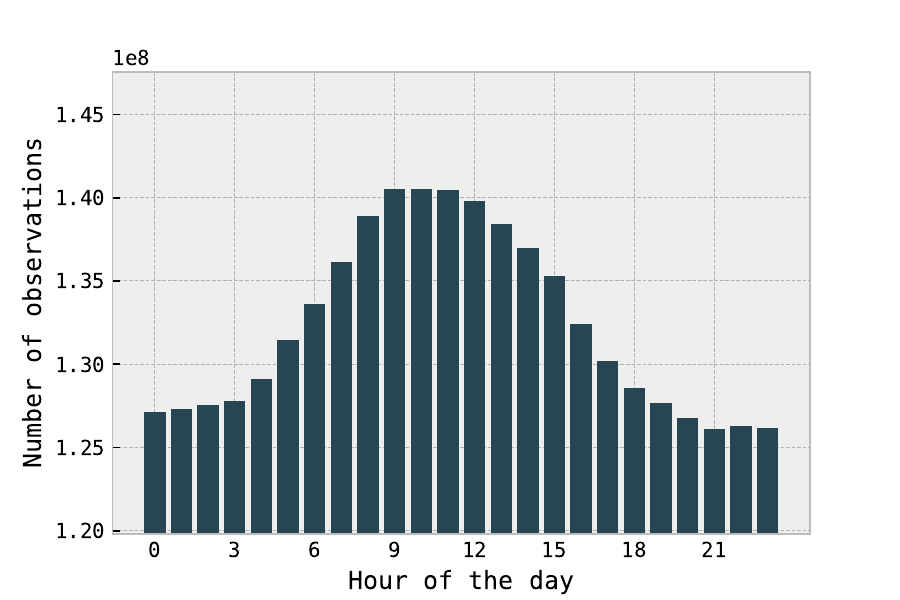}
  \caption{Absolute message count for each hour of the day, as calculated from receiving timestamps [UTC].}
  \label{msghour}
\end{subfigure}
\begin{subfigure}{.32\textwidth}
  \centering
  \includegraphics[width=\linewidth]{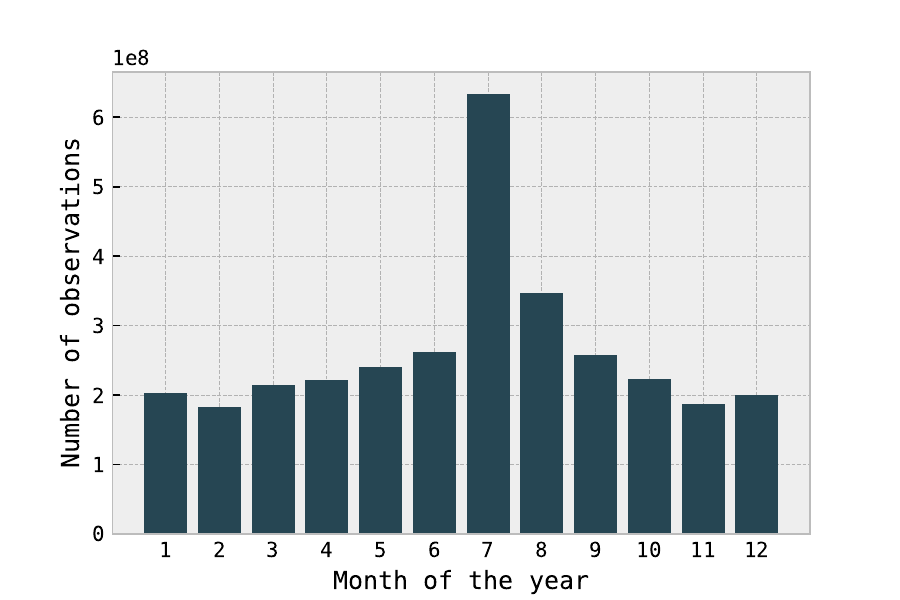}
  \caption{Absolute message counts per month of the year, as calculated from receiving timestamps.}
  \label{msgmonth}
\end{subfigure}
\caption{Descriptive statistics for the entire data set. Note that for the year 2022, the months from July until December are not part of the data set, and thus are not part of the absolute counts above.}
\label{descr}
\end{figure}

The AIS data used in this study has been provided by the European Maritime Safety Agency (EMSA) and spans the geographical coordinates from $51.85^{\circ}$N to $60.49^{\circ}$N latitude and $4.85^{\circ}$E to $14.3^{\circ}$E longitude (refer to Figure \ref{area}). Due to its geographical extent, the data set contains messages received by base stations of the following countries: Belgium, Denmark, Estonia, Finland, France, Germany, Iceland, Ireland, Latvia, Lithuania, Netherlands, Norway, Poland, Russia, Sweden, and the United States military bases in Europe. The temporal scope encompasses the period from January 1, 2020, to June 30, 2022, which leads to approximately 3.2 billion ($3\,175\,199\,499$) messages encoded as raw \texttt{AIVDM / AIVDO} sentences (refer to Table \ref{aismsg}) \citep{itu2001}, originating from 101\,166 unique MMSIs. Dynamic messages, including types 1/2/3/18, constitute around 81.7\% of the dataset, while the remaining messages pertain to static voyage reports of type 5, as presented in Figure \ref{msgtype}. 

Descriptively, intra-day message patterns demonstrate an area-specific and predictable ``workday'' trend, with a heightened message frequency observed during Coordinated Universal Time (UTC) daytime hours (see Figure \ref{msghour}). However, monthly messages present an atypical surge in message count during July (see Figure \ref{msgmonth}), whose origin remains unknown.

\begin{table}[]
    \centering
    \resizebox{\columnwidth}{!}{
    \begin{tabular}{lc}
        Message Type & AIVDM sentence(s)\\\midrule
        Type 1 & \texttt{!AIVDM,1,1,,,13`dUP0P000GqBjMw`im0wvD0000,0*01}\\
        Type 5 & \texttt{!AIVDM,2,1,7,B,539g?6T00000@8i6221HU=E8LU>222222222220j1h5334@P04hTQCADR,0*15}\\
          & \texttt{!AIVDM,2,2,7,B,0EQC`888888880,2*27}\\
    \end{tabular}}
    \caption{Example of raw AIVDM sentences from the data set for a type 1 and a type 5 AIS message. Note, that type 1 messages consist of a single sentence, while type 5 messages are transmitted in two distinct sentences. A sentence always starts with the introducer "\texttt{!AIVDM}".}
    \label{aismsg}
\end{table}

The data set was provided as one-day chunks in the \texttt{.csv} format, with messages of types 1/2/3/18 as one file and messages of type 5 in a separate file. The variables of the raw data set can be found in Table \ref{datavars}. 
\begin{table}[]
    \centering
    \begin{tabular}{lr}
        Variable name & Description \\\midrule
        \texttt{timestamp} & UTC timestamp of message reception\\
        \texttt{message\_id} & Message type\\
        \texttt{latitude} & Latitude of the message\\
        \texttt{longitude} & Longitude of the message\\
        \texttt{raw\_message} & \texttt{AIVDM} sentence\\
        \texttt{MMSI} & Maritime Mobile Service Identifier\\
        \texttt{originator} & Country abbreviation of the receiving base station
    \end{tabular}
    \caption{Variables of the raw data set.}
    \label{datavars}
\end{table}
The raw \texttt{AIVDM / AVIDO} sentences were decoded using the open-source \texttt{pyais} \citep{pyais} Python package according to IMO regulations \citep{IMO2010SN1Circ289}. In the decoding process of the AIS messages, an initial filtering stage was implemented to address the issue of duplicate data. This phenomenon arises when a vessel's AIS report is within the coverage area of multiple base stations, leading to the same message being received from different locations. To mitigate this, any messages that were identical in content and received from more than one base station within a time frame shorter than the AIS system's minimum sending frequency $f_{min} = 2s$ were excluded. 

\section{Individual message exclusion (\frbox{frpurple}{frpurplebo})}
\label{message-exclusion}

\subsection{Position reports}
In line with methodologies outlined by \citet{zhang2018novel}, \citet{sang2015novel}, \citet{yuan2019novel}, and \citet{chen2020ship}, we systematically exclude messages with geospatial coordinates falling outside the defined spatial region of the data set. This process also conveniently addresses the issue of erroneous positional data. By setting logical boundaries for latitude ($\latt$) and longitude ($\lon$) -- specifically, $51.85^{\circ} \leq \latt \leq 60.49^{\circ}$ and $4.85^{\circ} \leq \lon \leq 14.3^{\circ}$ -- any data points falling outside these ranges are automatically discarded. Notably, the data set sourced from the EMSA had already undergone this preliminary filtering step; therefore, applying this spatial boundary criterion resulted in removing zero additional messages from the data set.

\subsection{Velocity reports}
\label{velreps}
\begin{figure}[t]
        \centering
        \includegraphics[width=.45\textwidth]{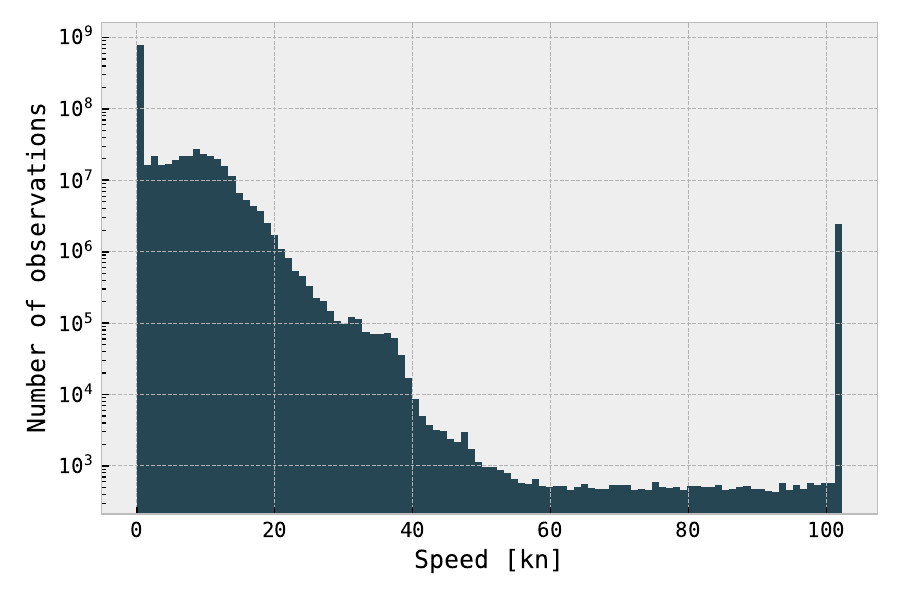}
        \captionof{figure}{Histogram of speeds for all messages received in 2021 ($n \approx 1.07\times 10^9$). Note that the ordinate is log-transformed to improve visualization due to the large range of counts.}
        \label{speedhist}
    \end{figure}

The data feature an extensive range of reported SOGs, of which only a fraction is relevant for trajectory extraction. The analysis of the SOG data from 2021, as presented in Figure \ref{speedhist}, indicates a notable number of observations registering a speed of approximately 102 knots. This observation aligns with established standards in the AIS protocol, in which a reported speed of 102.3 knots is conventionally used to signify that actual speed data is unavailable. This value does not represent an actual speed measurement but serves as a placeholder in the data set when speed information cannot be provided or is not applicable.

Figure \ref{speedhist} also reveals that a small fraction of messages ($\approx0.3\%$) report velocities exceeding 30 knots. This observation is significant given the economic ramifications of high-speed maritime travel, such as elevated fuel consumption and accelerated attrition of materials. As also documented by \citet{notteboom2009fuel}, commercial vessels rarely surpass the 30-knot threshold due to these constraints. Primarily, only specific categories of ships, such as high-speed crafts and certain military vessels, regularly operate beyond this speed limit. 

In light of these insights, in concordance with \citet{chen2020ship}, this study opts to exclude all messages reporting speeds above 30 knots, thus keeping 99.7\% of the data. This decision is grounded in the understanding that such high speeds are atypical for most commercial maritime traffic and, thus, may not represent typical operational behaviors and could potentially stem from errors in the ship's transceiving equipment.

    \begin{figure}[t]
        \centering
        \includegraphics[width=.45\textwidth]{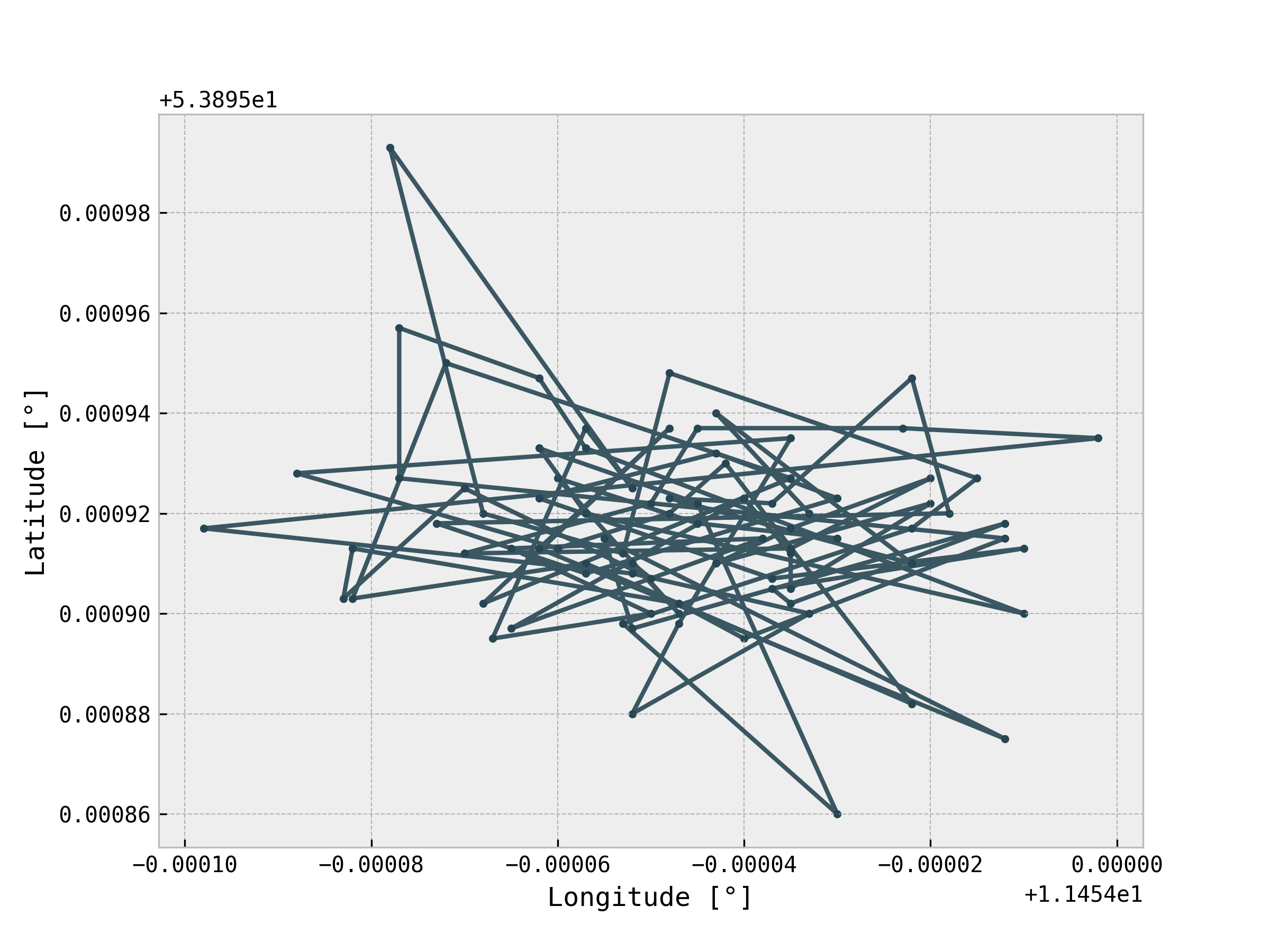}
        \captionof{figure}{Trajectory of the vessel with MMSI 211XXXX90 laying at anchor in the harbor of Wismar, Germany.}
        \label{jiggle}
    \end{figure}

A significant proportion of the recorded messages ($>71\%$) indicated speeds below one knot. This prevalence of low-speed signals in the data set can be attributed to several factors, the most notable of which is the inclusion of non-moving, i.e., anchored or moored vessels. Slow-moving construction vessels, which are mandated to transmit AIS records regardless of their motion status, also contribute to these messages. For instance, Figure \ref{jiggle} illustrates an instance of a vessel at anchor with its AIS transceiver operational. While minimal, the motion of this vessel is still captured and transmitted as part of the AIS data. Such behavior, characterized by minimal or no actual navigational movement, should be considered for exclusion from data analyses focusing on vessel trajectories. For this reason, in this study, we remove messages whose SOG is less than one knot.

\section{Trajectory construction (\frbox{frblue}{frbluebo})}
\label{trex}
After the preliminary cleaning, the next step is to assemble the remaining AIS records into trajectories using a two-step approach. In the first step, we identify potential \emph{split-points} within continuous vessel trajectories, defined as pairs of consecutive messages that appear to be more appropriately classified as belonging to separate trajectories. The separation criteria are based on empirical quantiles of several metrics detailed below. In the second step, the concept of \emph{absolute change of course} of a trajectory is introduced to provide a flexible examination tool of the split trajectories.

A series of definitions are introduced below to facilitate comprehension and minimize confusion in the subsequent sections of this paper.

\subsection{Definitions}

Let $m$ represent a single AIS message. Define the set $\mathcal{T}_{k} = \{m \,|\, \mmsi(m) = k\}$ as the collection of all messages from a vessel with MMSI being equal to $k$, where $\mmsi(m)$ is a mapping of the AIS message to its corresponding MMSI. For this study, we focus on a vessel's trajectory, i.e., ordered time-position tuples represented by the ordered set:
\begin{equation}
    \mathcal{T}_k^> = \{m_1,m_2,\dots,m_p\}, \, \unix(m_1) \leq \unix(m_2) \leq \dots \leq \unix(m_p), \, \forall m\in \mathcal{T}_k.
\end{equation}
This set consists of the vessel's transmitted messages, ordered by their UNIX timestamps in seconds, extracted from the messages via the $\unix(m)$ function. Furthermore, define $\latt(m)$, $\lon(m)$, and $\sog(m)$ functions returning latitude, longitude, and SOG, respectively, from a message $m$. 

Some analyses require the computation of velocities obtained from spatial and temporal data. Given the Earth's oblate spheroid geometry, distance estimation between geographical coordinates cannot be performed using Euclidean geometry. Predominantly, two methods are employed: the Haversine formula and the \cite{vincenty1975direct} algorithm, offering superior precision \citep{mahmoud2016shortest}. However, for the objectives of this study, such a level of accuracy is unnecessary, predominantly due to the significantly larger error margins inherent in original GPS data \citep{jankowski2021determination}. In addition, the Haversine formula's computational efficiency is notably higher, making it a more pragmatic choice. It is defined as 
\begin{equation}
    \hav(m_1,m_2) = 2r \arcsin\left[\sqrt{\sin^2\left(\frac{\Delta_\phi}{2}\right) + \cos\{\latt(m_1)\}\cos\{\latt(m_2)\} \sin^2\left(\frac{\Delta_\lambda}{2}\right)}\right],
\end{equation}
with lateral and longitudinal differences $\Delta_\phi = \latt(m_2)-\latt(m_1)$, $\Delta_\lambda = \lon(m_2)-\lon(m_1)$, and $r = 6371 km$ being earth's volumetric radius. The formulas' output is a distance in meters, while latitude and longitude are provided in radians.

Furthermore, let 
\begin{equation}
    \overbar{SOG}_{m_i}^{m_{i+1}}=\frac{\sog(m_i)+\sog(m_{i+1})}{2},
\end{equation}
be the average reported speed over the ground for two consecutive messages $\sog(m_i)$ and $\sog(m_{i+1})$, and 
\begin{equation}
    \widehat{SOG}_{m_i}^{m_{i+1}}=\frac{\hav(m_i,m_{i+1})}{\unix(m_{i+1})-\unix(m_i)} \cdot \frac{1852}{3600},
\end{equation}
be the estimated SOG in knots calculated from the change in position over time. The scalar constant $1852/3600$ is the conversion factor that translates $ms^{-1}$ to $kn$.

\subsection{Trajectory splitting}
\label{splitpoints}
\begin{figure}
    \centering
    \includegraphics[width=\textwidth]{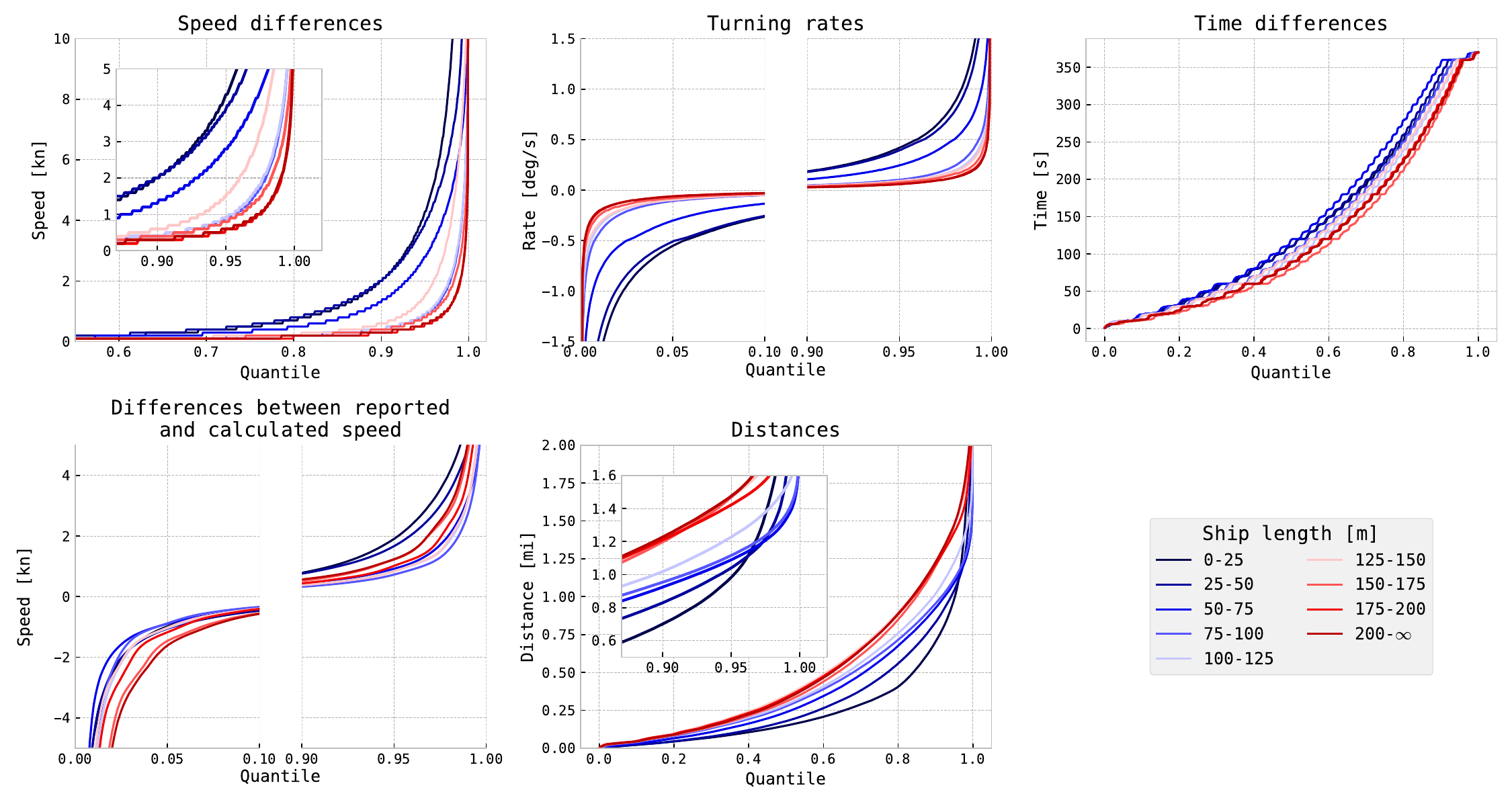}
    \caption{Length-dependent quantile values for the five different metrics used to determine split points in the constructed trajectories. Note that the time difference quantiles are not used ship-length dependent, and are only displayed for completeness.}
    \label{allquants}
\end{figure}

Consistent with the abovementioned research, we use the MMSI to assign AIS messages to their originating vessels and order these messages using the UTC timestamp to formulate a trajectory. Nevertheless, this methodology exhibits significant shortcomings without further processing, as it often results in trajectories with numerous undesirable characteristics, which will be outlined in this section. Accordingly, this study advances the trajectory segmentation methodology proposed by \citet{zhao2018ship}, applying it to ascertain the continuity of a trajectory by evaluating whether two successive AIS messages should be attributed to the same navigational path via a \emph{split-point} approach. Therefore, $\alpha$-quantiles of various distributions, detailed in the next Sections, are computed to decide whether to split the trajectory at a particular point. Additionally, we incorporate the ideas from \citet{zhang2018novel}, and \citet{chen2020ship} by explicitly taking the vessel maneuverability into account, by using the ship length as a proxy. Preliminary testing concluded that other proxies such as the ship type work worse, as there are ship types such as \texttt{CARGO} spanning the entire range of ship lengths and widths, therefore not representing maneuverability accurately. Ship mass, while being a promising proxy, was also discarded, as critical information for its calculation are either unreliable in the data (draft) or not part of it (block coefficient). Furthermore, \citet{hansen2022autonomous} related ship lentgths to thier heading change per meter and concluded that ship length is a valid proxy for maneuverability. To get a useful segmentation of the length, we introduce 9 distinct bins $[0,25,50,75,100,125,150,175,200,\infty] m$. We acknowledge, that these boundaries have no physical justification, but are a practical tradeoff between accuracy and computational complexity. Empirical distributions are then calculated for each of the bins. The procedure is consolidated in Algorithm \ref{algo:spp} and a visual representation of the length-dependent quantile values can be found in Figure \ref{allquants}. Throughout the next subsections, it is assumed that if a trajectory is split multiple times such that any of the sub-trajectories only consists of one message, this sub-trajectory is discarded. 

\subsubsection{Temporal difference}
\label{par:tempdiffs}
\begin{figure}
    \centering
    \includegraphics[width=\textwidth]{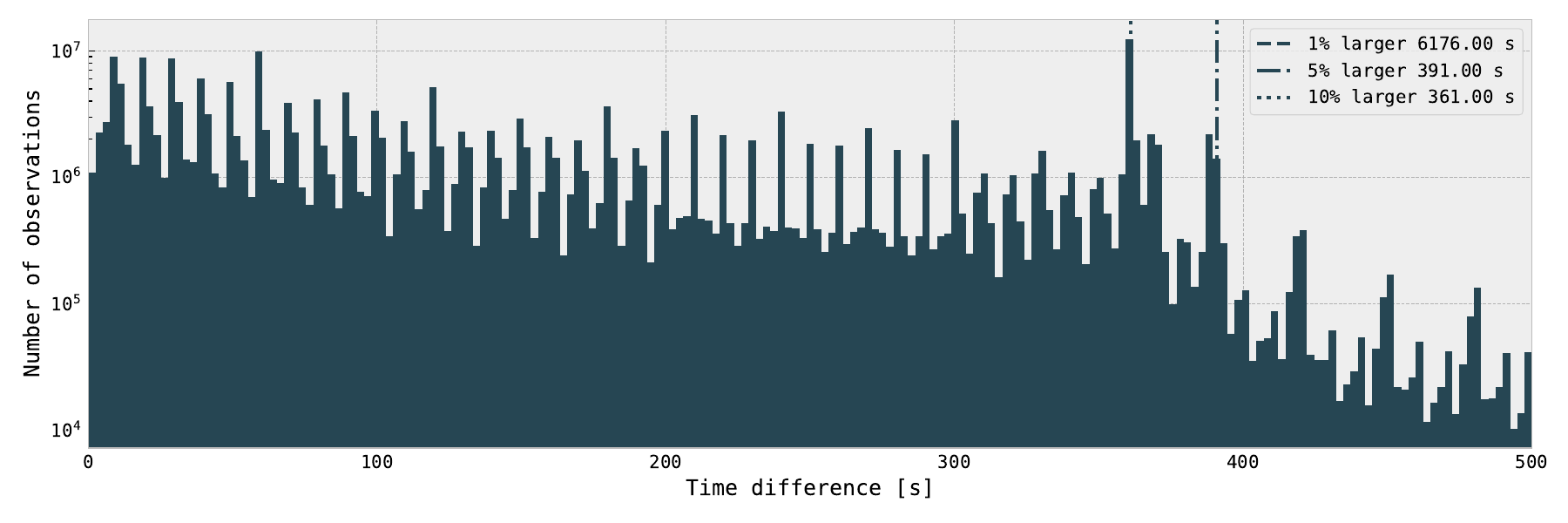}
    \caption{Histogram of the gap size for two consecutive dynamic messages in one trajectory. Gap sizes larger than 500s are truncated for better visualization. The histogram was built with speed-filtered data.}
    \label{gap-size}
\end{figure}

The most natural metric to investigate for split points is the temporal difference between two successive messages of one trajectory $\unix(m_{i+1}) - \unix(m_{i})$. Revisiting the AIS specifications, the defined transmission interval for dynamic AIS messages ranges from two to ten seconds for class A equipment ($2-30$ seconds for class B equipment) for vessels not at anchor and 180 seconds for vessels at anchor. Therefore, it is expected to find an accumulation of temporal gaps around these thresholds in the data, which could then be used to split trajectories. A histogram of the temporal differences between two successive messages inside each trajectory for 2021 is depicted in Figure \ref{gap-size} to verify the assumptions. A brief discussion on why one year of data was used in the calculation is provided in Appendix \ref{timescale}. It is noteworthy that we abstained from determining the distribution for each length interval, as the temporal distribution is not influenced by the ship's length but rather by technical or human factors such as delay or intent. The distribution clearly shows that our initial assumptions are untenable, as temporal differences vary significantly in the interval from $[2,360]s$ with an exceptional peak around the $360s$ mark, whose origin remains undisclosed. The only message types in the AIS specification transmitted at $360s$ intervals are safety-relevant messages, none in the data set, and static voyage reports, which had not been considered.

The findings of this study suggest that it is not advisable to segment vessel trajectories based on the transmission intervals prescribed by AIS standards. Instead the calculated values from the distribution of Figure \ref{gap-size} are used, depending on $\alpha$.

To determine all following empirical distribution functions, we only consider the value of the respective metric if the two messages under consideration are less than the 95\%-percentile of temporal gaps, namely $391s$, apart. We do this to explicitly exclude artificial outliers that are so far apart temporally that the significance of the obtained values is questionable.

\subsubsection{Velocity changes}
\label{par:vel-changes}
Given the defined transmission interval for dynamic AIS messages, the variance in reported speeds between two adjacent records is anticipated to be reasonably small. This expectation is based on the premise that substantial acceleration is atypical for commercial vessels. Therefore, a threshold value must determine whether two consecutive messages belong to the same trajectory.

Following the procedure from the preceding paragraph, the absolute speed change distribution for all vessel length intervals for 2021 was obtained. The results are illustrated in Table \ref{speed-quantiles} and show a clear pattern of receding speed and thus acceleration values for increasing ship lengths, reflecting the different maneuvering capabilities of different ship types.

\begin{table}[]
    \centering
    \resizebox{\textwidth}{!}{%
    \begin{tabular}{l|rlrlrll|rlrlrll|rlrlrl}
        \toprule
    \multirow{2}{*}{Length bin} & \multicolumn{5}{c}{$\alpha$} & \multirow{2}{*}{\begin{tabular}[c]{@{}l@{}}Accel. range\\ at $\alpha = 0.05$\end{tabular}} & \multirow{2}{*}{Length bin} & \multicolumn{5}{c}{$\alpha$} & \multirow{2}{*}{\begin{tabular}[c]{@{}l@{}}Accel. range\\ at $\alpha = 0.05$\end{tabular}} & \multirow{2}{*}{Length bin} & \multicolumn{5}{c}{$\alpha$} & \multirow{2}{*}{\begin{tabular}[c]{@{}l@{}}Accel. range\\ at $\alpha = 0.05$\end{tabular}} \\ \cline{2-6} \cline{9-13} \cline{16-20}
     & \multicolumn{1}{c}{0.1} &  & \multicolumn{1}{c}{0.05} &  & \multicolumn{1}{c}{0.01} &  &  & \multicolumn{1}{c}{0.1} &  & \multicolumn{1}{c}{0.05} &  & \multicolumn{1}{c}{0.01} &  &  & \multicolumn{1}{c}{0.1} &  & \multicolumn{1}{c}{0.05} &  & \multicolumn{1}{c}{0.01} &  \\ \cline{1-2} \cline{4-4} \cline{6-9} \cline{11-11} \cline{13-16} \cline{18-18} \cline{20-21} 
    $[0,25)m$ & 2.3 &  & 4.7 &  & 14.4 & $[0.012,2.350]$ & $[75,100)m$ & 0.4 &  & 0.9 &  & 3.4 & $[0.002,0.450]$ & $[150,175)m$ & 0.4 &  & 0.8 &  & 2.8 & $[0.002,0.400]$ \\
    $[25,50)m$ & 1.9 &  & 3.7 &  & 9.0 & $[0.009,1.850]$ & $[100,125)m$ & 0.4 &  & 0.9 &  & 3.5 & $[0.002,0.450]$ & $[175,200)m$ & 0.3 &  & 0.5 &  & 1.8 & $[0.001,0.250]$ \\
    $[50,75)m$ & 1.2 &  & 2.5 &  & 6.2 & $[0.006,1.250]$ & $[125,150)m$ & 0.5 &  & 1.3 &  & 5.9 & $[0.003,0.650]$ & $[200,\infty)m$ & 0.3 &  & 0.6 &  & 1.9 & $[0.002,0.300]$\\\bottomrule
    \end{tabular}%
    }
    \caption{Cutoff quantile values for the speed of two consecutive messages in miles for selected values of $\alpha$ across the different length intervals. The lower acceleration values are calculated by dividing the cutoff value by the $5\%$-percentile of the temporal difference distribution ($391 s$). The upper acceleration values are obtained by dividing by the minimum AIS sending frequency of $2s$. Units of acceleration are $kn/s$.}
    \label{speed-quantiles}
    \end{table}

\subsubsection{Turning rates}
\label{par:turning-rates}

\begin{figure}
    \centering
    \includegraphics[width = \textwidth]{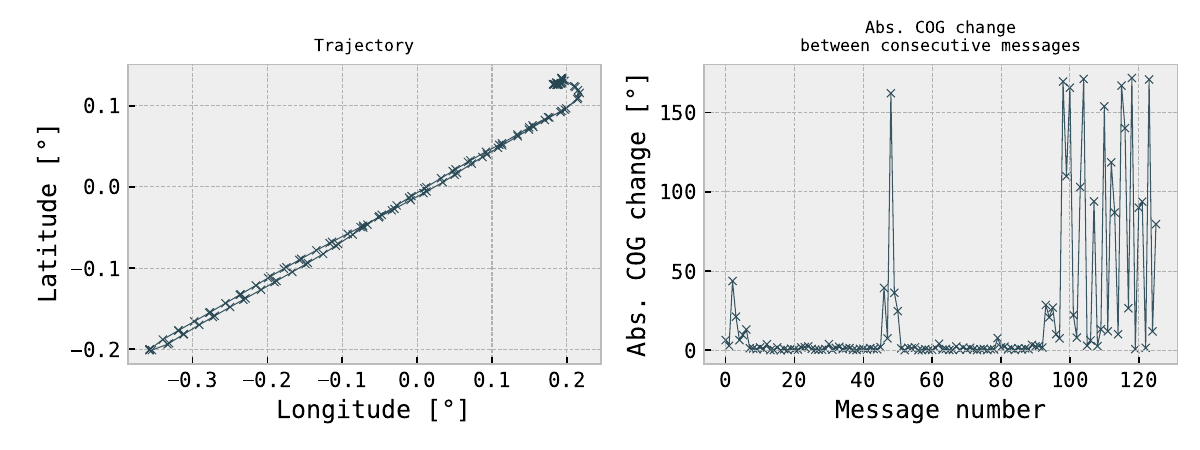}
    \caption{Example of a vessel's trajectory that will be cut into several sub-trajectories by the heading change split-point threshold. The route exerts a ferry-like pattern where most of its trajectory is inconspicuous except for the turning points (around message numbers 50 and 100), where the ferry presumably performs a turn-around maneuver or is moored.}
    \label{high-cog-change}
\end{figure}

It is generally not anticipated that significant deviations in a ship's turning rate will occur during navigation, as rapid directional shifts as those in Figure \ref{high-cog-change} not only increase material wear and tear but also contravene the IMO's routing rationale advocating for safe and straightforward routing \citep{imo2019shiprouteing}. Consequently, Table \ref{turning-rate-quantile} presents the tuning rates for two consecutive messages using all trajectories from 2021 and three selected $\alpha$ values. Turning rates are calculated by dividing the difference of COG changes of two consecutive messages by their temporal difference. As with the differences in speed, we also see a clear length-dependent pattern of inversely related turning rates and ship lengths.

The relatively low turning rates may be attributed to several factors that underline the principles of safe and efficient maritime navigation. As mentioned, the IMO's guidelines advocate for straightforward and safe routing. These guidelines are designed to minimize the risk of accidents, enhance the safety of the ship's voyage, and protect the marine environment. Ships often use predetermined routes optimized for safety and efficiency, reducing the need for frequent or sharp turns. Further, many modern ships have advanced navigation systems like autopilots and GPS-based tracking. These systems are programmed to execute changes in course in a controlled and gradual manner, adhering closely to the calculated optimal route. Using such technology helps maintain the consistency of movements and avoid abrupt directional changes. Thus, the small range of turning rates reflects the combined impact of ship design, navigational technology, regulatory compliance, and environmental strategy aimed at ensuring a smooth, safe, and efficient voyage.

\begin{table}[]
    \centering
    \resizebox{\textwidth}{!}{%
    \begin{tabular}{lrlrlrlllllll}
        \toprule
    \multicolumn{1}{l|}{\multirow{2}{*}{Length bin}} & \multicolumn{5}{c}{$[1-\alpha/2,\alpha/2]$} & \multirow{2}{*}{} & \multicolumn{1}{l|}{\multirow{2}{*}{Length bin}} & \multicolumn{5}{c}{$[1-\alpha/2,\alpha/2]$} \\ \cline{2-6} \cline{9-13} 
    \multicolumn{1}{l|}{} & \multicolumn{1}{c}{0.1} &  & \multicolumn{1}{c}{0.05} &  & \multicolumn{1}{c}{0.01} &  & \multicolumn{1}{l|}{} & \multicolumn{1}{c}{0.1} &  & \multicolumn{1}{c}{0.05} &  & \multicolumn{1}{c}{0.01} \\ \cline{1-2} \cline{4-4} \cline{6-9} \cline{11-11} \cline{13-13} 
    \multicolumn{1}{l|}{$[0,25)m$} & $[-0.619,0.473]$ &  & $[-1.067,0.840]$ &  & $[-1.949,1.600]$ &  & \multicolumn{1}{l|}{$[75,100)m$} & \multicolumn{1}{r}{$[-0.113,0.106]$} &  & \multicolumn{1}{r}{$[-0.218,0.200]$} &  & \multicolumn{1}{r}{$[-0.436,0.408]$} \\
    \multicolumn{1}{l|}{$[25,50)m$} & $[-0.521,0.402]$ &  & $[-0.880,0.700]$ &  & $[-1.557,1.300]$ &  & \multicolumn{1}{l|}{$[100,125)m$} & \multicolumn{1}{r}{$[-0.100,0.092]$} &  & \multicolumn{1}{r}{$[-0.183,0.169]$} &  & \multicolumn{1}{r}{$[-0.365,0.340]$} \\
    \multicolumn{1}{l|}{$[50,75)m$} & $[-0.305,0.251]$ &  & $[-0.510,0.459]$ &  & $[-0.860,0.785]$ &  & \multicolumn{1}{l|}{$[125,150)m$} & \multicolumn{1}{r}{$[-0.099,0.091]$} &  & \multicolumn{1}{r}{$[-0.173,0.160]$} &  & \multicolumn{1}{r}{$[-0.324,0.300]$} \\
        & \multicolumn{1}{l}{} &  & \multicolumn{1}{l}{} &  & \multicolumn{1}{l}{} &  &  &  &  &  &  &  \\
    \multicolumn{1}{l|}{\multirow{2}{*}{Length bin}} & \multicolumn{5}{c}{$[1-\alpha/2,\alpha/2]$} & \multicolumn{1}{c}{} & \multicolumn{1}{c}{} & \multicolumn{1}{c}{} & \multicolumn{1}{c}{} &  &  &  \\ \cline{2-6}
    \multicolumn{1}{l|}{} & \multicolumn{1}{l}{0.1} & \multicolumn{1}{c}{} & \multicolumn{1}{c}{0.05} & \multicolumn{1}{c}{} & \multicolumn{1}{c}{0.01} & \multicolumn{1}{c}{} &  &  &  &  &  &  \\ \cline{2-2} \cline{4-4} \cline{6-6}
    \multicolumn{1}{l|}{$[150,175)m$} & $[-0.083,0.080]$ &  & $[-0.148,0.142]$ &  & $[-0.148,0.142]$ &  &  &  &  &  &  &  \\
    \multicolumn{1}{l|}{$[175,200)m$} & $[-0.062,0.060]$ &  & $[-0.114,0.111]$ &  & $[-0.218,0.210]$ &  &  &  &  &  &  &  \\
    \multicolumn{1}{l|}{$[200,\infty)m$} & $[-0.065,0.062]$ &  & $[-0.113,0.109]$ &  & $[-0.205,0.200]$ &  &  &  &  &  &  & \\\bottomrule
    \end{tabular}%
    }
    \caption{Cutoff quantile intervals for the turning rate of two consecutive messages in $[{}^\circ/s]$ for selected values of $\alpha$ across the different length intervals.}
    \label{turning-rate-quantile}
    \end{table}

\subsubsection{Reported against positional velocity}
\label{par:repagpos}

\begin{figure}
    \centering
    \includegraphics[width = \textwidth]{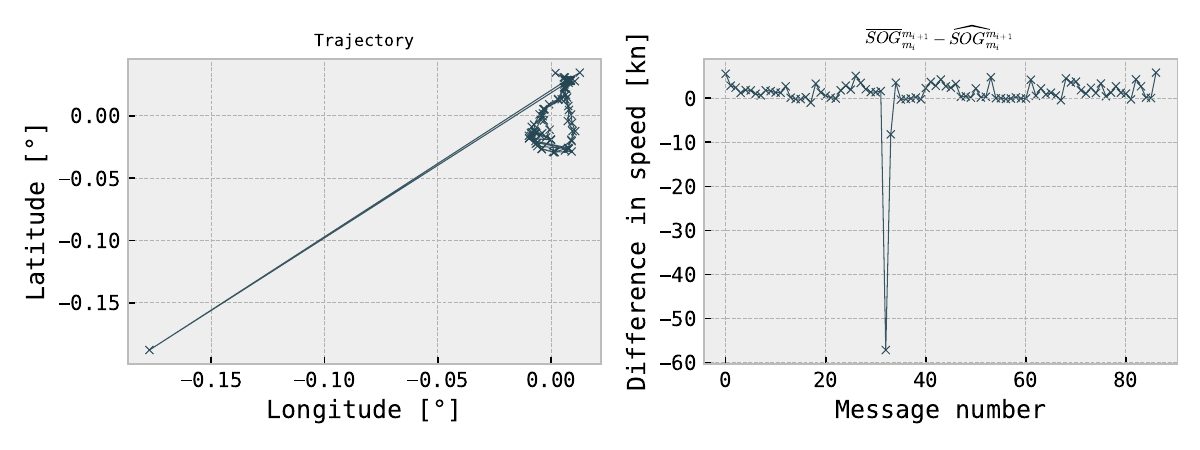}
    \caption{Erroneous positional outlier detected via the difference between the average reported SOG from the messages and the calculated speed from positional and temporal data. Latitude and longitude had been normalized for data protection}
    \label{repvscalc-outlier}
\end{figure}

\begin{figure}
    \centering
    \includegraphics[width=\linewidth]{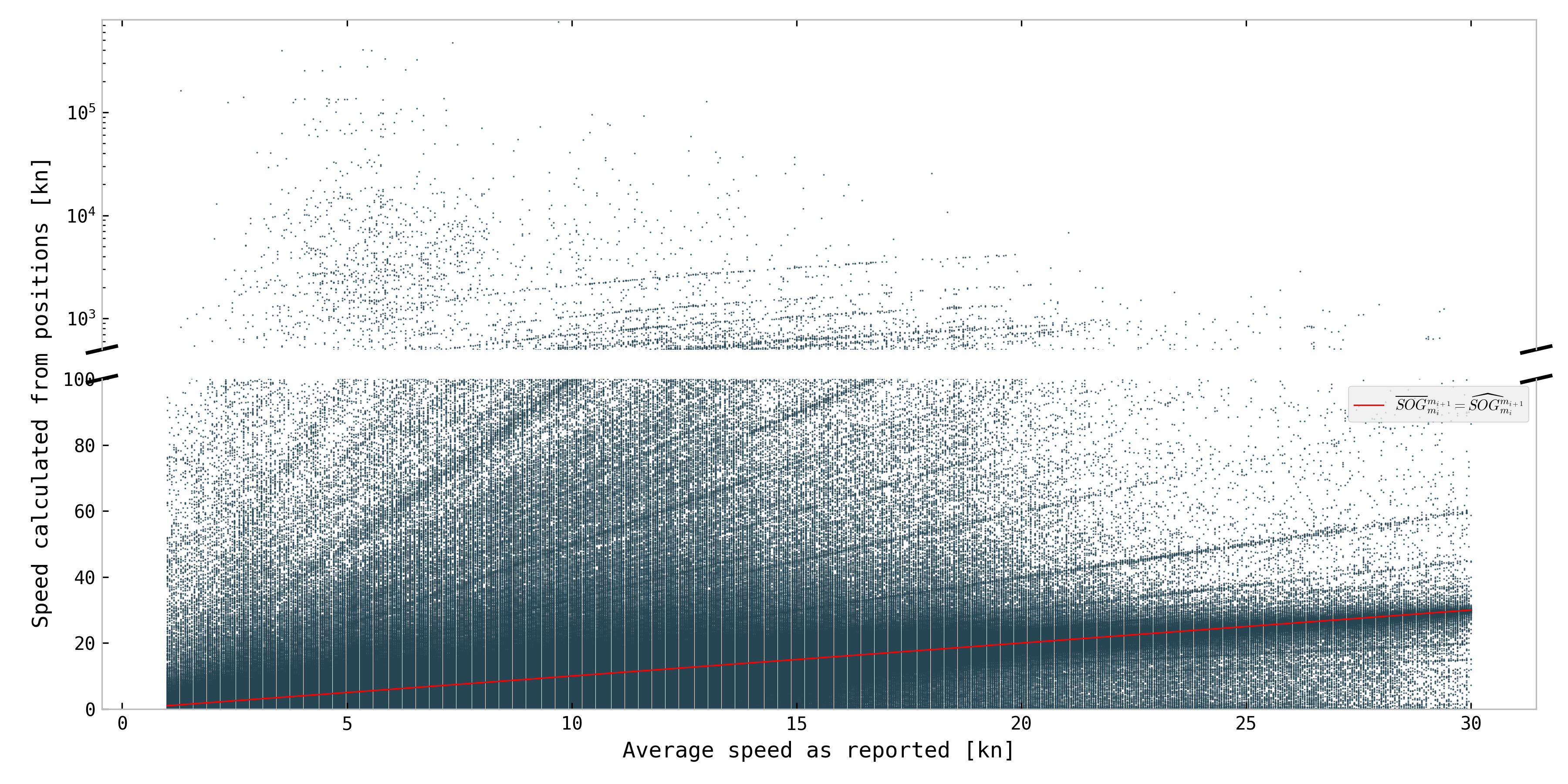}
    \caption{Scatter plot of the average observed SOG ($\overbar{SOG}_{m_i}^{m_{i+1}}$) and the calculated SOG from positional and temporal data ($\widehat{SOG}_{m_i}^{m_{i+1}})$ for all trajectories of July 2021.}
    \label{repvscalc_scatter}
\end{figure}

Another deficiency of trajectories that cannot be captured with the three metrics above are erroneous position reports that lead to significant outliers, as seen in Figure \ref{repvscalc-outlier}. In those cases, there is usually no anomaly detectable in terms of speed, course, or temporal differences, which is why this study follows \citet{zhao2018ship} and uses the difference between $\overbar{SOG}_{m_i}^{m_{i+1}}$ and $\widehat{SOG}_{m_i}^{m_{i+1}}$ to determine the outliers. Figure \ref{repvscalc_scatter} provides context on the magnitude of differences across the admitted speed range. It can be observed that there are substantial deviations of different magnitudes in both directions. As with the other metrics, some of the determined threshold values can be viewed in Table \ref{repvscalc-quantiles}. The skewness of the intervals is expected, as $\overbar{SOG}_{m_i}^{m_{i+1}}$ is bounded at $30 kn$ due to the speed filtering employed in Section \ref{velreps}, while $\widehat{SOG}_{m_i}^{m_{i+1}}$ is unbounded.

\begin{table}[]
    \centering
    \resizebox{\textwidth}{!}{%
    \begin{tabular}{lrlrlrlllllll}
        \toprule
    \multicolumn{1}{l|}{\multirow{2}{*}{Length bin}} & \multicolumn{5}{c}{$[1-\alpha/2,\alpha/2]$} & \multirow{2}{*}{} & \multicolumn{1}{l|}{\multirow{2}{*}{Length bin}} & \multicolumn{5}{c}{$[1-\alpha/2,\alpha/2]$} \\ \cline{2-6} \cline{9-13} 
    \multicolumn{1}{l|}{} & \multicolumn{1}{c}{0.1} &  & \multicolumn{1}{c}{0.05} &  & \multicolumn{1}{c}{0.01} &  & \multicolumn{1}{l|}{} & \multicolumn{1}{c}{0.1} &  & \multicolumn{1}{c}{0.05} &  & \multicolumn{1}{c}{0.01} \\ \cline{1-2} \cline{4-4} \cline{6-9} \cline{11-11} \cline{13-13} 
    \multicolumn{1}{l|}{$[0,25)m$} & $[-1.434,2.332]$ &  & $[-3.237,4.022]$ &  & $[-8.412,6.699]$ &  & \multicolumn{1}{l|}{$[75,100)m$} & \multicolumn{1}{r}{$[-1.074,0.822]$} &  & \multicolumn{1}{r}{$[-2.701,1.507]$} &  & \multicolumn{1}{r}{$[-10.197,3.338]$} \\
    \multicolumn{1}{l|}{$[25,50)m$} & $[-1.426,1.806]$ &  & $[-3.233,3.240]$ &  & $[-9.794,5.381]$ &  & \multicolumn{1}{l|}{$[100,125)m$} & \multicolumn{1}{r}{$[-1.294,0.975]$} &  & \multicolumn{1}{r}{$[-3.353,1.861]$} &  & \multicolumn{1}{r}{$[-12.804,4.014]$} \\
    \multicolumn{1}{l|}{$[50,75)m$} & $[-1.032,0.995]$ &  & $[-2.110,1.876]$ &  & $[-6.892,3.633]$ &  & \multicolumn{1}{l|}{$[125,150)m$} & \multicolumn{1}{r}{$[-1.294,0.987]$} &  & \multicolumn{1}{r}{$[-3.041,1.948]$} &  & \multicolumn{1}{r}{$[-11.222,4.233]$} \\
     & \multicolumn{1}{l}{} &  & \multicolumn{1}{l}{} &  & \multicolumn{1}{l}{} &  &  &  &  &  &  &  \\
    \multicolumn{1}{l|}{\multirow{2}{*}{Length bin}} & \multicolumn{5}{c}{$[1-\alpha/2,\alpha/2]$} & \multicolumn{1}{c}{} & \multicolumn{1}{c}{} & \multicolumn{1}{c}{} & \multicolumn{1}{c}{} &  &  &  \\ \cline{2-6}
    \multicolumn{1}{l|}{} & \multicolumn{1}{l}{0.1} & \multicolumn{1}{c}{} & \multicolumn{1}{c}{0.05} & \multicolumn{1}{c}{} & \multicolumn{1}{c}{0.01} & \multicolumn{1}{c}{} &  &  &  &  &  &  \\ \cline{2-2} \cline{4-4} \cline{6-6}
    \multicolumn{1}{l|}{$[150,175)m$} & $[-2.023,1.393]$ &  & $[-5.134,2.780]$ &  & $[-19.192,5.877]$ &  &  &  &  &  &  &  \\
    \multicolumn{1}{l|}{$[175,200)m$} & $[-1.534,1.188]$ &  & $[-4.032,2.433]$ &  & $[-15.734,5.155]$ &  &  &  &  &  &  &  \\
    \multicolumn{1}{l|}{$[200,\infty)m$} & $[-2.176,1.371]$ &  & $[-5.907,2.850]$ &  & $[-21.707,6.044]$ &  &  &  &  &  &  & \\\bottomrule
    \end{tabular}%
    }
    \caption{Cutoff quantile intervals for the difference between reported and calculated speed of two consecutive messages in $[kn]$ for selected values of $\alpha$ across the different length intervals.}
    \label{repvscalc-quantiles}
    \end{table}

\subsubsection{Distance between messages}
\label{par:dist}

The final metric to be considered in our analysis is the distance between two successive AIS messages. The speed filtering process described in Section \ref{velreps} involves removing trajectory segments where the SOG falls below one knot. This exclusion results in significant spatial gaps between successive messages within a trajectory, potentially leading to substantial temporal gaps if the vessel continues sailing. If the vessel increases its SOG to over one knot, its messages appear again in the data. In these cases, the discrepancy between the calculated speed (derived from positional and temporal data) and the average reported speed might be insufficiently large to be identified as split points based on their percentile criterion.

To additionally capture these special cases, we also label two consecutive messages a split-point if their distance in miles, as calculated via the Haversine formula, is greater than the $(1-\alpha)$ quantile of the distance distribution obtained for the year 2021, which are depicted in Figure \ref{dist-quantiles}.

\begin{table}[]
    \centering
    \resizebox{\textwidth}{!}{%
    \begin{tabular}{l|rlrlrll|rlrlrll|rlrlr}
        \toprule
    \multirow{2}{*}{Length bin} & \multicolumn{5}{c}{$\alpha$} &  & \multirow{2}{*}{Length bin} & \multicolumn{5}{c}{$\alpha$} &  & \multirow{2}{*}{Length bin} & \multicolumn{5}{c}{$\alpha$} \\ \cline{2-6} \cline{9-13} \cline{16-20} 
        & \multicolumn{1}{c}{0.1} &  & \multicolumn{1}{c}{0.05} &  & \multicolumn{1}{c}{0.01} &  &  & \multicolumn{1}{c}{0.1} &  & \multicolumn{1}{c}{0.05} &  & \multicolumn{1}{c}{0.01} &  &  & \multicolumn{1}{c}{0.1} &  & \multicolumn{1}{c}{0.05} &  & \multicolumn{1}{c}{0.01} \\ \cline{1-2} \cline{4-4} \cline{6-6} \cline{8-9} \cline{11-11} \cline{13-13} \cline{15-16} \cline{18-18} \cline{20-20} 
    $[0,25)m$ & 0.727 &  & 1.001 &  & 1.969 &  & $[75,100)m$ & 0.922 &  & 1.095 &  & 1.352 &  & $[150,175)m$ & 1.148 &  & 1.424 &  & 1.879 \\
    $[25,50)m $& 0.771 &  & 1.004 &  & 1.574 &  & $[100,125)m$ & 0.992 &  & 1.205 &  & 1.517 &  & $[175,200)m$ & 1.162 &  & 1.391 &  & 1.775 \\
    $[50,75)m$ & 0.895 &  & 1.072 &  & 1.332 &  & $[125,150)m$ & 1.160 &  & 1.409 &  & 1.808 &  & $[200,\infty)m$ & 1.177 &  & 1.444 &  & 1.994\\\bottomrule
    \end{tabular}%
    }
    \caption{Cutoff quantile values for the distance of two consecutive messages in miles for selected values of $\alpha$ across the different length intervals.}
    \label{dist-quantiles}
    \end{table}

\subsection{Re-joining trajectories}
\label{sec:rejoin}

\begin{figure}
    \centering
    \begin{subfigure}[t]{0.25\textwidth}
        \includegraphics[width=\linewidth]{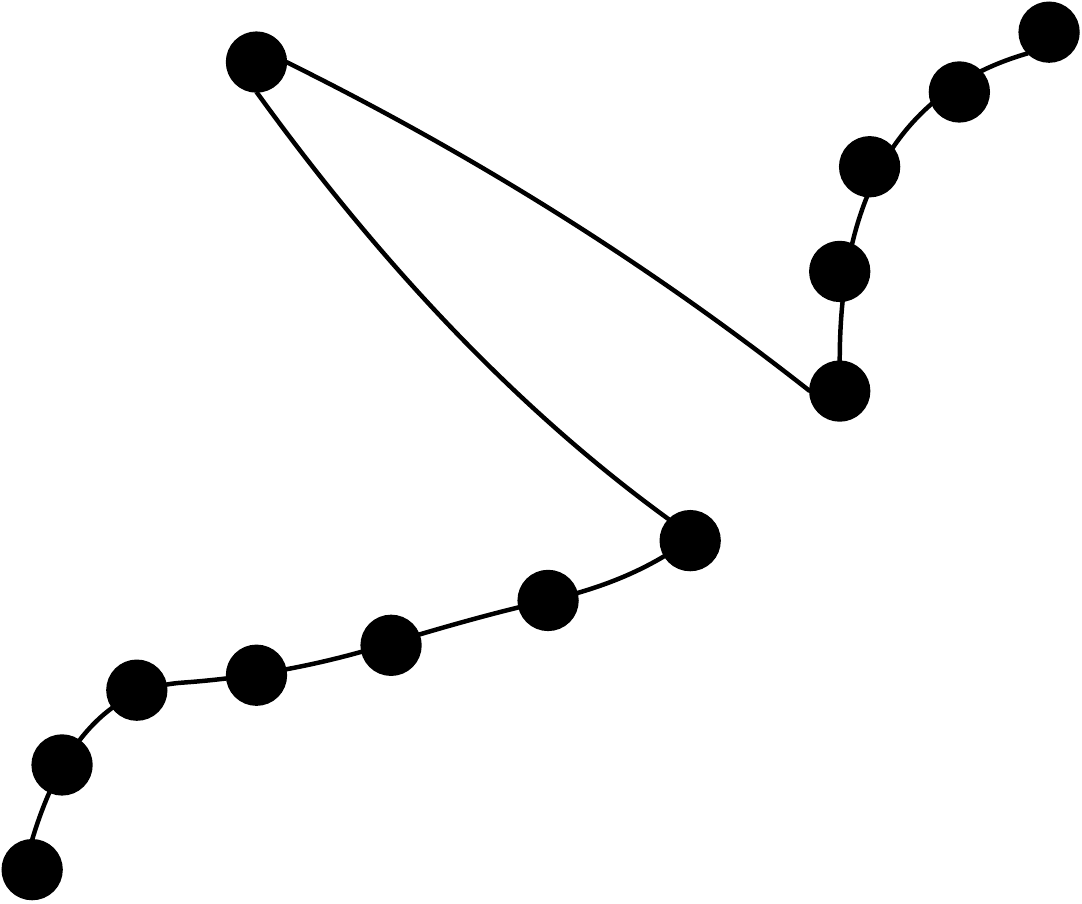}
        \caption{Original trajectory.}
        \label{rejoin-orig}
    \end{subfigure}
    \begin{subfigure}[t]{0.25\textwidth}
        \includegraphics[width=\linewidth]{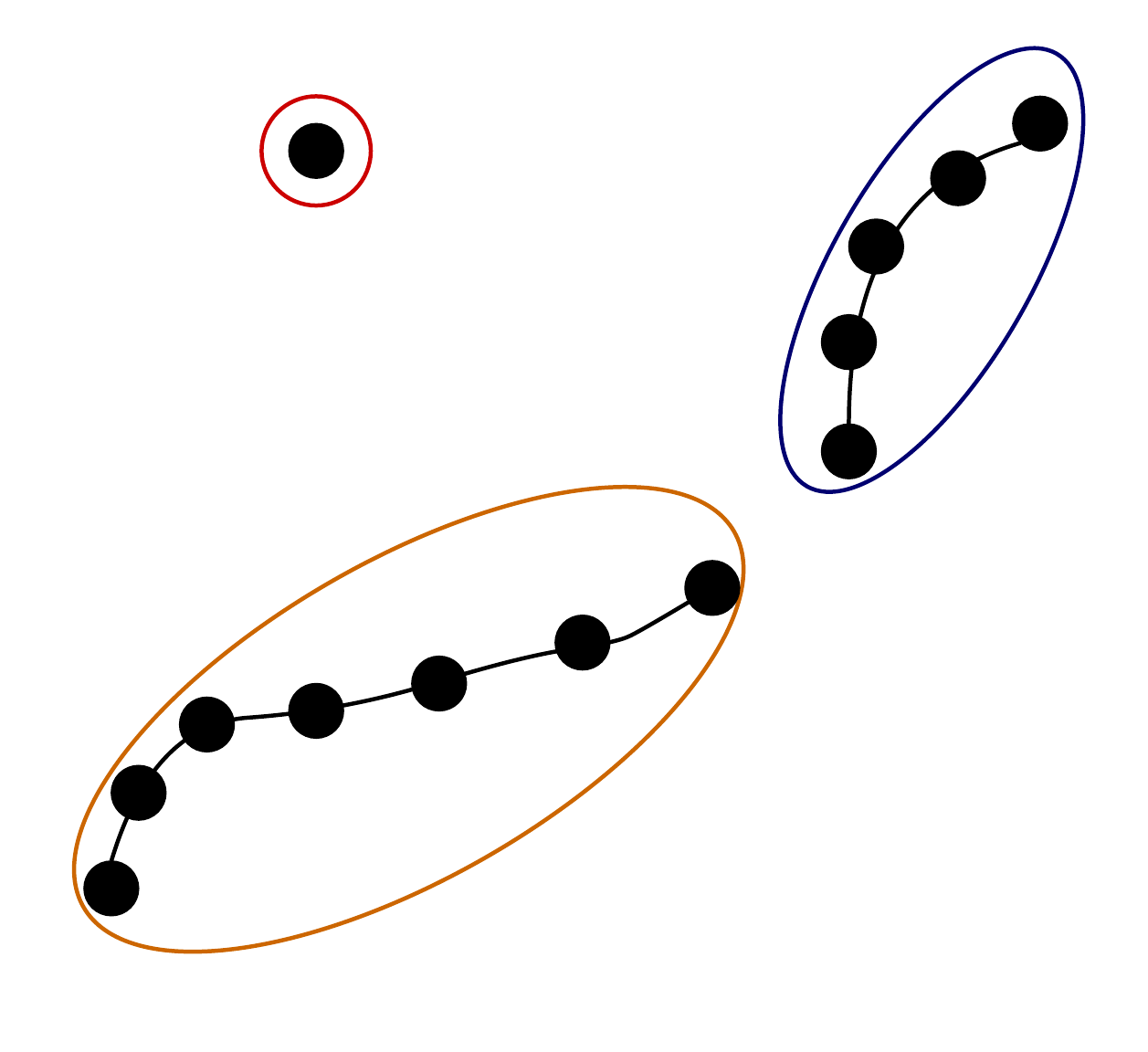}
        \caption{After split-point procedure applied.}
        \label{rejoin-after}
    \end{subfigure}
    \begin{subfigure}[t]{0.25\textwidth}
        \includegraphics[width=\linewidth]{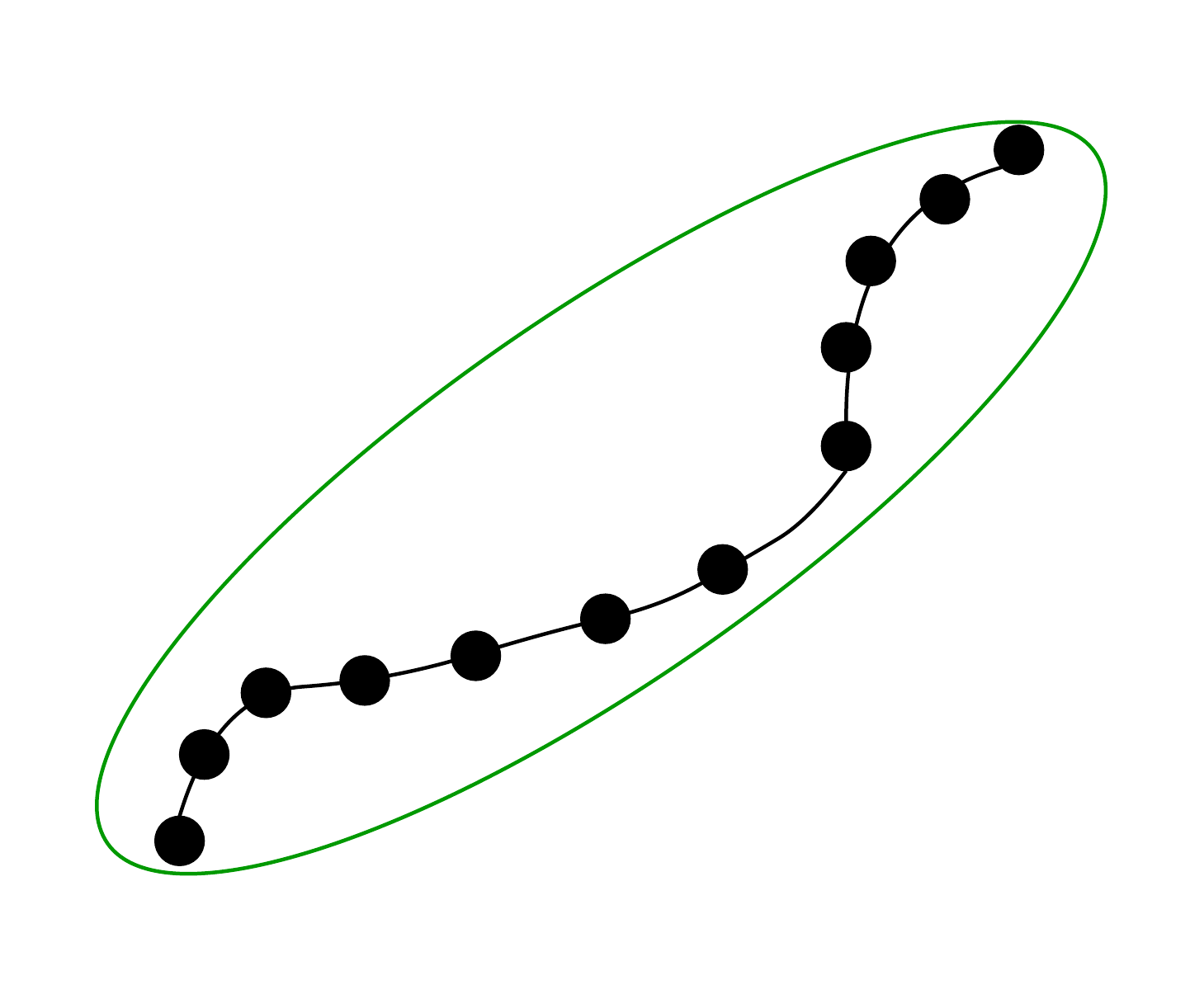}
        \caption{After re-joining.}
        \label{rejoin-rejoined}
    \end{subfigure}
    \caption{Exemplary rejoin procedure. The split-point method splits the original, erroneous trajectory (a) into three separate ones, of which one only has a single message (b). Trajectories are judged again by the derived thresholds from the split-point method and possibly get rejoined (c)—illustration inspired by \citet{zhang2018novel}.}
    \label{rejoin}
\end{figure}

As pointed out by \citet{zhao2018ship}, rejecting singular anomalous points may lead to an accidental separation of a trajectory that logically belongs together. The example in Figure \ref{rejoin} illustrates such a case. The original trajectory (Figure \ref{rejoin-orig}) consists of a single anomalous point that should be removed from the data. Our split-point procedure, therefore, looks at consecutive overlapping pairs of messages and decides to split them based on the empirical distribution functions of the above-derived metrics. In the example, this leads to the trajectory being split into three separate trajectories, of which one only consists of a single message (Figure \ref{rejoin-after}). Since single-message-trajectories cannot sensibly count as such, they get removed automatically. In an attempt to rejoin the two remaining trajectories, the last message of the first trajectory and the first message of the second trajectory are again judged by the split-point procedure. If no split point is found, the two trajectories are rejoined into one (Figure \ref{rejoin-rejoined}). 

\begin{algorithm}
	\DontPrintSemicolon
	\SetAlgoLined
	\KwData{Trajectory to split $\mathcal{T}^>$, Ship length $l$, Length-dependent empirical quantile functions for change in speed $\hat{q}_{l}^{\Delta_{SOG}}$, tuning rate $\hat{q}_{l}^{{ROT}}$, time difference $\hat{q}_{l}^{\Delta_t}$, distance $\hat{q}_{l}^{dist}$, and reported against calculated speed $\hat{q}_{l}^{\overbar{SOG} - \widehat{SOG}}$, quantile threshold $\alpha$.}
    \tcc*[l]{Set the thresholds}
    $s \leftarrow \hat{q}_{l}^{\Delta_{SOG}}(1-\alpha)$\\
    $r \leftarrow [\hat{q}_{l}^{{ROT}}(\alpha/2),\hat{q}_{l}^{{ROT}}(1-\alpha/2)]$ \\
    $t \leftarrow \hat{q}_{l}^{\Delta_t}(1-\alpha)$\\
    $d \leftarrow \hat{q}_{l}^{dist}(1-\alpha)$\\
    $b \leftarrow [\hat{q}_{l}^{\overbar{SOG} - \widehat{SOG}}(\alpha/2),\hat{q}_{l}^{\overbar{SOG} - \widehat{SOG}}(1-\alpha/2)]$\\
	\SetKwBlock{Traj}{for $(m_i,m_{i+1}) \in \mathcal{T}^>$}{end}
	\Traj{
        \If{$\sog(m_{i+1})-\sog(m_i) > s$ (Section \ref{par:vel-changes}) \Condition{
            \Or $\frac{\cog(m_{i+1})-\cog(m_i)}{\unix(m_{i+1})-\unix(m_{i})} \notin r$ (Section \ref{par:turning-rates}) \\
            \Or $\unix(m_{i+1})-\unix(m_i) > t$ (Section \ref{par:tempdiffs}) \\
            \Or $\hav(m_{i+1},m_i) > d$ (Section \ref{par:dist}) \\
            \Or $\overbar{SOG}_{m_i}^{m_{i+1}}-\widehat{SOG}_{m_i}^{m_{i+1}}\notin b$ (Section \ref{par:repagpos})}}
        {
            Split the trajectory between $m_i$ and $m_{i+1}$
        } 
    }
	\caption{Split-point procedure.}
	\label{algo:spp}
\end{algorithm}

\section{Applications of the split-point procedure to data}
\label{applications}

\begin{figure}[h!]
\centering
    \begin{subfigure}[t]{.85\textwidth}
      \centering
      \includegraphics[trim={0 0cm 1cm 0cm},clip,width=\linewidth]{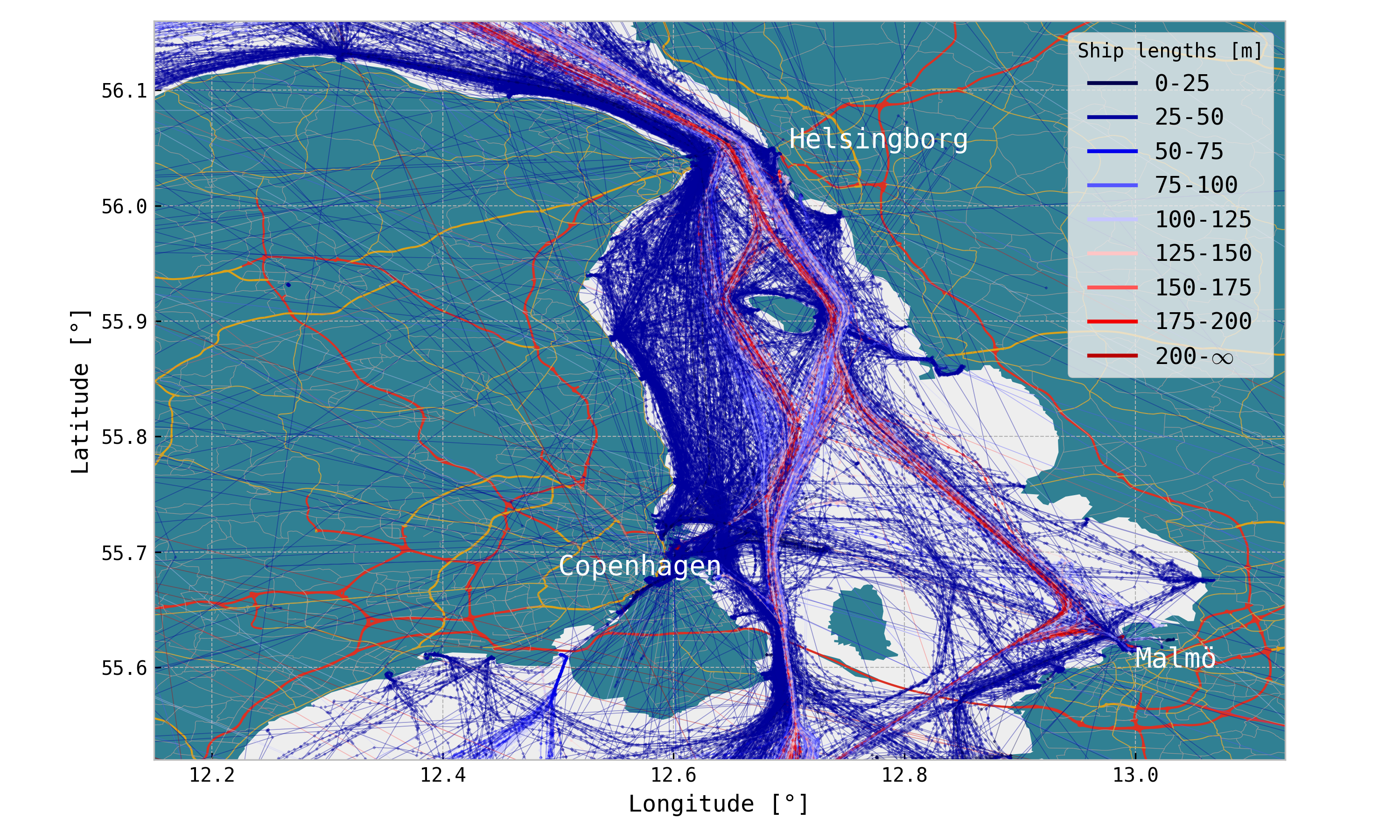}
      \caption{Trajectories without split-point filtering. }
      \label{nosplitpointfilter}
    \end{subfigure}
    \begin{subfigure}[t]{.85\textwidth}
      \centering
      \includegraphics[trim={0 0cm 1cm 0cm},clip,width=\linewidth]{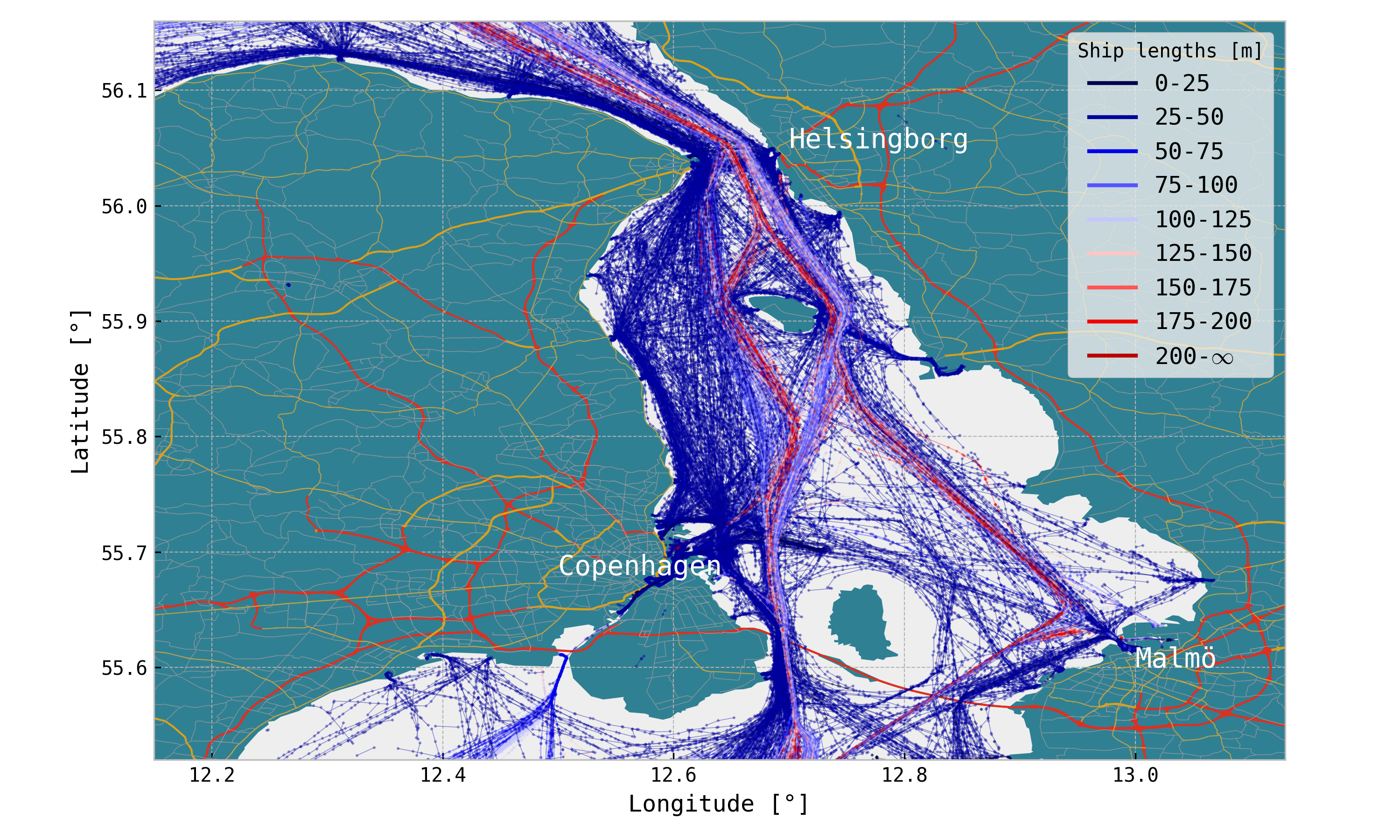}
      \caption{Trajectories after applying the split-point procedure from Section \ref{splitpoints} with $\alpha=0.03$.}
      \label{splitpointfilter}
    \end{subfigure}
    \begin{subfigure}[t]{\textwidth}
      \centering
      \footnotesize
      \begin{tabular}{lr|lr|lr}
      \toprule
          Number of Raw Messages & \num{26612047} & Percentage of Duplicates & 0.46\% & Number of Trajectories & \num{722624} \\
          Number of Split-Points & \num{1444724} & Number of Rejoined Tracks & \num{20558} & Average Trajectory length & \num{4.53} $nm$\\
          \bottomrule
      \end{tabular}
      \caption{Quantitative information on the split-point procedure. }
      \label{splitpointstats}
    \end{subfigure}
    \caption{Effects of the split-point filter applied to an area in the \emph{Øresund} around Copenhagen in Denmark. Data from 10/08/2024 - 19/08/24 is used.}
    \label{applicationofsplitpointfilter}
\end{figure}

To demonstrate the effectiveness of the proposed split-point methodology, Figure \ref{applicationofsplitpointfilter} showcases a collection of trajectories derived from segments of the \emph{\O{}resund} and harbors of Copenhagen in Denmark and Malmö and Helsingborg in Sweden, both with and without the application of the split-point procedure. In the absence of this procedure (as depicted in Figure \ref{nosplitpointfilter}), numerous trajectories exhibit anomalously large distances between consecutive messages. This results in unrealistic jumps in the trajectory paths, leading to instances where the trajectories erroneously appear to traverse over land. This behavior can arise if multiple vessels transmit messages from a single MMSI, which is a common occurrence \citep{wu2017mapping,zhao2018ship,yan2020analysis}, if trajectories are interrupted by the SOG-filter constraint or if vessels deliberately disable their AIS transceiver and re-enable it at a different location. In the filtered version, as depicted in Figure \ref{splitpointfilter}, the application of the split-point procedure effectively eliminates these anomalies, thereby accurately reconstructing the underlying structure of the waterway network and, due to the different coloring by the length of the ship, one can also gain valuable insights into the distribution of ship sizes along the waterway, enabling a more comprehensive analysis of traffic patterns and the potential identification of areas with specific navigational challenges or congestion points based on vessel dimensions.Quantitative information about the procedure can be found in Figure \ref{splitpointstats}.

The split-point procedure also allows for significant insights into the navigational behaviors of vessels across different regions and types. This technique is particularly effective in generating ship-type dependent density maps, which illustrate the varying concentrations of vessel traffic along established routes and can also be used to discover new routes. This procedure allows key navigational routes and traffic flow dynamics specific to various ship types to be analyzed effectively. The methodology's utility and effectiveness are demonstrated in the Appendix \ref{timescale} in Figures \ref{heatmaps-1} and \ref{heatmaps-2}, which present detailed density maps of vessel traffic routes in the area of this research.

\begin{figure}
    \centering
    \includegraphics[width=\textwidth]{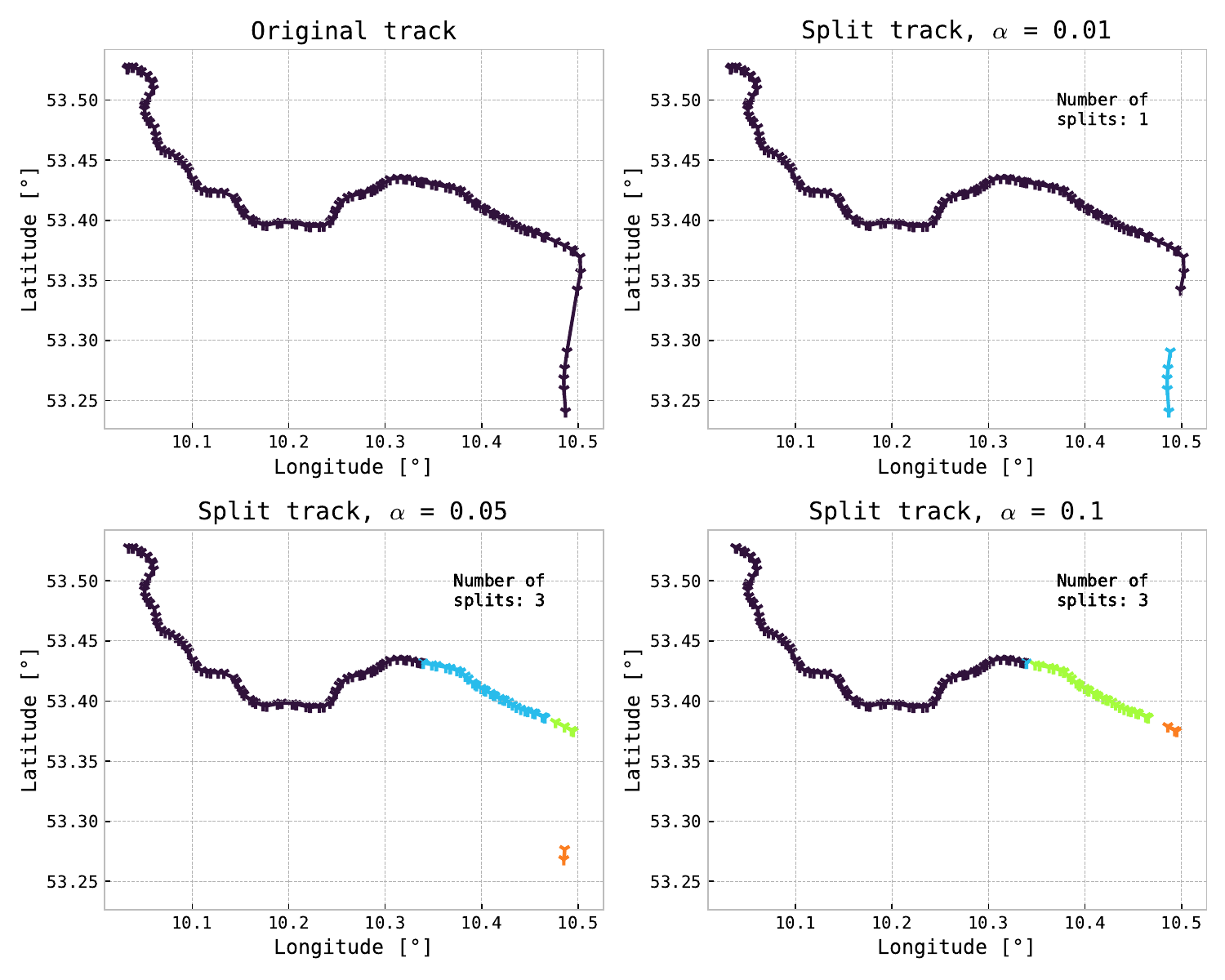}
    \caption{Results of our proposed split-point procedure under different values of $\alpha$, where lower values split less aggressively at the cost of larger margins of accepted values between messages. Differently colored parts of a trajectory indicate sub-trajectories. On the rightmost illustration, the lower-right part of the trajectory gets filtered out entirely as it was decomposed into several single-message trajectories, which automatically get discarded. The most significant positional gap in the original trajectory is $3.11nm$.}
    \label{alpha-comp}
\end{figure}

On an individual trajectory level Figure \ref{alpha-comp} demonstrates the effects of choosing different values of $\alpha$ on an unfiltered trajectory by showing how different values of $\alpha$ impact the number of splits in the trajectory. As $\alpha$ decreases, the trajectory becomes less interrupted with fewer splits. This demonstrates the trade-off between the level of data retention and the degree of realism, highlighting how lower $\alpha$ values result in a more generalized path with increasingly unreliable jumps in the metrics between two consecutive messages. The properties of trajectories extracted with a given value of $\alpha$ are easily interpretable by guaranteeing that all consecutive messages within the trajectory lie inside the navigational bounds of $(1-\alpha)\cdot 100\%$ of all vessels used to determine the thresholds. This ensures that the extracted trajectories represent common navigational patterns, making them reliable and consistent with the overall traffic flow while filtering out outliers or unusual movements.

Moreover, due to its data-driven nature, the split-point methodology is highly generalizable and can be effectively applied to any geographical region. Whether dealing with complex networks of inland waterways, coastal regions, or open ocean environments, the underlying principles of the split-point procedure remain consistent and adaptable. The flexibility of this approach allows for its application in diverse maritime contexts, making it a valuable tool for global maritime navigation analysis but also as a precursor for algorithmic testing procedures that rely on realistic real-world trajectories. Practitioners can leverage the accompanying Python package, as detailed in Appendix \ref{implementation}, to use and adapt Algorithm \ref{algo:spp} and generate results similar to those presented, regardless of the geographical region under study. This generalizability ensures that the insights gained from this methodology are not confined to the specific regions of the North Sea but are applicable worldwide.

\section{Comparison}
\label{comparison}

When comparing trajectory extraction processes from AIS data, it is essential to recognize the inherent limitations in quantitatively evaluating the performance of different methods. As mentioned in the introduction, AIS data, while a valuable source of real-time maritime information, is subject to various errors and inconsistencies. These can include inaccuracies in positional data due to GPS errors, data loss from signal gaps, or intentional manipulation of AIS transmissions. Consequently, no definitive \emph{ground truth} can be used as a baseline for comparison.

Given this lack of an absolute reference, direct quantitative comparisons of trajectory extraction methods become problematic. Traditional metrics like accuracy or error rates, which rely on a known correct answer, are not applicable in this context. Instead, comparisons between methods must often be based on qualitative assessments or indirect metrics that reflect the performance of the extraction processes under realistic conditions.

Considering the above restrictions, we compare our approach to two commonly used extraction methods, namely the ones by \citet{zhao2018ship} and \citet{guo2021improved}. Both are popular approaches, whereby \citet{guo2021improved} is a pure exclusion process that does not split trajectories but removes erroneous messages. \citet{zhao2018ship}, on the other hand, acts as the base of our proposed framework, as it also splits trajectories. It is to be mentioned that the approach from \citet{guo2021improved} also features an extended trajectory interpolation scheme, which we will not implement in our comparison because neither the approach of \citet{zhao2018ship} nor ours interpolate filtered trajectories, rendering the comparison skewed.

For our comparison, we will investigate the maximum jump in metrics such as time, distance, and velocity inside each trajectory, examine the values, and view them in a maritime context. For all compared methods we start with a time-sorted collection of AIS messages coming from one MMSI.

\paragraph{\textbf{Setup}}

\citet{zhao2018ship} uses three steps to filter messages. (1) All messages whose absolute difference between receiving time and sending time is more than $5s$. (2) For a pair of consecutive messages $m_1,m_2$, inside a trajectory, a split point is defined if $\widehat{SOG}_{m_1}^{m_{2}}>15kn$ or $\unix(m_2) - \unix(m_1) > 600s$. (3) All sub-tracks constructed in (2) are rejoined similarly as described in Section \ref{sec:rejoin}, with the exception that only the condition $\widehat{SOG}_{m_1}^{m_{2}}>15kn$ is taken for a decision.

The exclusion process of \citet{guo2021improved} consists of two steps. (1)  Turning rates are calculated in the same way as in Section \ref{par:turning-rates}, and a turning rate limit $\Delta c_{lim}$ is defined. If the turning rate of any two messages, $m_1$ and $m_2$, is higher than the limit, the message $m_2$ is discarded. (2) If for any two messages, $\widehat{SOG}_{m_1}^{m_{2}}> v_{lim}  kn$, for a velocity limit $v_{lim}$, message $m_2$ is discarded. Unfortunately, the authors did not provide specific values for $\Delta c_{lim}$ and $v_{lim}$ but rather reference the ship's maneuverability under consideration and the speed limit in the area.

\paragraph{\textbf{Example trajectory}}

\begin{figure}
    \centering
    \includegraphics[width=\textwidth]{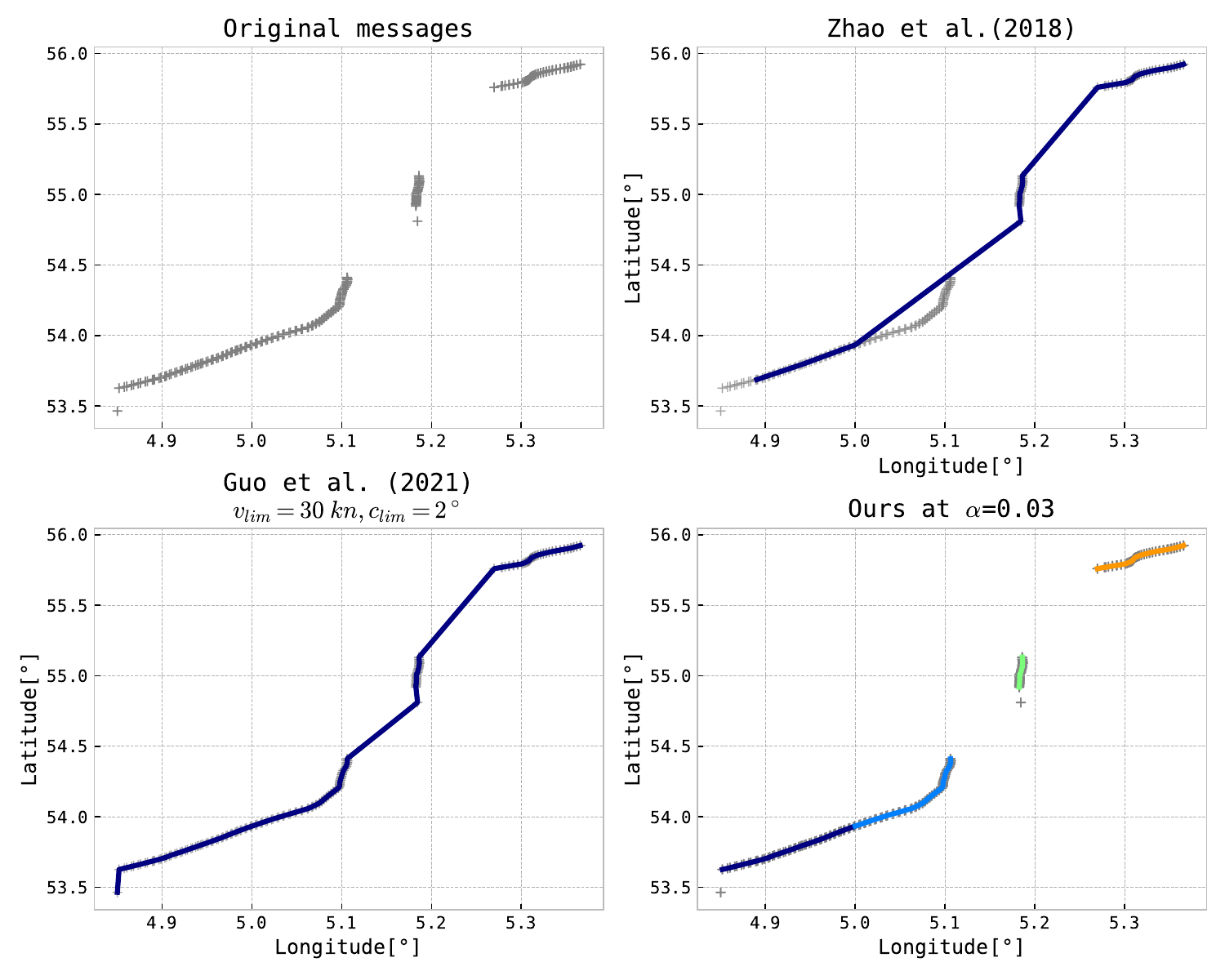}
    \caption{Method comparison for different trajectory extraction approaches.}
    \label{trex-comparison}
\end{figure}

A 223-messages-long trajectory with distinct jumps driven by a $19m$ long fishing vessel, roughly $80km$, off the coast of the Netherlands, is investigated. The grey line in Figure \ref{trex-comparison} shows the raw trajectory. Since there is no speed limit this far off the coast, we set $v_{lim}=30kn$, and for $c_{lim}=2^\circ$ which is a bit more the 99-percentile of the turning rate distribution for vessels with lengths $[0,25)m$ as calculated in Section \ref{par:turning-rates}.

Visually comparing the different approaches in Figure \ref{trex-comparison} reveals that no other approach was able to either effectively filter out or separate the trajectory at the two large spatial jumps that span distances of $37.85nm$ and $24.15nm$ each. As seen in Table \ref{tbl:trex-comp}, the approach of \citet{guo2021improved} discarded a single message or $0.45\%$ of the messages, the one of \citet{zhao2018ship} discarded $54.71\%$ of the messages mainly due to mismatching sending and receiving timestamps. No splitting was performed. \citet{guo2021improved} additionally allowed for a quite extensive turning rate of $5.10^\circ/s$.

\begin{table}[]
    \centering
    \begin{tabular}{l|ccccc}
        \multicolumn{1}{c|}{} & \begin{tabular}[c]{@{}c@{}}\# of discarded\\ messages\end{tabular} & \begin{tabular}[c]{@{}c@{}}\# of split\\ points\end{tabular} & \begin{tabular}[c]{@{}c@{}}max. turning\\ rate $[{}^\circ/s]$\end{tabular} & \begin{tabular}[c]{@{}c@{}}max. velocity\\ change $[kn/s]$\end{tabular} & \begin{tabular}[c]{@{}c@{}}max. distance\\ $[nm]$\end{tabular} \\ \cline{2-6} 
        \citet{zhao2018ship} & 1 & 0 & 0.58 & 0.04 & 53.06 \\
        \citet{guo2021improved} & 122 & - & 5.10 & 0.10 & 37.87 \\
        Proposed & 2 & 4 & 0.4 & 0.04 & 1.57
    \end{tabular}
    \caption{Maximum values attained for consecutive messages inside a trajectory or any of its split sub-trajectories.}
    \label{tbl:trex-comp}
\end{table}

In contrast, the proposed method demonstrates a balanced approach with minimal message loss and more optimal spatial filtering. As highlighted in Table \ref{tbl:trex-comp}, the proposed approach discarded just two messages, introduced four split points to avoid large spatial jumps, and maintained a realistic maximum turning rate of $0.4^\circ/s$. These results indicate that the proposed method achieves accurate and efficient trajectory filtering, ensuring superior spatial consistency and precision.

While this approach demonstrates superior performance in the specific scenario analyzed, it is important to acknowledge that there may be better choices for some applications. The setup process for the proposed method is considerably more complex and time-consuming compared to the more straightforward approaches of \citet{zhao2018ship} and \citet{guo2021improved}. This complexity arises from the need to pre-compute all threshold values from large amounts of data, which may require substantial computational resources. For scenarios where quick deployment or lower computational overhead is prioritized, the simplicity and ease of setup offered by other methods may outweigh the benefits of our more intricate approach. Thus, while the proposed method achieves high accuracy and precision, its applicability may be limited in contexts with critical ease of use and rapid implementation. Consequently, both methods are also implemented in the accompanying software package.

\section{Spatial properties of split trajectories (\frbox{fryellow}{fryellowbo})}
\label{spatial-properties}

From the approach in Section \ref{splitpoints} we extracted trajectories, each comprising a reliable and consistent series of messages, notably free from anomalies. Nonetheless, we have no tool to assess the spatial properties of the extracted trajectories, especially concerning our original aim of extracting long, clean, and uninterrupted trajectories.

Therefore, we need a tool that measures these three quantities. Upon closer inspection, we find that all trajectories that had undergone the split-point procedure in Section \ref{splitpoints} are already uninterrupted. Addressing the length and cleanliness of trajectories, we adopt a dual approach to control the length and the spatial properties of the trajectories. Consistent with methodologies employed in prior studies \citep{yuan2019novel,zhao2018ship,mao2018automatic}, the length of the trajectories is quantified using the count of messages per track, $n_{\mathrm{msg}}$. While effective in gauging trajectory length, this metric does not provide insights into the trajectory's shape or spatial extent.

To bridge this gap, we investigate the convex hull each trajectory forms, or more precisely, its area. Since the calculation of a convex hull's area relies on Euclidean geometry to apply, we first have to project the spheroid space of Earth to a Euclidean space, which is achieved using the Universal Transverse Mercator (UTM) projection \citep{kruger1912konforme}. While no isometric map from the sphere to the plane exists, the UTM projection still provides high accuracy if not used directly at the poles \citep{karney2011transverse}. Now, let $U = \{u_1,u_2,\dots,u_n\}$ be the set of UTM-projected positions of a trajectory. The convex hull $C(U)$ is the set of all convex combinations of positions from $U$. Formally, it is defined as
\begin{equation}
    C(U) = \left\{\sum_{i=1}^n \lambda_iu_i \bigm\vert \lambda_i \geq 0, \sum_{i=1}^n\lambda_i = 1 \right\},
\end{equation}
and in practice, is obtained using the quickhull algorithm by \cite{barber1996quickhull}.  In two-dimensional space, the convex hull $C(U)$ forms a regular polygon whose area $A^C$ can be easily obtained.

From investigating the number of messages in and the convex hull area of a trajectory, it is not immediately clear how these metrics connect to its \emph{cleanliness}. To measure this influence, this study adapts the idea of \emph{average complexity} by \citet{mao2018automatic}. In its original form, it is defined as 

\begin{equation}
    \bar{c}(\mathcal{T}_k^>) = |\mathcal{T}_k^>|^{-1}\sum_{m\in\mathcal{T}_k^>} \frac{\mathbf{p}^{\top} \mathbf{q}}{\|\mathbf{p}\|_2 \|\mathbf{q}\|_2},
\label{eq:avg_cmplx}
\end{equation}
with

\begin{equation*}
    \mathbf{p} = 
    \begin{bmatrix} 
        \latt(m_i) - \latt(m_{i-1}) \\ 
        \lon(m_i) - \lon(m_{i-1}) 
    \end{bmatrix}, \quad 
    \mathbf{q} = 
    \begin{bmatrix} 
        \latt(m_{i+1}) - \latt(m_{i}) \\ 
        \lon(m_{i+1}) - \lon(m_{i}) 
    \end{bmatrix},
\end{equation*}
which can be interpreted as the average cosine of the angle that the connecting lines of three consecutive messages enclose over the entire trajectory $\mathcal{T}_k^>$. The usage of this formula, however, has some drawbacks. In the original paper by \citet{mao2018automatic}, only routes with an average complexity of more than 0.8 were included in the selection of accepted trajectories. We argue that the term complexity in this application is misleading as higher values of \emph{average complexity} correspond to trajectories with smaller average course changes between messages. Additionally, the resulting values can be difficult to interpret in a maritime context as they merely describe a mathematical property of the examined trajectory. 

Conveniently, a small adaption allows us to mitigate these shortcomings and construct an easily interpretable metric for assessing the \emph{cleanliness} of a trajectory. We define the \emph{average absolute change of the course} as

\begin{equation}
    \bar{\Delta}(\mathcal{T}^>_k) = \pi^{-1} \cos^{-1}[\bar{c}(\mathcal{T}^>_k)]\times 180, \quad \text{for}  \quad |\mathcal{T}^>_k| > 3,
\end{equation}
which gives us a good intuition about the magnitude of directional changes in one trajectory in degrees. For a value of $\bar{c}(\mathcal{T}^>_k)=0$, i.e., minimum complexity, the average absolute change of the course calculates to $180^\circ$ reflecting the U-turn behavior, while for $\bar{c}(\mathcal{T}^>_k)=1$ the average absolute change of the course $\bar{\Delta}(\mathcal{T}^>_k)=0^\circ$, encoding a perfectly straight trajectory.

\begin{figure}[t]
\centering
    \begin{subfigure}[t]{.48\textwidth}
      \centering
      \includegraphics[trim={0 1cm 0 1cm},clip,width=\linewidth]{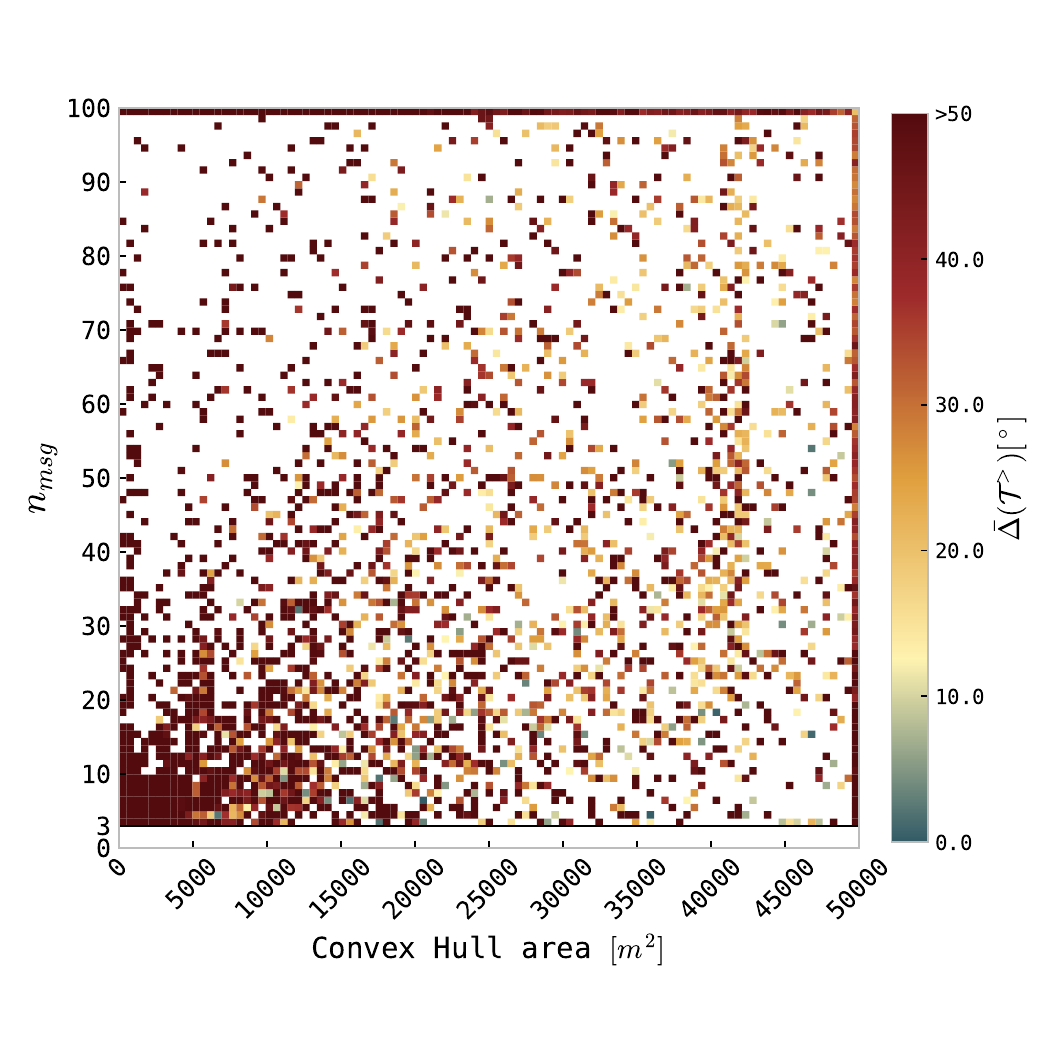}
      \caption{Average values of $\bar{\Delta}(\mathcal{T}^>)$ for unfiltered trajectories.}
      \label{avgsmth_before}
    \end{subfigure}
    \begin{subfigure}[t]{.48\textwidth}
        \centering
        \includegraphics[trim={0 1cm 0 1cm},clip,width=\linewidth]{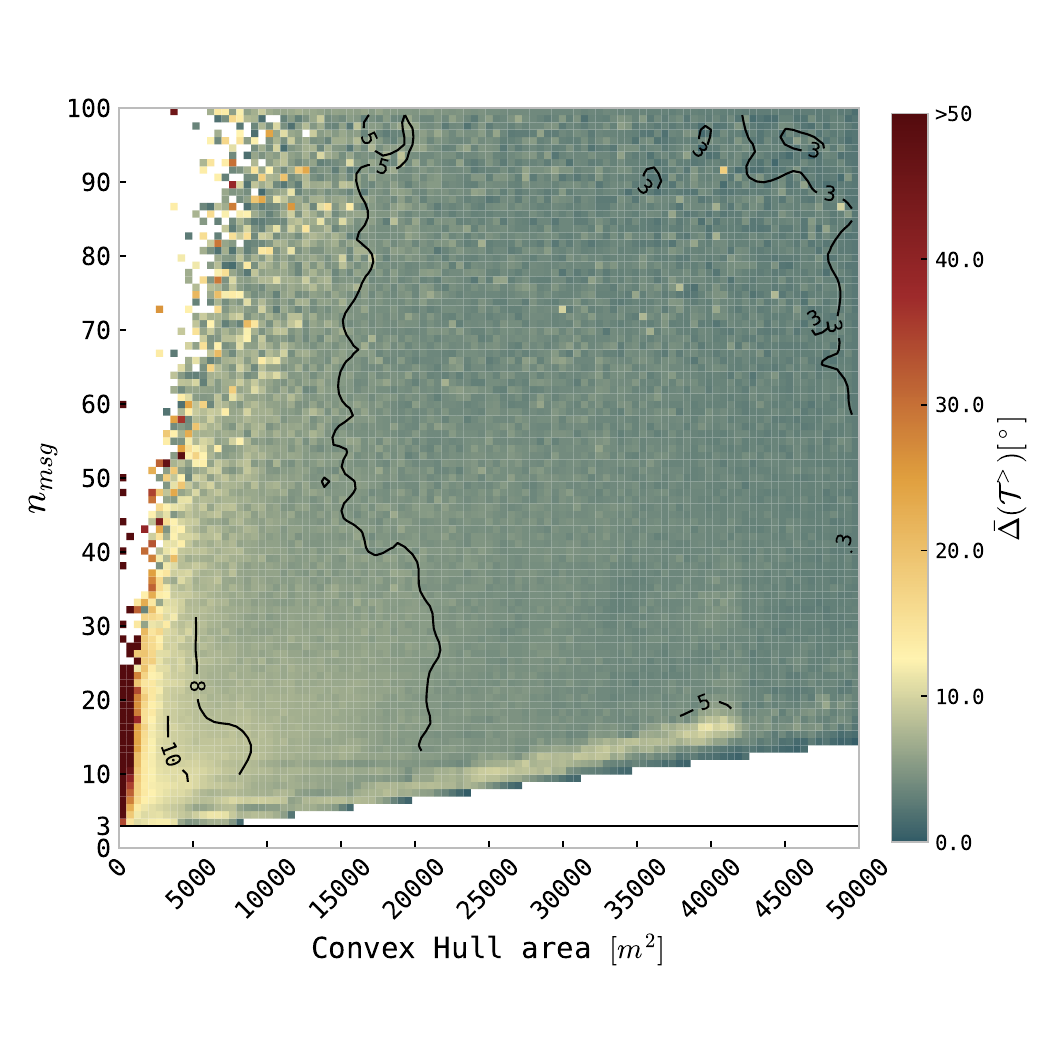}
        \caption{Average values of $\bar{\Delta}(\mathcal{T}^>)$ for trajectories after split-point filtering.}
        \label{avgsmth_after}
    \end{subfigure}
    \caption{$100\times100$ pixel map of $\bar{\Delta}(\mathcal{T}^>)$ in degrees as a function of the number of messages per track ($n_{\mathrm{msg}}$) and convex hull area ($A^C$). Each pixel resembles a bin, showing the average of $\bar{\Delta}(\mathcal{T}^>)$ for all trajectories falling inside it. (a) considers unfiltered trajectories, (b) used the same data but with the split-point procedure applied. White pixels indicate the absence of data. The black line at $n_{\mathrm{msg}}=3$ is the minimum number of messages a trajectory needs to comprise to calculate $\bar{\Delta}(\mathcal{T}^>)$. The convex hull area is restricted to $5\times 10^4$ for this visualization, as there are no changes in patterns above this threshold.}
    \label{influence}
\end{figure}

Figure \ref{influence} presents a pixel map illustrating the impact of the number of messages ($n_{\mathrm{msg}}$) and the convex hull area ($A^C$) on $\bar{\Delta}(\mathcal{T}_k^>)$ of trajectories in degrees. This analysis aggregates trajectories from 2021, both before (a) and following (b) the implementation of the split-point procedure, calculating $\bar{\Delta}(\mathcal{T}_k^>)$ for each pixel. Each pixel represents a bin that averages $\bar{\Delta}(\mathcal{T}_k^>)$ for trajectories within specific $n_{\mathrm{msg}}$ and $A^C$ ranges. The analysis is limited to $n_{\mathrm{msg}} \leq 100$ and $A^C \leq 5\times 10^4$ to maintain relevancy, as values beyond these thresholds yield negligible additional insights.

The results indicate that unprocessed trajectories exhibit no discernible pattern, suggesting that $n_{\mathrm{msg}}$ and $A^C$ are ineffective metrics for understanding unprocessed trajectory data. Conversely, applying the split-point procedure enhances the diversity of observed routes (fewer white spots) and demonstrates a fast trend towards higher smoothness values, particularly as $A^C$ increases.

\section{Discussion}
\label{discussion}

This study provides a framework for trajectory extraction from raw, encoded AIVDM sentences. A multi-level filtering and splitting procedure has been proposed to construct long, clean, uninterrupted trajectories. Particular emphasis lies on the data-driven nature of the splitting process of this study. While other approaches \citep{sang2015novel,zhao2018ship,zhang2018novel,yuan2019novel,chen2020ship}, too, propose functioning extraction processes, many threshold values used in those studies are still expert-derived, questioning their generalizability beyond the locally tested data. This study, instead, proposes to generate an empirical cumulative distribution function for the metric under consideration and control the statistical properties of the extracted trajectories freely by using the $1-\alpha$ quantile bounds to determine the threshold values. 

The methodology developed in this study demonstrates effective performance and aligns to establish a self-explanatory, data-driven approach for trajectory extraction. However, it is acknowledged that there exists no specific guidance on the selection of the quantile parameter $\alpha$, thus the final decision must be carefully considered by the practitioner, especially as many derived thresholds have no specific meaning regarding established AIS transceiving regulations or the maritime domain in general.

\section{Conclusion}
\label{conclusion}

Maritime AIS data is inherently noisy due to technical inaccuracies, seafarer inattention, or lacking compliance. However, trajectories extracted from AIS messages are valuable for maritime surveillance, domain awareness, and algorithmic testing, for which we propose a data-driven, open-source trajectory extraction and refinement framework. The extraction process is based on a split-point procedure that uses quantile boundaries of five empirically derived distribution functions to decide whether two consecutive messages belong to the same trajectory or should be split. We found that this procedure effectively eliminates anomalies in the data without resorting to expert-derived thresholds, making it universally adaptable to any AIS data set. Nevertheless, carefully scrutinizing the quantile boundary and threshold values is recommended, as they may vary depending on data availability. For spatial understanding, we introduced the concept of \emph{average absolute change of course}, $\bar{\Delta}(\cdot)$, which allows us to assess the presence or absence of erratic positional movement in a trajectory and recommended the number of messages per trajectory ($n_{\mathrm{msg}}$) as well as the convex hull area ($A^C$) as means to control $\bar{\Delta}(\cdot)$. The entire extraction and assessment pipeline, described by this article, is publicly available as an open-source Python package at \citet{pytsa2024}.

\section{Acknowledgments}

We thank the Center for Information Services and High-Performance Computing at TU Dresden for providing its facilities for high throughput calculations. We thank the European Maritime Safety Agency for providing the raw AIS records analyzed in this study. Further, we would also like to thank Martin Waltz for his valuable support throughout this project.

\bibliographystyle{cas-model2-names}

\begin{thebibliography}{45}
\expandafter\ifx\csname natexlab\endcsname\relax\def\natexlab#1{#1}\fi
\providecommand{\url}[1]{\texttt{#1}}
\providecommand{\href}[2]{#2}
\providecommand{\path}[1]{#1}
\providecommand{\DOIprefix}{doi:}
\providecommand{\ArXivprefix}{arXiv:}
\providecommand{\URLprefix}{URL: }
\providecommand{\Pubmedprefix}{pmid:}
\providecommand{\doi}[1]{\href{http://dx.doi.org/#1}{\path{#1}}}
\providecommand{\Pubmed}[1]{\href{pmid:#1}{\path{#1}}}
\providecommand{\bibinfo}[2]{#2}
\ifx\xfnm\relax \def\xfnm[#1]{\unskip,\space#1}\fi
\bibitem[{Amigo~Herrero et~al.(2019)Amigo~Herrero, Sanchez~Pedroche,
  Garcia~Herrero and Molina~Lopez}]{herrero2019ais}
\bibinfo{author}{Amigo~Herrero, D.}, \bibinfo{author}{Sanchez~Pedroche, D.},
  \bibinfo{author}{Garcia~Herrero, J.}, \bibinfo{author}{Molina~Lopez, J.M.},
  \bibinfo{year}{2019}.
\newblock \bibinfo{title}{Ais trajectory classification based on imm data}, in:
  \bibinfo{booktitle}{2019 22ND INTERNATIONAL CONFERENCE ON INFORMATION FUSION
  (FUSION 2019)}, \bibinfo{organization}{Int Soc Informat Fus; Thales Grp; Syst
  \& Technol Res; Univ Windsor}.
\newblock \DOIprefix\doi{10.23919/fusion43075.2019.9011384}.
  \bibinfo{note}{22nd International Conference on Information Fusion (FUSION),
  Ottawa, CANADA, JUL 02-05, 2019}.
\bibitem[{Bailey(2005)}]{bailey2005training}
\bibinfo{author}{Bailey, N.J.}, \bibinfo{year}{2005}.
\newblock \bibinfo{title}{Training, technology and ais: looking beyond the
  box}.
\newblock \bibinfo{journal}{Proceedings of the Seafarers International Research
  Centre’s Fourth International Symposium} .
\bibitem[{Barber et~al.(1996)Barber, Dobkin and
  Huhdanpaa}]{barber1996quickhull}
\bibinfo{author}{Barber, C.B.}, \bibinfo{author}{Dobkin, D.P.},
  \bibinfo{author}{Huhdanpaa, H.}, \bibinfo{year}{1996}.
\newblock \bibinfo{title}{The quickhull algorithm for convex hulls}.
\newblock \bibinfo{journal}{ACM Transactions on Mathematical Software (TOMS)}
  \bibinfo{volume}{22}, \bibinfo{pages}{469--483}.
\bibitem[{Capobianco et~al.(2021)Capobianco, Millefiori, Forti, Braca and
  Willett}]{capobianco2021deep}
\bibinfo{author}{Capobianco, S.}, \bibinfo{author}{Millefiori, L.M.},
  \bibinfo{author}{Forti, N.}, \bibinfo{author}{Braca, P.},
  \bibinfo{author}{Willett, P.}, \bibinfo{year}{2021}.
\newblock \bibinfo{title}{Deep learning methods for vessel trajectory
  prediction based on recurrent neural networks}.
\newblock \bibinfo{journal}{IEEE Transactions on Aerospace and Electronic
  Systems} \bibinfo{volume}{57}, \bibinfo{pages}{4329--4346}.
\bibitem[{Chen et~al.(2018)Chen, Shi, Liu and Gao}]{chen2018pattern}
\bibinfo{author}{Chen, P.}, \bibinfo{author}{Shi, G.}, \bibinfo{author}{Liu,
  S.}, \bibinfo{author}{Gao, M.}, \bibinfo{year}{2018}.
\newblock \bibinfo{title}{Pattern knowledge discovery of ship collision
  avoidance based on ais data analysis}.
\newblock \bibinfo{journal}{International Journal of Performability
  Engineering} \bibinfo{volume}{14}, \bibinfo{pages}{2449}.
\bibitem[{Chen et~al.(2020)Chen, Ling, Yang, Zheng, Xiong, Postolache and
  Xiong}]{chen2020ship}
\bibinfo{author}{Chen, X.}, \bibinfo{author}{Ling, J.}, \bibinfo{author}{Yang,
  Y.}, \bibinfo{author}{Zheng, H.}, \bibinfo{author}{Xiong, P.},
  \bibinfo{author}{Postolache, O.}, \bibinfo{author}{Xiong, Y.},
  \bibinfo{year}{2020}.
\newblock \bibinfo{title}{Ship trajectory reconstruction from ais sensory data
  via data quality control and prediction}.
\newblock \bibinfo{journal}{Mathematical Problems in Engineering}
  \bibinfo{volume}{2020}, \bibinfo{pages}{1--9}.
\bibitem[{Duan et~al.(2022)Duan, Ma, Miao and Zhang}]{duan2022asemi}
\bibinfo{author}{Duan, H.}, \bibinfo{author}{Ma, F.}, \bibinfo{author}{Miao,
  L.}, \bibinfo{author}{Zhang, C.}, \bibinfo{year}{2022}.
\newblock \bibinfo{title}{A semi-supervised deep learning approach for vessel
  trajectory classification based on ais data}.
\newblock \bibinfo{journal}{OCEAN \& COASTAL MANAGEMENT} \bibinfo{volume}{218}.
\newblock \DOIprefix\doi{10.1016/j.ocecoaman.2021.106015}.
\bibitem[{Gu et~al.(2023)Gu, Zhen, Liu and Li}]{gu2023improved}
\bibinfo{author}{Gu, Q.}, \bibinfo{author}{Zhen, R.}, \bibinfo{author}{Liu,
  J.}, \bibinfo{author}{Li, C.}, \bibinfo{year}{2023}.
\newblock \bibinfo{title}{An improved rrt algorithm based on prior ais
  information and dp compression for ship path planning}.
\newblock \bibinfo{journal}{Ocean Engineering} \bibinfo{volume}{279},
  \bibinfo{pages}{114595}.
\bibitem[{Guo et~al.(2021)Guo, Mou, Chen and Chen}]{guo2021improved}
\bibinfo{author}{Guo, S.}, \bibinfo{author}{Mou, J.}, \bibinfo{author}{Chen,
  L.}, \bibinfo{author}{Chen, P.}, \bibinfo{year}{2021}.
\newblock \bibinfo{title}{Improved kinematic interpolation for ais trajectory
  reconstruction}.
\newblock \bibinfo{journal}{Ocean Engineering} \bibinfo{volume}{234},
  \bibinfo{pages}{109256}.
\bibitem[{Hansen et~al.(2022)Hansen, Enevoldsen, Papageorgiou and
  Blanke}]{hansen2022autonomous}
\bibinfo{author}{Hansen, P.N.}, \bibinfo{author}{Enevoldsen, T.T.},
  \bibinfo{author}{Papageorgiou, D.}, \bibinfo{author}{Blanke, M.},
  \bibinfo{year}{2022}.
\newblock \bibinfo{title}{Autonomous navigation in confined waters-a colregs
  rule 9 compliant framework}.
\newblock \bibinfo{journal}{IFAC-PapersOnLine} \bibinfo{volume}{55},
  \bibinfo{pages}{222--228}.
\bibitem[{Harati-Mokhtari et~al.(2007)Harati-Mokhtari, Wall, Brooks and
  Wang}]{harati2007automatic}
\bibinfo{author}{Harati-Mokhtari, A.}, \bibinfo{author}{Wall, A.},
  \bibinfo{author}{Brooks, P.}, \bibinfo{author}{Wang, J.},
  \bibinfo{year}{2007}.
\newblock \bibinfo{title}{Automatic identification system (ais): data
  reliability and human error implications}.
\newblock \bibinfo{journal}{the Journal of Navigation} \bibinfo{volume}{60},
  \bibinfo{pages}{373--389}.
\bibitem[{He et~al.(2019)He, Zhang, Zhang, Zhang and Li}]{he2019ship}
\bibinfo{author}{He, Y.K.}, \bibinfo{author}{Zhang, D.},
  \bibinfo{author}{Zhang, J.}, \bibinfo{author}{Zhang, M.},
  \bibinfo{author}{Li, T.}, \bibinfo{year}{2019}.
\newblock \bibinfo{title}{Ship route planning using historical trajectories
  derived from ais data}.
\newblock \bibinfo{journal}{TransNav, International Journal on Marine
  Navigation and Safety od Sea Transportation} \bibinfo{volume}{13},
  \bibinfo{pages}{69--76}.
\bibitem[{{IMO}(2003)}]{imo-ais}
\bibinfo{author}{{IMO}}, \bibinfo{year}{2003}.
\newblock \bibinfo{title}{Regulations for carriage of ais}.
\newblock
  \bibinfo{howpublished}{\url{https://www.imo.org/en/OurWork/Safety/Pages/AIS.aspx}}.
\newblock \bibinfo{note}{Accessed: March 14, 2024}.
\bibitem[{{IMO}(2010)}]{IMO2010SN1Circ289}
\bibinfo{author}{{IMO}}, \bibinfo{year}{2010}.
\newblock \bibinfo{title}{Guidelines for the Presentation of
  Navigational-Related Symbols, Terms and Abbreviations}.
\newblock \bibinfo{type}{Circular} \bibinfo{number}{SN.1/Circ.289}.
  International Maritime Organization.
\newblock \URLprefix
  \url{https://vislab-ccom.unh.edu/~schwehr/papers/2010-IMO-SN.1-Circ.289.pdf}.
\bibitem[{IMO(2019)}]{imo2019shiprouteing}
\bibinfo{author}{IMO}, \bibinfo{year}{2019}.
\newblock \bibinfo{title}{Ships' Routeing}.
\newblock \bibinfo{publisher}{International Maritime Organization}.
\bibitem[{{ITU}(2001)}]{itu2001}
\bibinfo{author}{{ITU}}, \bibinfo{year}{2001}.
\newblock \bibinfo{title}{Technical characteristics for an automatic
  identification system using time-division multiple access in the VHF maritime
  mobile frequency band}.
\newblock \bibinfo{type}{Technical Report} \bibinfo{number}{ITU-R M.1371-1}.
  ITU-R.
\newblock \URLprefix
  \url{https://www.itu.int/dms_pubrec/itu-r/rec/m/R-REC-M.1371-1-200108-S!!PDF-E.pdf}.
\bibitem[{Jankowski et~al.(2021)Jankowski, Lamm and
  Hahn}]{jankowski2021determination}
\bibinfo{author}{Jankowski, D.}, \bibinfo{author}{Lamm, A.},
  \bibinfo{author}{Hahn, A.}, \bibinfo{year}{2021}.
\newblock \bibinfo{title}{Determination of ais position accuracy and evaluation
  of reconstruction methods for maritime observation data}.
\newblock \bibinfo{journal}{IFAC-PapersOnLine} \bibinfo{volume}{54},
  \bibinfo{pages}{97--104}.
\bibitem[{Karney(2011)}]{karney2011transverse}
\bibinfo{author}{Karney, C.F.}, \bibinfo{year}{2011}.
\newblock \bibinfo{title}{Transverse mercator with an accuracy of a few
  nanometers}.
\newblock \bibinfo{journal}{Journal of Geodesy} \bibinfo{volume}{85},
  \bibinfo{pages}{475--485}.
\bibitem[{Kr{\"u}ger(1912)}]{kruger1912konforme}
\bibinfo{author}{Kr{\"u}ger, L.}, \bibinfo{year}{1912}.
\newblock \bibinfo{title}{Konforme Abbildung des Erdellipsoids in der Ebene}.
\newblock \bibinfo{number}{52}, \bibinfo{publisher}{BG Teubner}.
\bibitem[{Luo et~al.(2023)Luo, Chen, Yang, Li and Zhao}]{luo2023anew}
\bibinfo{author}{Luo, D.}, \bibinfo{author}{Chen, P.}, \bibinfo{author}{Yang,
  J.}, \bibinfo{author}{Li, X.}, \bibinfo{author}{Zhao, Y.},
  \bibinfo{year}{2023}.
\newblock \bibinfo{title}{A new classification method for ship trajectories
  based on ais data}.
\newblock \bibinfo{journal}{JOURNAL OF MARINE SCIENCE AND ENGINEERING}
  \bibinfo{volume}{11}.
\newblock \DOIprefix\doi{10.3390/jmse11091646}.
\bibitem[{Mahmoud and Akkari(2016)}]{mahmoud2016shortest}
\bibinfo{author}{Mahmoud, H.}, \bibinfo{author}{Akkari, N.},
  \bibinfo{year}{2016}.
\newblock \bibinfo{title}{Shortest path calculation: a comparative study for
  location-based recommender system}, in: \bibinfo{booktitle}{2016 world
  symposium on computer applications \& research (WSCAR)},
  \bibinfo{organization}{IEEE}. pp. \bibinfo{pages}{1--5}.
\bibitem[{Mao et~al.(2018)Mao, Tu, Zhang, Rachmawati, Rajabally and
  Huang}]{mao2018automatic}
\bibinfo{author}{Mao, S.}, \bibinfo{author}{Tu, E.}, \bibinfo{author}{Zhang,
  G.}, \bibinfo{author}{Rachmawati, L.}, \bibinfo{author}{Rajabally, E.},
  \bibinfo{author}{Huang, G.B.}, \bibinfo{year}{2018}.
\newblock \bibinfo{title}{An automatic identification system (ais) database for
  maritime trajectory prediction and data mining}, in:
  \bibinfo{booktitle}{Proceedings of ELM-2016},
  \bibinfo{organization}{Springer}. pp. \bibinfo{pages}{241--257}.
\bibitem[{Morien(2023)}]{pyais}
\bibinfo{author}{Morien, L.}, \bibinfo{year}{2023}.
\newblock \bibinfo{title}{pyais: Ais message decoding and encoding in python
  (aivdm/aivdo)}.
\newblock \bibinfo{howpublished}{\url{https://github.com/M0r13n/pyais}}.
\bibitem[{Mou et~al.(2010)Mou, Van~der Tak and Ligteringen}]{mou2010study}
\bibinfo{author}{Mou, J.M.}, \bibinfo{author}{Van~der Tak, C.},
  \bibinfo{author}{Ligteringen, H.}, \bibinfo{year}{2010}.
\newblock \bibinfo{title}{Study on collision avoidance in busy waterways by
  using ais data}.
\newblock \bibinfo{journal}{Ocean Engineering} \bibinfo{volume}{37},
  \bibinfo{pages}{483--490}.
\bibitem[{Notteboom and Cariou(2009)}]{notteboom2009fuel}
\bibinfo{author}{Notteboom, T.}, \bibinfo{author}{Cariou, P.},
  \bibinfo{year}{2009}.
\newblock \bibinfo{title}{Fuel surcharge practices of container shipping lines:
  Is it about cost recovery or revenue making}, in:
  \bibinfo{booktitle}{Proceedings of the 2009 international association of
  maritime economists (IAME) conference}, \bibinfo{organization}{IAME
  Copenhagen, Denmark}. pp. \bibinfo{pages}{24--26}.
\bibitem[{Pallotta et~al.(2013)Pallotta, Vespe and Bryan}]{pallotta2013vessel}
\bibinfo{author}{Pallotta, G.}, \bibinfo{author}{Vespe, M.},
  \bibinfo{author}{Bryan, K.}, \bibinfo{year}{2013}.
\newblock \bibinfo{title}{Vessel pattern knowledge discovery from ais data: A
  framework for anomaly detection and route prediction}.
\newblock \bibinfo{journal}{Entropy} \bibinfo{volume}{15},
  \bibinfo{pages}{2218--2245}.
\bibitem[{Paulig(2024)}]{pytsa2024}
\bibinfo{author}{Paulig, N.}, \bibinfo{year}{2024}.
\newblock \bibinfo{title}{{PyTSA}: Python trajectory splitting and assessment
  agent for ais data}.
\newblock \bibinfo{howpublished}{\url{https://github.com/nikpau/pytsa}}.
\bibitem[{Rong et~al.(2020)Rong, Teixeira and Soares}]{rong2020data}
\bibinfo{author}{Rong, H.}, \bibinfo{author}{Teixeira, A.},
  \bibinfo{author}{Soares, C.G.}, \bibinfo{year}{2020}.
\newblock \bibinfo{title}{Data mining approach to shipping route
  characterization and anomaly detection based on ais data}.
\newblock \bibinfo{journal}{Ocean Engineering} \bibinfo{volume}{198},
  \bibinfo{pages}{106936}.
\bibitem[{Rong et~al.(2022)Rong, Teixeira and Soares}]{rong2022ship}
\bibinfo{author}{Rong, H.}, \bibinfo{author}{Teixeira, A.},
  \bibinfo{author}{Soares, C.G.}, \bibinfo{year}{2022}.
\newblock \bibinfo{title}{Ship collision avoidance behaviour recognition and
  analysis based on ais data}.
\newblock \bibinfo{journal}{Ocean Engineering} \bibinfo{volume}{245},
  \bibinfo{pages}{110479}.
\bibitem[{Sanchez~Pedroche et~al.(2020)Sanchez~Pedroche, Amigo, Garcia and
  Manuel~Molina}]{sanchez2020arch}
\bibinfo{author}{Sanchez~Pedroche, D.}, \bibinfo{author}{Amigo, D.},
  \bibinfo{author}{Garcia, J.}, \bibinfo{author}{Manuel~Molina, J.},
  \bibinfo{year}{2020}.
\newblock \bibinfo{title}{Architecture for trajectory-based fishing ship
  classification with ais data}.
\newblock \bibinfo{journal}{SENSORS} \bibinfo{volume}{20}.
\newblock \DOIprefix\doi{10.3390/s20133782}.
\bibitem[{Sang et~al.(2015)Sang, Wall, Mao, Yan and Wang}]{sang2015novel}
\bibinfo{author}{Sang, L.z.}, \bibinfo{author}{Wall, A.}, \bibinfo{author}{Mao,
  Z.}, \bibinfo{author}{Yan, X.p.}, \bibinfo{author}{Wang, J.},
  \bibinfo{year}{2015}.
\newblock \bibinfo{title}{A novel method for restoring the trajectory of the
  inland waterway ship by using ais data}.
\newblock \bibinfo{journal}{Ocean Engineering} \bibinfo{volume}{110},
  \bibinfo{pages}{183--194}.
\bibitem[{Silveira et~al.(2013)Silveira, Teixeira and Soares}]{silveira2013use}
\bibinfo{author}{Silveira, P.}, \bibinfo{author}{Teixeira, A.},
  \bibinfo{author}{Soares, C.G.}, \bibinfo{year}{2013}.
\newblock \bibinfo{title}{Use of ais data to characterise marine traffic
  patterns and ship collision risk off the coast of portugal}.
\newblock \bibinfo{journal}{The Journal of Navigation} \bibinfo{volume}{66},
  \bibinfo{pages}{879--898}.
\bibitem[{Vincenty(1975)}]{vincenty1975direct}
\bibinfo{author}{Vincenty, T.}, \bibinfo{year}{1975}.
\newblock \bibinfo{title}{Direct and inverse solutions of geodesics on the
  ellipsoid with application of nested equations}.
\newblock \bibinfo{journal}{Survey review} \bibinfo{volume}{23},
  \bibinfo{pages}{88--93}.
\bibitem[{Waltz et~al.(2023)Waltz, Paulig and Okhrin}]{waltz2023}
\bibinfo{author}{Waltz, M.}, \bibinfo{author}{Paulig, N.},
  \bibinfo{author}{Okhrin, O.}, \bibinfo{year}{2023}.
\newblock \bibinfo{title}{2-level reinforcement learning for ships on inland
  waterways}.
\newblock \bibinfo{journal}{arXiv preprint arXiv:2307.16769} .
\bibitem[{Wilcox et~al.(2014)Wilcox, Erceg-Hurn, Clark and
  Carlson}]{wilcox2014comparing}
\bibinfo{author}{Wilcox, R.R.}, \bibinfo{author}{Erceg-Hurn, D.M.},
  \bibinfo{author}{Clark, F.}, \bibinfo{author}{Carlson, M.},
  \bibinfo{year}{2014}.
\newblock \bibinfo{title}{Comparing two independent groups via the lower and
  upper quantiles}.
\newblock \bibinfo{journal}{Journal of Statistical Computation and Simulation}
  \bibinfo{volume}{84}, \bibinfo{pages}{1543--1551}.
\bibitem[{Wolsing et~al.(2022)Wolsing, Roepert, Bauer and
  Wehrle}]{wolsing2022anomaly}
\bibinfo{author}{Wolsing, K.}, \bibinfo{author}{Roepert, L.},
  \bibinfo{author}{Bauer, J.}, \bibinfo{author}{Wehrle, K.},
  \bibinfo{year}{2022}.
\newblock \bibinfo{title}{Anomaly detection in maritime ais tracks: A review of
  recent approaches}.
\newblock \bibinfo{journal}{Journal of Marine Science and Engineering}
  \bibinfo{volume}{10}, \bibinfo{pages}{112}.
\bibitem[{Wu et~al.(2017)Wu, Xu, Wang, Wang and Xu}]{wu2017mapping}
\bibinfo{author}{Wu, L.}, \bibinfo{author}{Xu, Y.}, \bibinfo{author}{Wang, Q.},
  \bibinfo{author}{Wang, F.}, \bibinfo{author}{Xu, Z.}, \bibinfo{year}{2017}.
\newblock \bibinfo{title}{Mapping global shipping density from ais data}.
\newblock \bibinfo{journal}{The Journal of Navigation} \bibinfo{volume}{70},
  \bibinfo{pages}{67--81}.
\bibitem[{Xu et~al.(2019)Xu, Rong and Soares}]{xu2019use}
\bibinfo{author}{Xu, H.}, \bibinfo{author}{Rong, H.}, \bibinfo{author}{Soares,
  C.G.}, \bibinfo{year}{2019}.
\newblock \bibinfo{title}{Use of ais data for guidance and control of
  path-following autonomous vessels}.
\newblock \bibinfo{journal}{Ocean Engineering} \bibinfo{volume}{194},
  \bibinfo{pages}{106635}.
\bibitem[{Yan et~al.(2020)Yan, Xiao, Cheng, Chen, Zhou, Ruan, Li, He and
  Ran}]{yan2020analysis}
\bibinfo{author}{Yan, Z.}, \bibinfo{author}{Xiao, Y.}, \bibinfo{author}{Cheng,
  L.}, \bibinfo{author}{Chen, S.}, \bibinfo{author}{Zhou, X.},
  \bibinfo{author}{Ruan, X.}, \bibinfo{author}{Li, M.}, \bibinfo{author}{He,
  R.}, \bibinfo{author}{Ran, B.}, \bibinfo{year}{2020}.
\newblock \bibinfo{title}{Analysis of global marine oil trade based on
  automatic identification system (ais) data}.
\newblock \bibinfo{journal}{Journal of Transport Geography}
  \bibinfo{volume}{83}, \bibinfo{pages}{102637}.
\bibitem[{Yuan et~al.(2019)Yuan, Liu, Liu and Li}]{yuan2019novel}
\bibinfo{author}{Yuan, Z.}, \bibinfo{author}{Liu, J.}, \bibinfo{author}{Liu,
  Y.}, \bibinfo{author}{Li, Z.}, \bibinfo{year}{2019}.
\newblock \bibinfo{title}{A novel approach for vessel trajectory reconstruction
  using ais data}, in: \bibinfo{booktitle}{ISOPE International Ocean and Polar
  Engineering Conference}, \bibinfo{organization}{ISOPE}. pp.
  \bibinfo{pages}{ISOPE--I}.
\bibitem[{Zhang et~al.(2023)Zhang, Liu, Hirdaris, Zhang and
  Tian}]{zhang2023interpretable}
\bibinfo{author}{Zhang, J.}, \bibinfo{author}{Liu, J.},
  \bibinfo{author}{Hirdaris, S.}, \bibinfo{author}{Zhang, M.},
  \bibinfo{author}{Tian, W.}, \bibinfo{year}{2023}.
\newblock \bibinfo{title}{An interpretable knowledge-based decision support
  method for ship collision avoidance using ais data}.
\newblock \bibinfo{journal}{Reliability Engineering \& System Safety}
  \bibinfo{volume}{230}, \bibinfo{pages}{108919}.
\bibitem[{Zhang et~al.(2018)Zhang, Meng, Xiao and Fu}]{zhang2018novel}
\bibinfo{author}{Zhang, L.}, \bibinfo{author}{Meng, Q.}, \bibinfo{author}{Xiao,
  Z.}, \bibinfo{author}{Fu, X.}, \bibinfo{year}{2018}.
\newblock \bibinfo{title}{A novel ship trajectory reconstruction approach using
  ais data}.
\newblock \bibinfo{journal}{Ocean Engineering} \bibinfo{volume}{159},
  \bibinfo{pages}{165--174}.
\bibitem[{Zhang et~al.(2016)Zhang, Goerlandt, Kujala and
  Wang}]{zhang2016advanced}
\bibinfo{author}{Zhang, W.}, \bibinfo{author}{Goerlandt, F.},
  \bibinfo{author}{Kujala, P.}, \bibinfo{author}{Wang, Y.},
  \bibinfo{year}{2016}.
\newblock \bibinfo{title}{An advanced method for detecting possible near miss
  ship collisions from ais data}.
\newblock \bibinfo{journal}{Ocean Engineering} \bibinfo{volume}{124},
  \bibinfo{pages}{141--156}.
\bibitem[{Zhao et~al.(2018)Zhao, Shi and Yang}]{zhao2018ship}
\bibinfo{author}{Zhao, L.}, \bibinfo{author}{Shi, G.}, \bibinfo{author}{Yang,
  J.}, \bibinfo{year}{2018}.
\newblock \bibinfo{title}{Ship trajectories pre-processing based on ais data}.
\newblock \bibinfo{journal}{The Journal of Navigation} \bibinfo{volume}{71},
  \bibinfo{pages}{1210--1230}.
\bibitem[{Zhen et~al.(2017)Zhen, Jin, Hu, Shao and
  Nikitakos}]{zhen2017maritime}
\bibinfo{author}{Zhen, R.}, \bibinfo{author}{Jin, Y.}, \bibinfo{author}{Hu,
  Q.}, \bibinfo{author}{Shao, Z.}, \bibinfo{author}{Nikitakos, N.},
  \bibinfo{year}{2017}.
\newblock \bibinfo{title}{Maritime anomaly detection within coastal waters
  based on vessel trajectory clustering and na{\"\i}ve bayes classifier}.
\newblock \bibinfo{journal}{The Journal of Navigation} \bibinfo{volume}{70},
  \bibinfo{pages}{648--670}.

\end{thebibliography}

\pagebreak
\appendix

\section{$A^C_v$ for different ship types}

Given that $ship\_type(MMSI)$ is a function mapping an MMSI to its corresponding ship type $v$, and $A^{C_{\mathcal{T}_k^>}}$ is the convex hull area for trajectory $\mathcal{T}_k^>$, the ship-type-average of their convex hull area can be found via 

\begin{equation}
    \bar{A}^C_v=|\mathcal{V}|^{-1}\sum_{\mathcal{T}_k^>\in\mathcal{V}} A^{C_{\mathcal{T}_k^>}},
\end{equation}
where $\mathcal{V} = \{\mathcal{T}_k^> | ship\_type(k)=v\}$ is the set of all trajectories belonging to one ship type.

\begin{table}[h]
    \centering
    \begin{tabular}{l|r}
        $v$ & $\bar{A}^C_v$ [\si{\metre\squared}] \\\hline
        \texttt{CARGO} & \num{174100} \\
        \texttt{TANKER} & \num{173443} \\
        \texttt{NOTAVAILABLE} & \num{141221} \\
        \texttt{TUGTOW} & \num{126509} \\
        \texttt{PASSENGER} & \num{115021} \\
        \texttt{MILITARY} & \num{109111} \\
        \texttt{OTHER} & \num{99771} \\
        \texttt{PLEASURE} & \num{93902} \\
        \texttt{SAILING} & \num{79730} \\
        \texttt{WIG} & \num{78384} \\
        \texttt{FISHING} & \num{60610} \\
        \texttt{HSC} & \num{37618} \\
    \end{tabular}
    \caption{Average convex hull area for all trajectories with $n_{\mathrm{msg}}>50$ for all ship types for the year 2021. The ship types are specified according to ITU standards \citep{itu2001}. For this analysis, only the base ship type had been distinguished; for example, any tanker (types 80 - 89) is included in the \texttt{TANKER} category. \texttt{NOTAVAILABLE} refers to the default if no ship type has been transmitted. \texttt{OTHER} are all ships not fitting into the above categories.}
    \label{avgsd}
\end{table}

Table \ref{avgsd} collects these averages. In line with our intuition, cargo ships' routes cover the largest average area of all ship types, as one of their primary goals is to ship goods effectively over long distances. Interestingly, Tug and tow boats cover the 3rd highest convex hull area. This finding might initially seem paradoxical given their role in assistive navigation in ports or terminals, typically associated with short and compressed trajectories. However, this observation can be rationalized by considering that our analysis only includes trajectories with more than 50 messages and a minimum speed of $1kn$. A visual inspection using heatmaps in Figures \ref{heatmaps-1} and \ref{heatmaps-2} revealed that Tug and Towboats travel longer routes than expected under these circumstances. On the other hand, High-Speed-Crafts (\texttt{HSC}) are primarily used as high-speed ferries for short passages in commercial contexts, which explains their relatively small area covered.

\begin{figure}[htp!]
\centering
    \begin{subfigure}[t]{.48\textwidth}
      \centering
      \includegraphics[trim={1cm 1cm 1cm 1cm},clip,width=\linewidth]{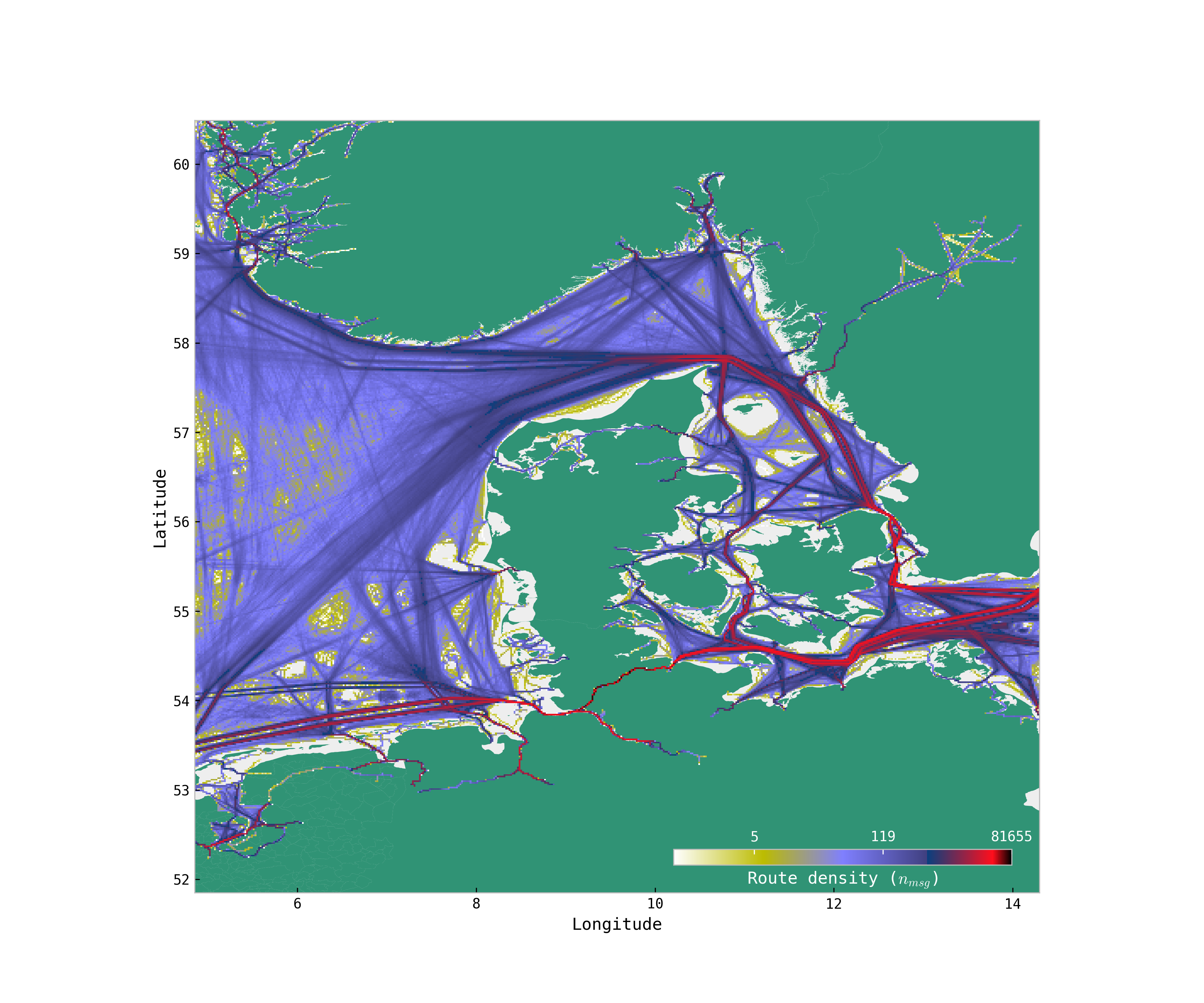}
      \caption{Cargo}
    \end{subfigure}
    \begin{subfigure}[t]{.48\textwidth}
      \centering
      \includegraphics[trim={1cm 1cm 1cm 1cm},clip,width=\linewidth]{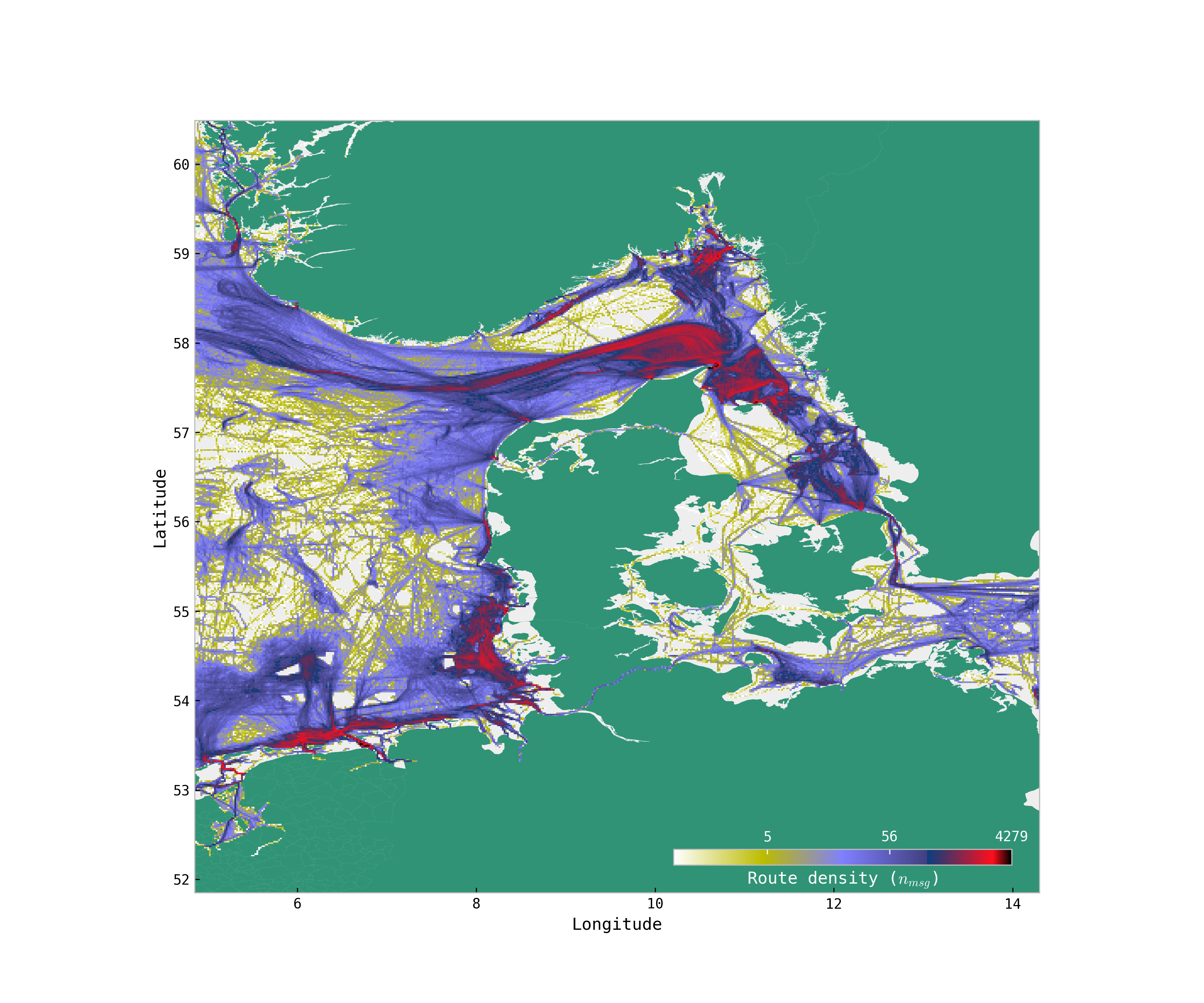}
      \caption{Fishing}
    \end{subfigure}
    \begin{subfigure}[t]{.48\textwidth}
      \centering
      \includegraphics[trim={1cm 1cm 1cm 1cm},clip,width=\linewidth]{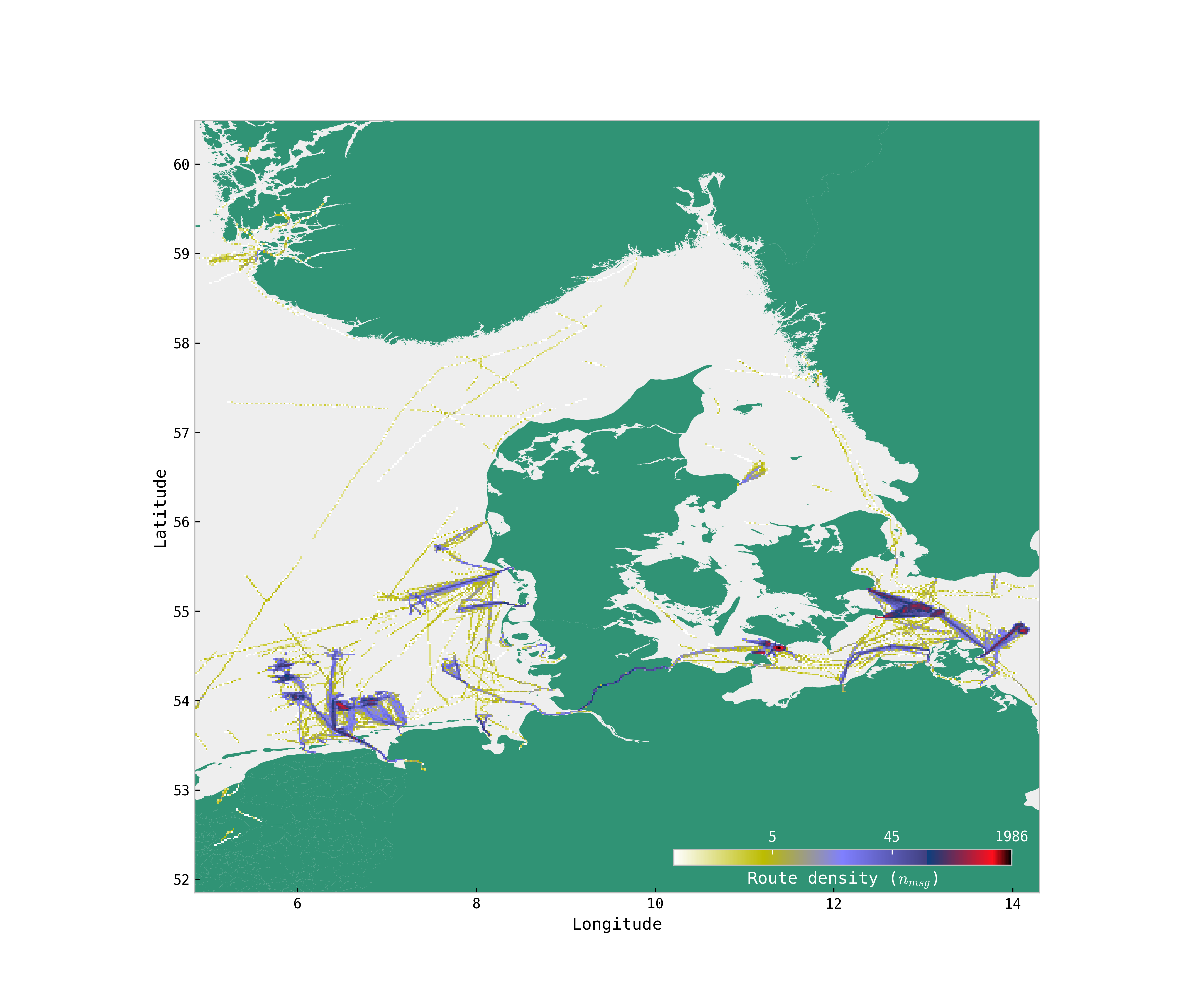}
      \caption{High Speed Crafts}
    \end{subfigure}
    \begin{subfigure}[t]{.48\textwidth}
      \centering
      \includegraphics[trim={1cm 1cm 1cm 1cm},clip,width=\linewidth]{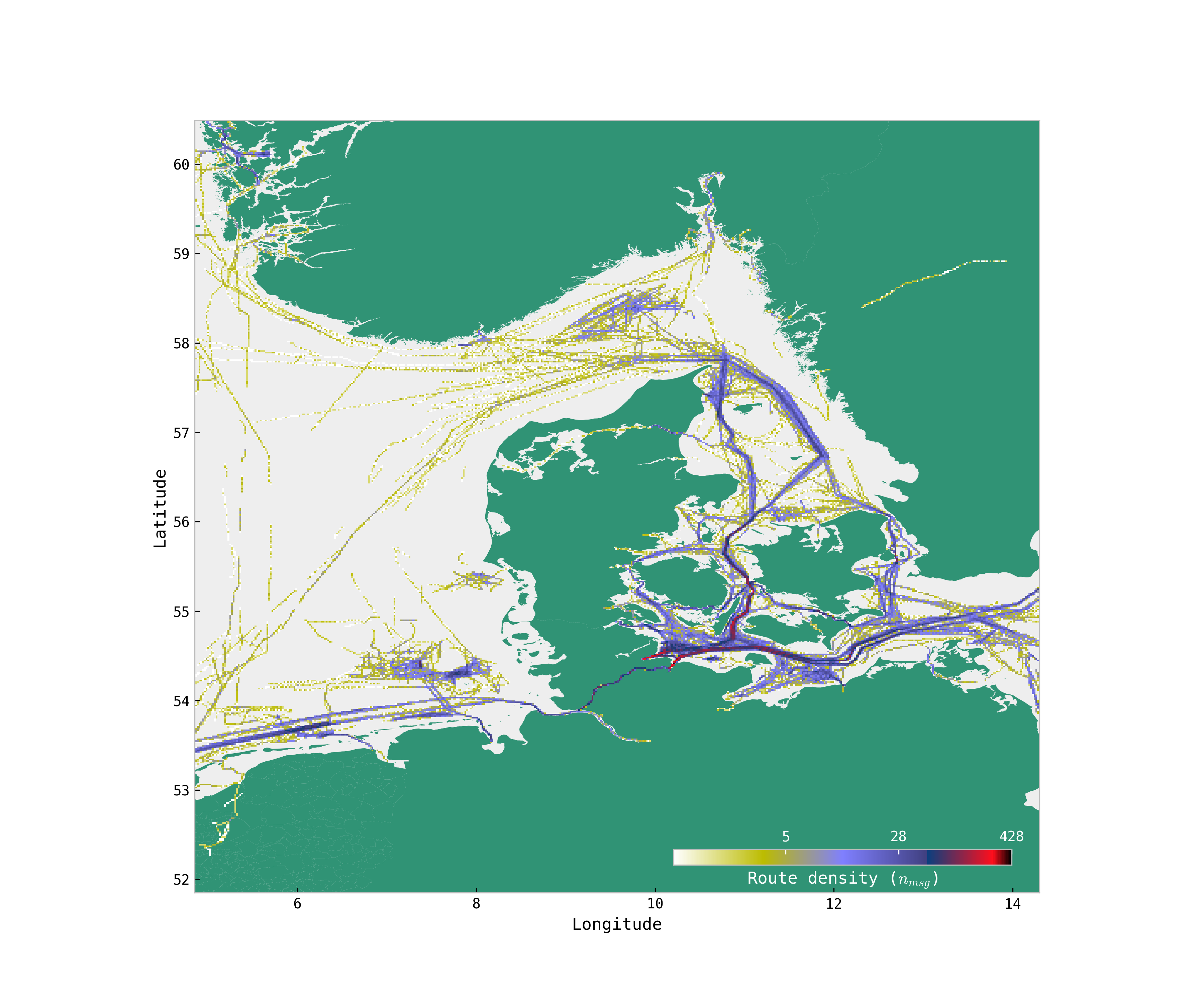}
      \caption{Military}
    \end{subfigure}
    \begin{subfigure}[t]{.48\textwidth}
      \centering
      \includegraphics[trim={1cm 1cm 1cm 1cm},clip,width=\linewidth]{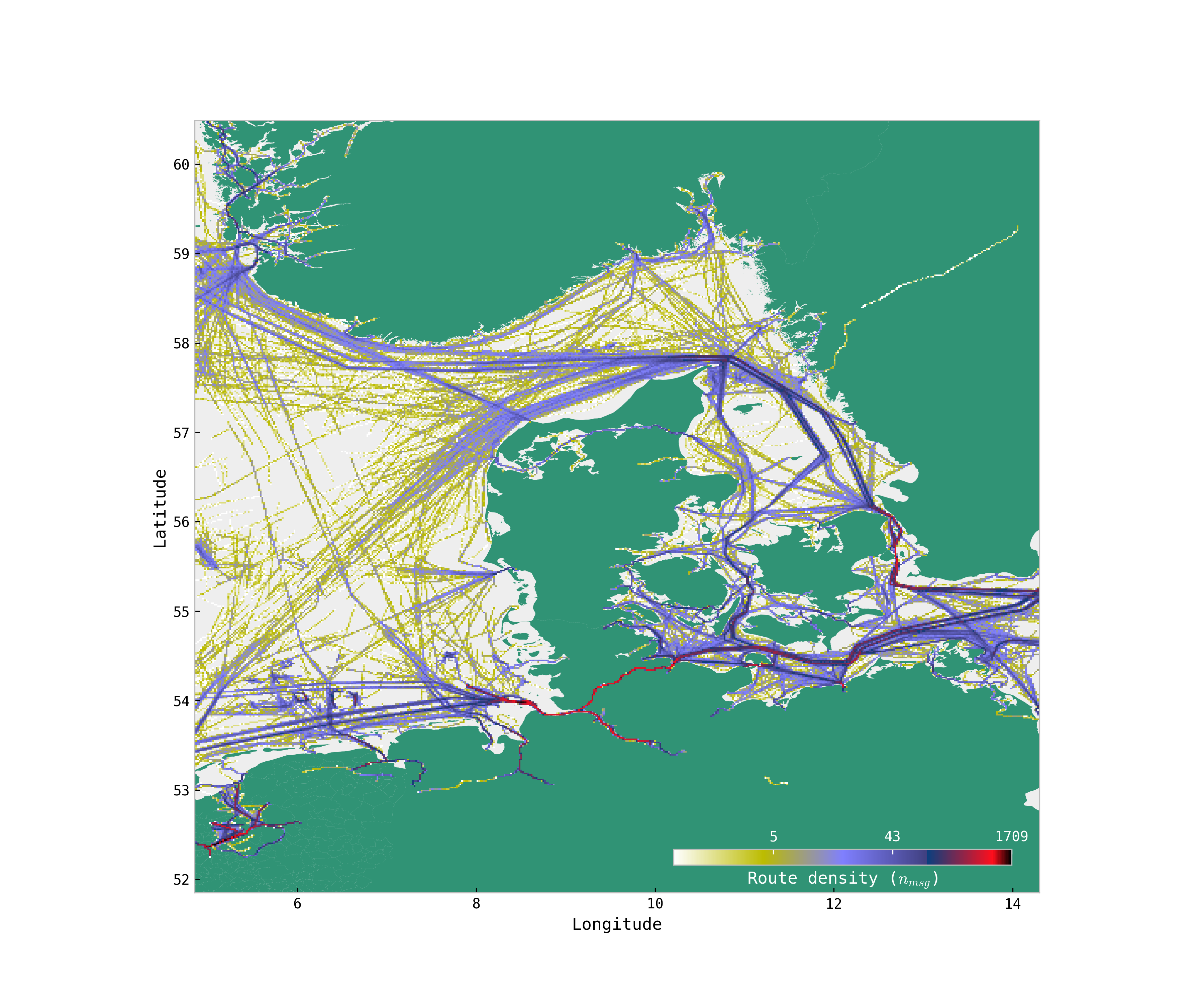}
      \caption{Not available}
    \end{subfigure}
    \begin{subfigure}[t]{.48\textwidth}
      \centering
      \includegraphics[trim={1cm 1cm 1cm 1cm},clip,width=\linewidth]{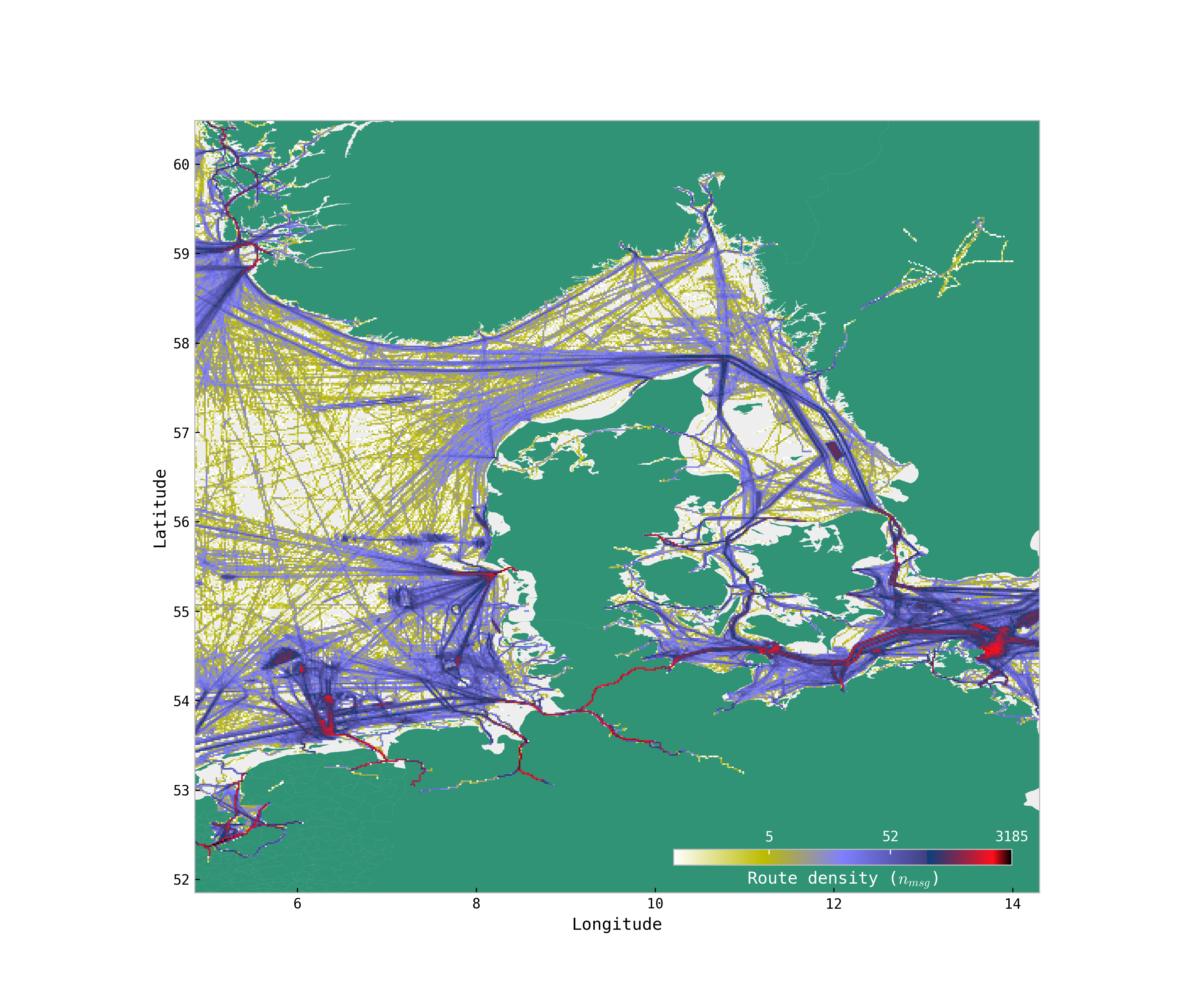}
      \caption{Other}
    \end{subfigure}
    \caption{Heatmap of route densities for different ship types for the year 2021. Plots generated from raw AIS records using the split-point method and $A^C>3\times 10^4 m^2$ and $n_{\mathrm{msg}}>50$-refinement}
    \label{heatmaps-1}
\end{figure}

\begin{figure}[htp!]
\centering
    \begin{subfigure}[t]{.49\textwidth}
      \centering
      \includegraphics[trim={1cm 1cm 1cm 1cm},clip,width=\linewidth]{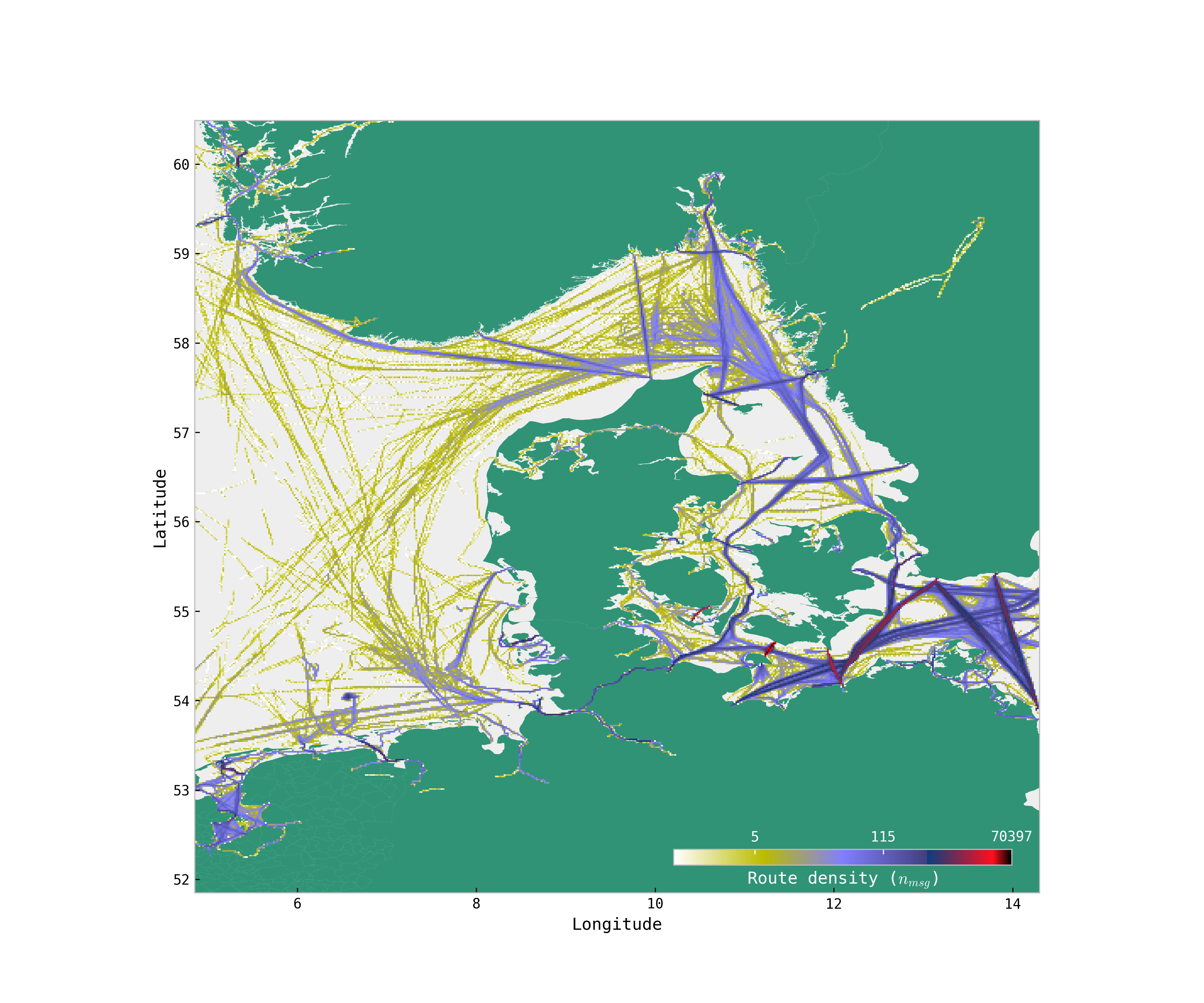}
      \caption{Passenger}
    \end{subfigure}
    \begin{subfigure}[t]{.49\textwidth}
      \centering
      \includegraphics[trim={1cm 1cm 1cm 1cm},clip,width=\linewidth]{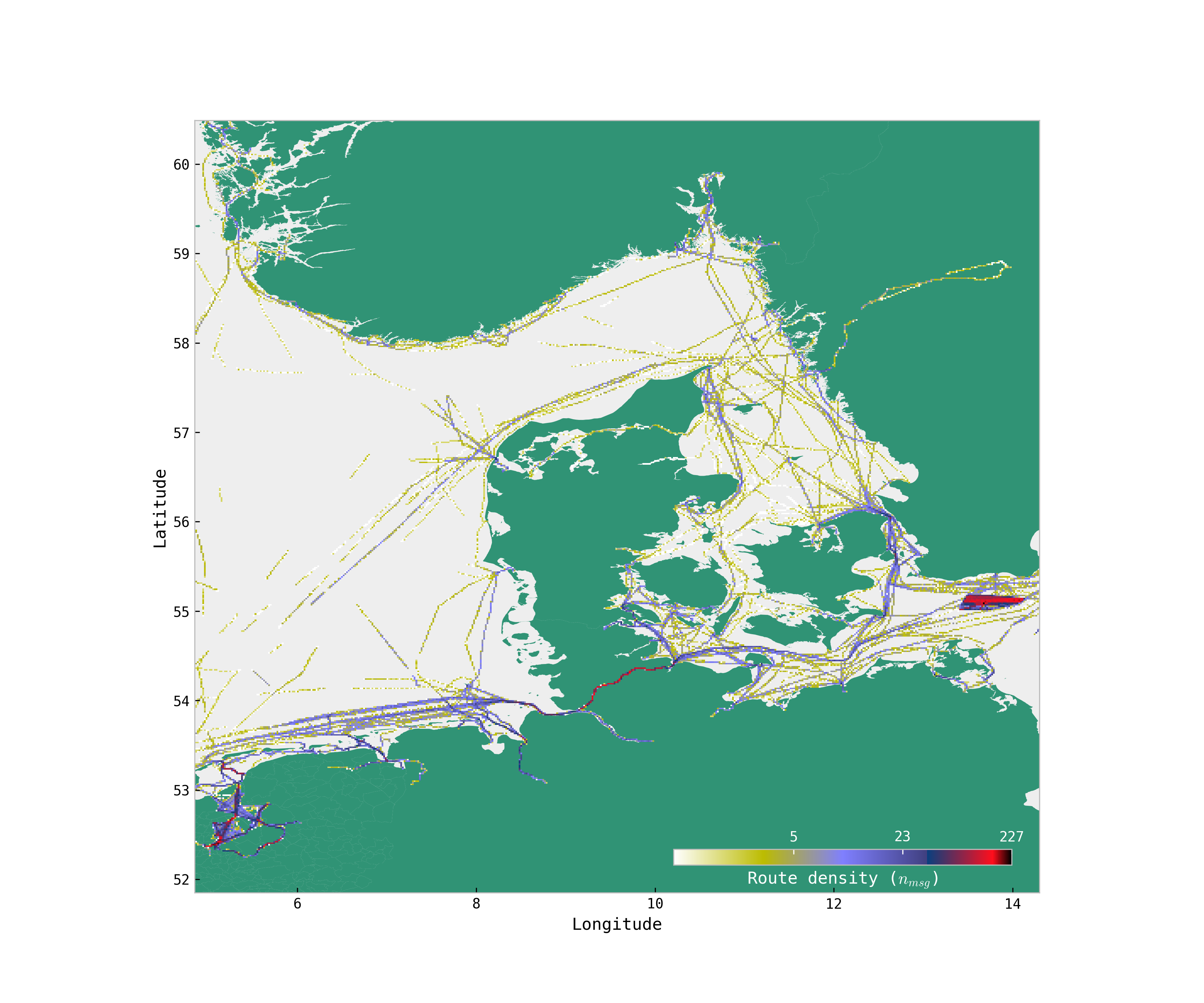}
      \caption{Pleasure}
    \end{subfigure}
    \begin{subfigure}[t]{.49\textwidth}
      \centering
      \includegraphics[trim={1cm 1cm 1cm 1cm},clip,width=\linewidth]{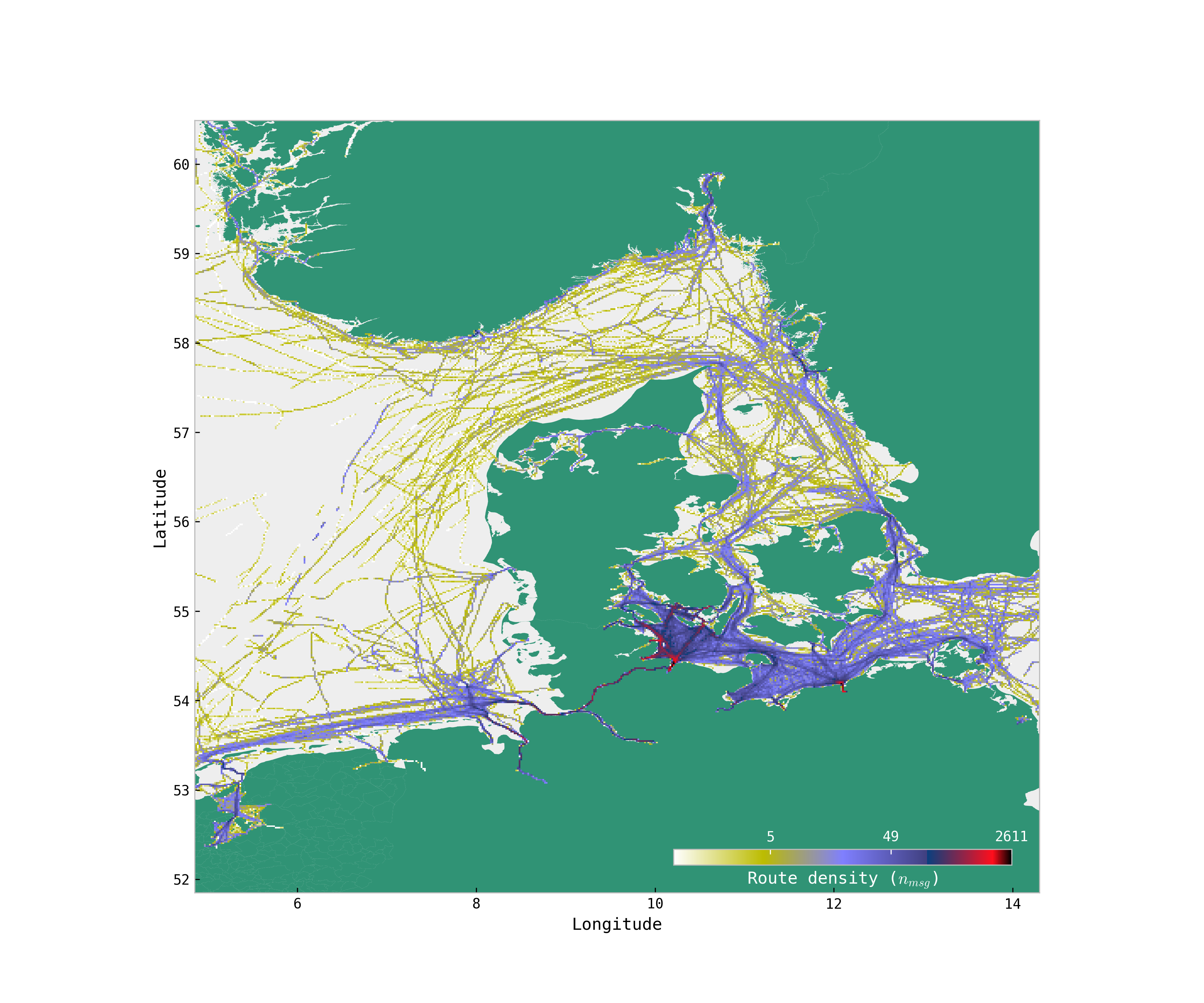}
      \caption{Sailing}
    \end{subfigure}
    \begin{subfigure}[t]{.49\textwidth}
      \centering
      \includegraphics[trim={1cm 1cm 1cm 1cm},clip,width=\linewidth]{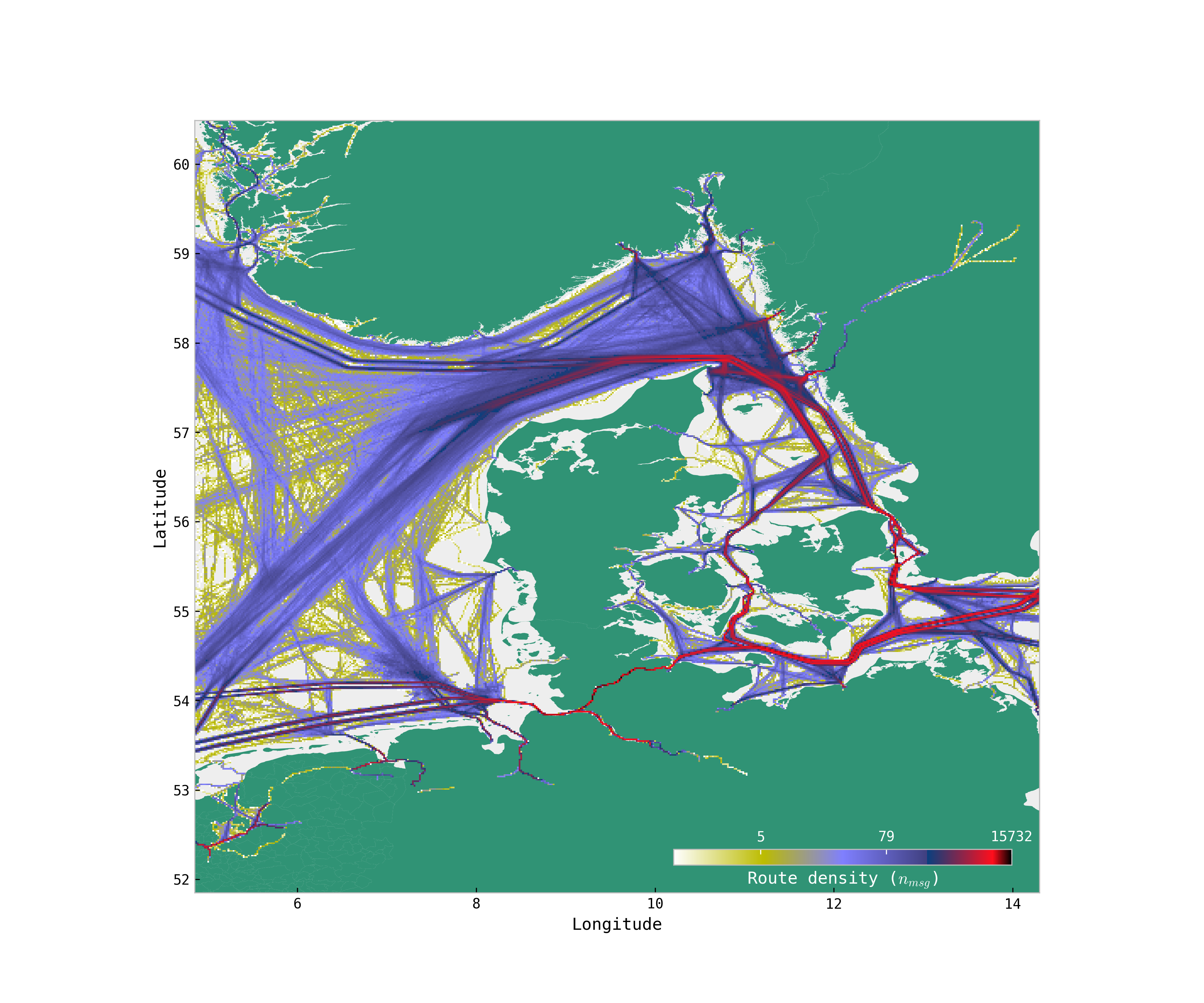}
      \caption{Tanker}
    \end{subfigure}
    \begin{subfigure}[t]{.49\textwidth}
      \centering
      \includegraphics[trim={1cm 1cm 1cm 1cm},clip,width=\linewidth]{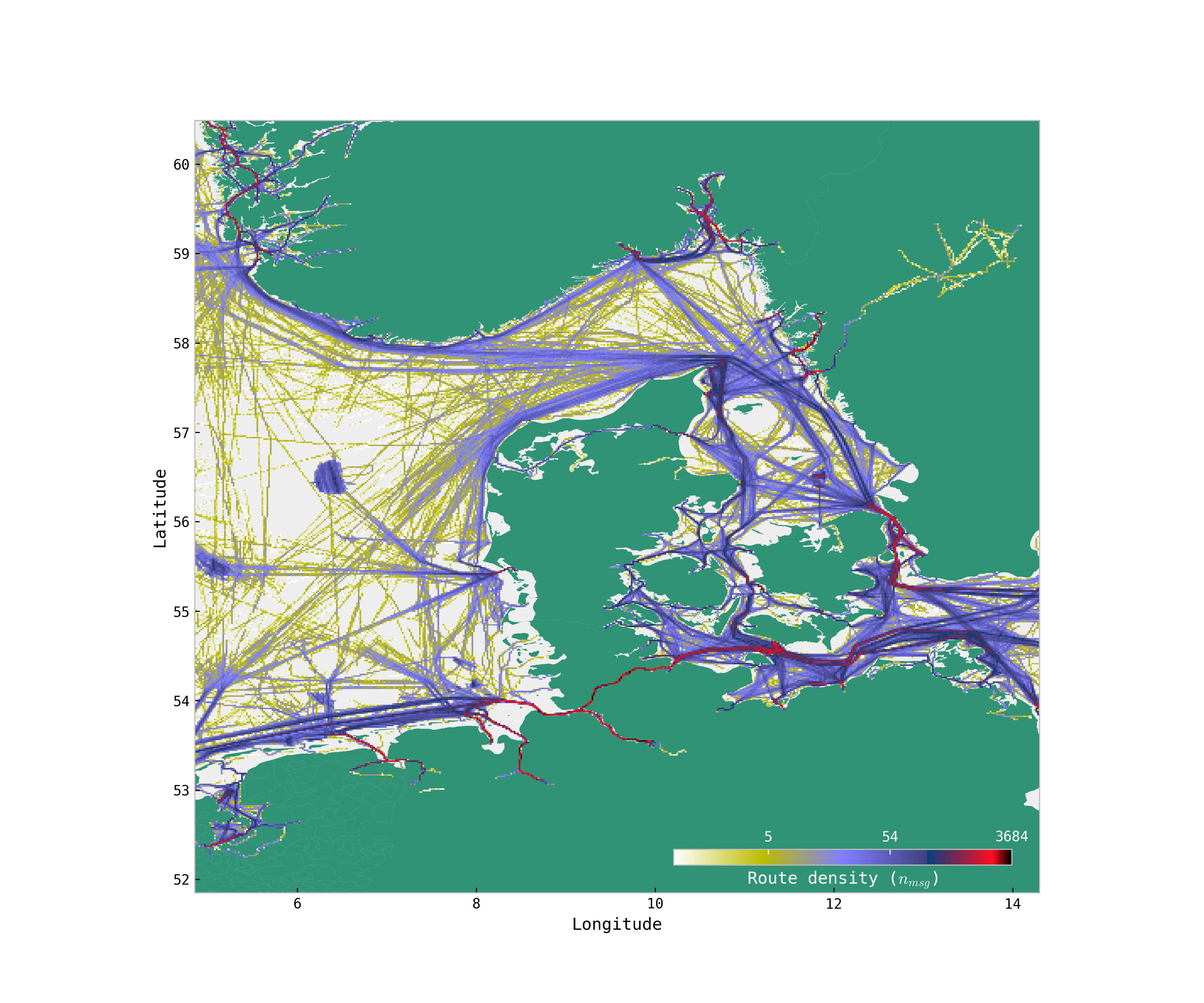}
      \caption{Tug Tow}
    \end{subfigure}
    \begin{subfigure}[t]{.49\textwidth}
      \centering
      \includegraphics[trim={1cm 1cm 1cm 1cm},clip,width=\linewidth]{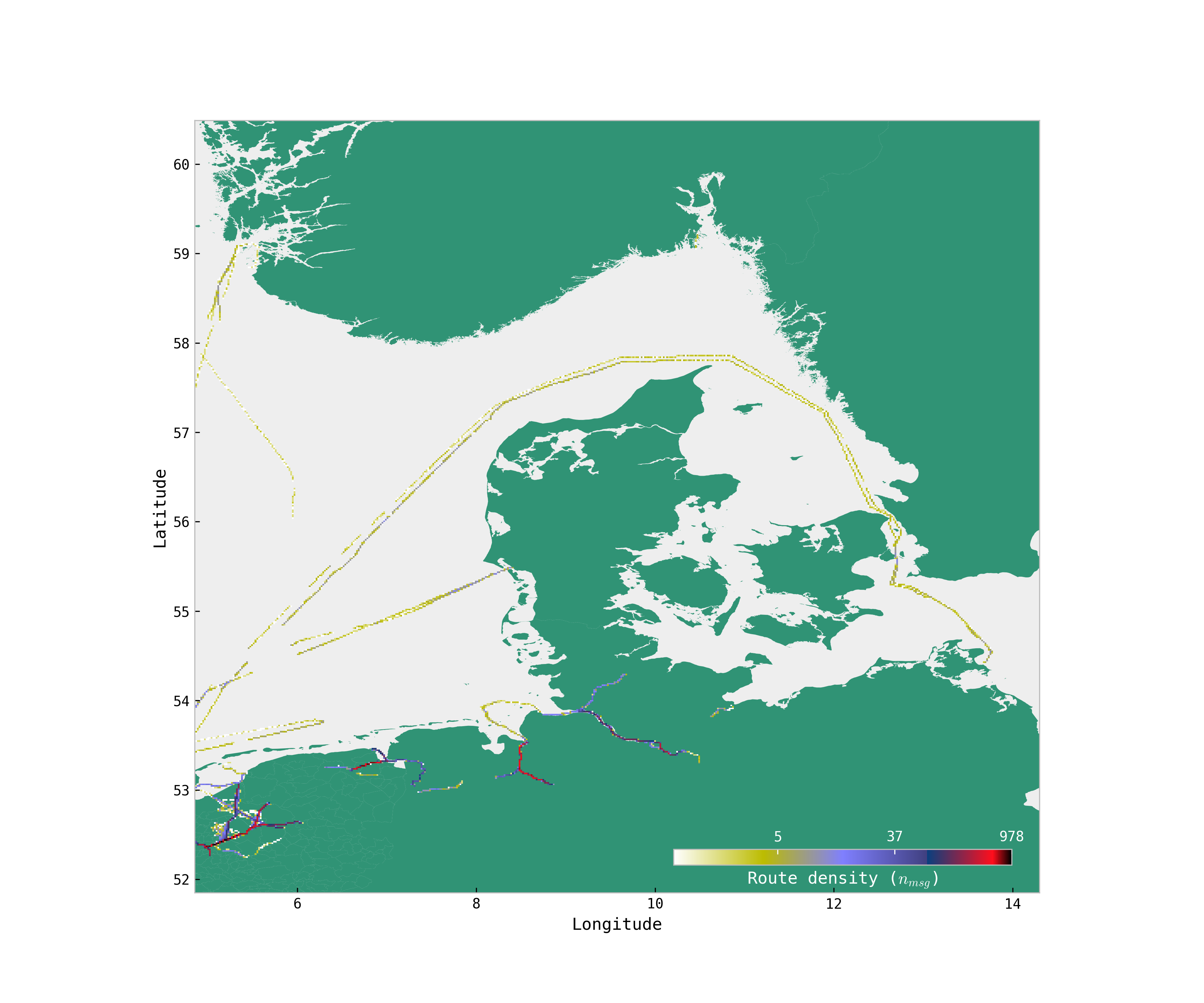}
      \caption{Wing in Ground (WIG)}
    \end{subfigure}
    \caption{Continuation of Figure \ref{heatmaps-1}}
    \label{heatmaps-2}
\end{figure}

\section{Time scale of empirical distributions}
\label{timescale}

\begin{figure}
    \centering
    \includegraphics[width = \textwidth]{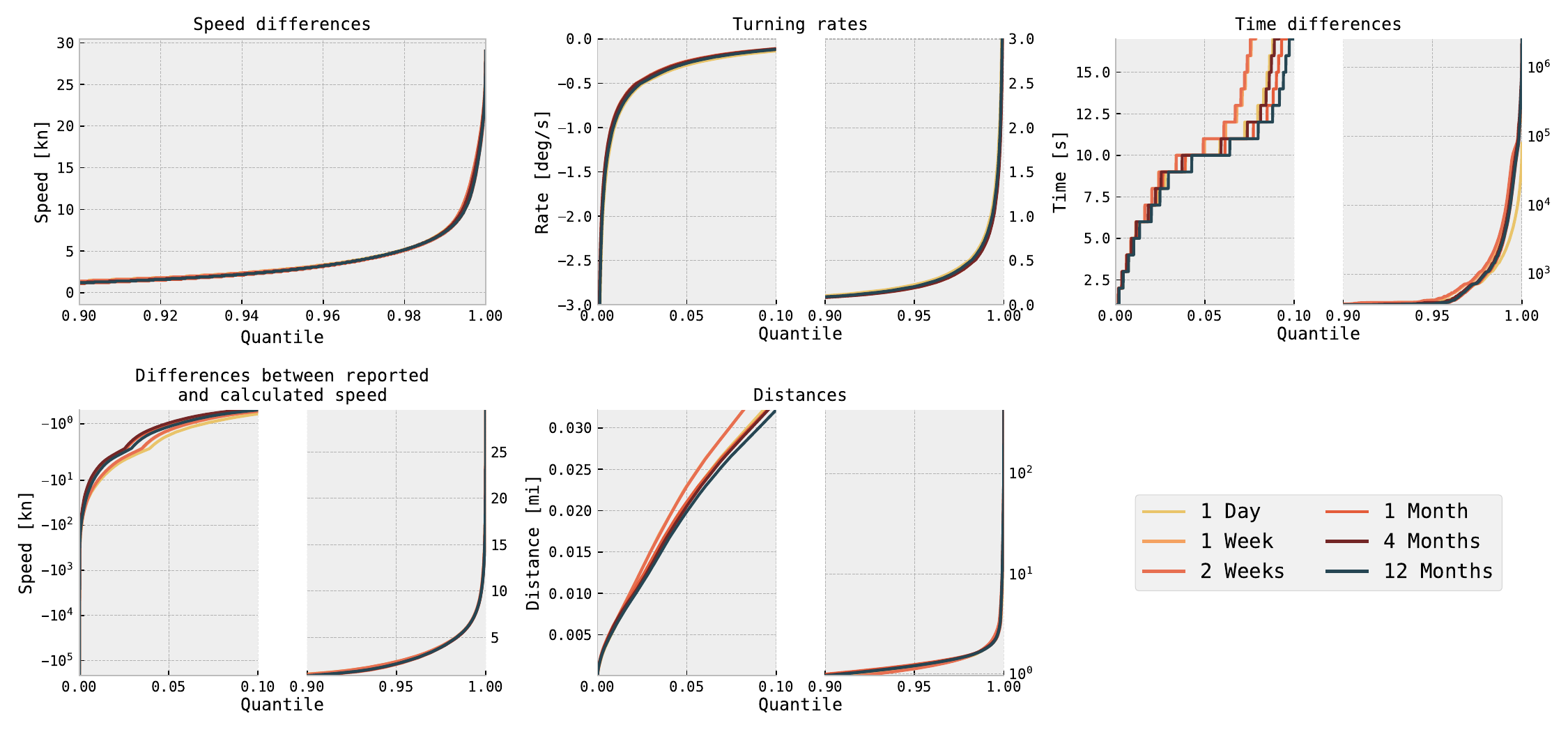}
    \caption{Comparison of quantile functions for the split-points metrics using different amounts of data. Please note the different ordinate and abscissa scalings.}
    \label{quantcomp}
\end{figure}

The split-point methodology outlined in Section \ref{splitpoints} requires the computation of five distinct empirical distributions, using their quantiles to establish thresholds. In this process, we explicitly used trajectory data from one year of AIS records as the basis for each distribution. To investigate how much data is needed to accurately calculate the empirical distribution functions \ref{quantcomp} displays the (ship-length-independent) quantile functions for all metrics, each calculated using different amounts of trajectory data (given as days and months). Visually, these distributions appear somewhat similar, leading to the assumption that the data follows identical distributions over various time horizons.

To substantiate this visual assumption, we compare the quantiles calculated from various time horizons against each other using the procedure outlined in \cite{wilcox2014comparing} using $10^6$ samples and $1000$ bootstrap replications each.
We deliberately chose this setup to test how much information is pertained when using less data compared to one-year data. Contrary to initial expectations, our tests indicate that the distributions of different time horizons differ. 

The computed quantile values in Table \ref{ksres} only show minor differences in absolute terms, which leaves leeway to the choice of the amount of data used to calculate the quantiles. However, due to the huge sample sizes (over $10^6$ observations) that lead to very narrow confidence intervals, almost all quantiles are tested statistically differently.  Ultimately, we use one year of data throughout the study to maximize generalizability.

Nonetheless, since variations may emerge for distributions derived from shorter or longer time spans, additional adjustments by practitioners are potentially necessary.

\begin{center}
\resizebox{\textwidth}{!}{%
\begin{tabular}{c|lcccclcccc}
\hline
\multirow{3}{*}{Metric} & \multicolumn{10}{c}{Quantiles used in Section \ref{splitpoints}} \\
 &  & \multicolumn{9}{c}{Upper and lower quantiles} \\ \cline{2-11} 
 & \multicolumn{1}{c}{} & \multicolumn{4}{c}{$q=0.025$} &  & \multicolumn{4}{c}{$q=0.975$} \\ \cline{3-6} \cline{8-11} 
\multicolumn{1}{l|}{} &  & $D=1$ & $D=7$ & $D=30$ & $D=120$ &  & $D=1$ & $D=7$ & $D=30$ & $D=120$ \\ \cline{1-1}
\begin{tabular}[c]{@{}c@{}}Difference between reported\\ and calculated speed $[kn]$\end{tabular} & \multicolumn{1}{c}{} & -3.577 & -3.149 & -2.866  & -4.078 & \multicolumn{1}{c}{} & 4.041  & 4.080 & 4.053 & 4.334  \\ \cline{1-1}
Turning rates $[{}^\circ / s]$ & \multicolumn{1}{c}{} & -0.507 & -0.466  & -0.466  & -0.491  & \multicolumn{1}{c}{} & 0.421 & 0.376  & 0.372 & 0.400  \\ \hline
\multicolumn{1}{l|}{\multirow{3}{*}{}} &  &  & \multicolumn{1}{l}{} & \multicolumn{1}{l}{} & \multicolumn{1}{l}{} &  & \multicolumn{4}{c|}{Only upper quantiles} \\ \cline{8-11} 
\multicolumn{1}{l|}{} &  &  &  &  &  &  & \multicolumn{4}{c|}{$q=0.95$} \\ \cline{8-11} 
\multicolumn{1}{l|}{} &  &  &  &  &  &  & $D=1$ & $D=7$ & $D=30$ & \multicolumn{1}{c|}{$D=120$} \\ \cline{1-1}
\begin{tabular}[c]{@{}c@{}}Time difference\\ between messages $[s]$\end{tabular} &  &  &  &  &  & \multicolumn{1}{c}{} & 397.0 & 420.0 & 430.0 & \multicolumn{1}{c|}{419.0}\\ \cline{1-1}
Speed changes $[m/s]$ &  &  &  &  &  &  & 2.800 & 2.800  & 2.799 & \multicolumn{1}{c|}{2.700} \\ \cline{1-1}
\begin{tabular}[c]{@{}c@{}}Distance between\\ consecutive messages $[nm]$\end{tabular} &  &  &  &  &  &  & 1.147  & 1.142  & 1.110 & \multicolumn{1}{c|}{1.123} \\ \cline{1-1}
\end{tabular}%
}
\captionof{table}{Quantile values for the metrics used in Section \ref{splitpoints}. $q$ is the value of the quantile, $D$ is the number of days of data used to obtain the quantiles.}
\label{ksres}
\end{center}

\section{Python package implementation details}
\label{implementation}
\begin{codelisting}[t]
    \centering
    \begin{lstlisting}[language=Python, numbers=left, frame=lines, basicstyle=\footnotesize]
    import pandas as pd
    import pytsa
    
    from pathlib import Path
    
    # Column name for SOG in the decoded 
    # csv data.
    _speed = ...
    
    # Filter out all vessels that are slower than 1 knot
    # and faster than 30 knots. 
    # Any pre-processor is mandated to take in a single
    # pandas DataFrame and return a single pandas DataFrame
    def speed_filter(df: pd.DataFrame) -> pd.DataFrame:
        return df[(df[_speed] > 1) & (df[_speed] < 30)]
    
    # Bounding Box for the area of study in this research
    frame = pytsa.BoundingBox(
        LATMIN = 52.2, # [°N]
        LATMAX = 56.9, # [°N]
        LONMIN = 6.3,  # [°E]
        LONMAX = 9.5,  # [°E]
    )
    # This can be either a single file or a 
    # list of files
    dynamic_data = Path("/path/to/dynamic.csv")
    static_data = Path("/path/to/static.csv")
    
    search_agent = pytsa.SearchAgent(
        msg12318file = dynamic_data,
        msg5file = static_data
        frame = frame,
        preprocessor = speed_filter # Apply the speed filter using the "preprocessor" keyword
    )
    \end{lstlisting}
    \caption{Basic instantiantion of the \texttt{SearchAgent} class with a custom pre-processing function.}
    \label{fig:instatiation}
\end{codelisting}

All pre-processing, filtering, extracting, splitting, assessment and visualization procedures outlined in this article are available as the open-source Python package \emph{PyTSA} (\textbf{Py}thon \textbf{T}rajectory \textbf{S}plitting and Assessment \textbf{A}gent, \citet{pytsa2024}). This Section is intended to provide implementation details and practitioners guidelines on how to use \emph{PyTSA} to reproduce the results presented in this research. For installation instructions, please refer to the package documentation.

\paragraph{Decoding raw AIS messages}
If the data to be analyzed is present as raw \texttt{AVIDM/AVIDO} sentences, they must first be decoded. To do this with \emph{PyTSA}, the input files must be provided as \texttt{.csv} files sorted such that one group of files only contains dynamic messages of types 1/2/3/18 and the other only messages of type 5 (static voyage reports). Both files must be named identically by the day for which they hold AIS messages, separated by underscores ("YYYY\_MM\_DD.csv").

Subsequently, decoding can be performed via

\begin{lstlisting}[language=Python, numbers=left, frame=lines, basicstyle=\footnotesize]
from pytsa import decode

decode(
    source = "path/to/raw_dir",
    dest = "path/to/decoded_dir",
    njobs = 1
)
\end{lstlisting}

which will take all \texttt{.csv} files in the "source" folder, decode them, and store the decoded files in the \texttt{.csv} format in the "dest" folder. For more information on the "njobs" keyword and the structure of the data files, the user is deferred to the package documentation.

\paragraph{Trajectory extraction using the \texttt{SearchAgent}}
\emph{PyTSA}'s central object, the \texttt{SearchAgent}, provides a convenient tool for constructing, filtering, and splitting trajectories extracted from the messages decoded earlier. The \texttt{SearchAgent} class is instantiated using at least three components:

\begin{enumerate}
    \item \texttt{mgs12318file} - A single file path or a list of file paths for dynamic messages
    \item \texttt{mgs5file} - A single file path or a list of file paths for static messages
    \item \texttt{frame} - A \emph{PyTSA} \texttt{BoundingBox} object determining the spatial extent of the extraction process.
\end{enumerate}

If, instead, it is desired for the data to undergo a pre-processing step, a pre-processing function that all messages must pass through before extraction can be defined. For example, to apply the speed constraint from Section \ref{velreps}, the instantiation can be performed as described in Code Listing \ref{fig:instatiation}.
After creating the instance, the \texttt{extract\_all()} method can be used to construct and, if desired, also split the trajectories according to the split-point procedure from Section \ref{splitpoints}. The method returns a Python dictionary with every unique MMSI as keys and their corresponding \texttt{TargetShip} objects as values. These objects contain information about the ship type and length and all trajectories identified as belonging to this MMSI. The method has an integrated switch to turn the split-point procedure on or off. If turned off, every returned \texttt{TargetShip} only contains a single trajectory, consisting of all AIS messages sent from its MMSI sorted by time.

We look at the code used to produce Figure \ref{applicationofsplitpointfilter} to demonstrate this approach's capabilities. We also utilize the visualization module shipped with \emph{PyTSA} for this approach.
\begin{lstlisting}[language=Python, numbers=left, frame=lines, basicstyle=\footnotesize]
#
# This listing continues Code Listing $\ref{fig:instatiation}$
#
# To perform only trajectory construction without
# using the split-point procedure, we use the 
# `skip_tsplit=True` option.
ships = search_agent.extract_all(njobs=16,skip_tsplit=True)

# Define the area around Aabenraa
AABENRAA = pytsa.BoundingBox(
    LATMIN = 54.9, # [°N]
    LATMAX = 55.5, # [°N]
    LONMIN = 9.2,  # [°E]
    LONMAX = 10.0,  # [°E]
)

# Plot trajectories on the map for the area
# around AABENRAA
plot_trajectories_on_map(
    ships = ships,
    extent = AABENRAA
)
    
# With split-point procedure applied
ships = search_agent.extract_all(njobs=4,skip_tsplit=False)
plot_trajectories_on_map(
    ships = ships,
    extent = AABENRAA
)
\end{lstlisting}

\paragraph{Assessing spatial properties of extracted trajectories}
With the help of the \texttt{SearchAgent} class, we could easily construct and split trajectories from raw AIS messages. The spatial assessment from Section \ref{spatial-properties} is implemented into \emph{PyTSA} via the \texttt{Inspector} class, which filters the extracted trajectories based on a set of user-defined \texttt{Rules}. All \texttt{Rules} must have a function signature of \lstinline[language=Python]|def my_rule(track: Track) -> bool|, where \lstinline[language=Python]|Track = list[AISMessage]| is a list of AIS message objects defined in \texttt{pytsa.structs}. \texttt{Rules} must be defined such that they take in a single trajectory and return \texttt{True} if the trajectory is to be \emph{rejected}. \texttt{Rules} must be combined to create a recipe given to the inspector for trajectory assessment. To showcase the functionality of the \texttt{Inspector} class and its usage with pre-defined \texttt{Rules}, we reconstruct the heatmaps from Figures \ref{heatmaps-1} and \ref{heatmaps-2} in the following code example.
\begin{lstlisting}[language=Python, numbers=left, frame=lines, basicstyle=\footnotesize]
#
# This listing continues Code Listing $\ref{fig:instatiation}$
#
# Construct and split trajectories in parallel
ships = search_agent.extract_all(njobs=4)

# Construct a Recipe object to assess every trajectory based on the rule functions
# given to it. In this case, we use the partial class to coerce the rules to only take a single argument, 
# which is the trajectory given to it. 
assessment = Recipe(
    partial(too_few_obs, n = 50),
    partial(convex_hull_area, area = 3e5)
)

# Instantiate the Inspector and split the trajectories into accepted and rejected ones.
# In this example, the rejected trajectories are discarded.
accepted, rejected = pytsa.Inspector(data = ships, recipe = assessment).inspect()

# Save a 500x500 px heatmap for every ship type
types = [t for t in pytsa.ShipType]
names = [t.name for t in types]
expanded = []
for t in types:
    if isinstance(t.value, int):
        expanded.append([t.value])
    else:
        expanded.append(list(t.value))
for i,t in enumerate(expanded):
    a = {mmsi:s for mmsi,s in accepted.items() if any(st in t for st in s.ship_type)}
    binned_heatmap(targets = a, bb = frame, npixels = 500, title = f"Heatmap for {names[i]}")

\end{lstlisting}

\end{document}